\documentclass[prd,twocolumn,nofootinbib,nobibnotes,noeprint,preprintnumbers,floatfix]{revtex4-2}
\usepackage{amssymb,amsmath,graphicx}
\usepackage{bm}
\usepackage[dvipsnames]{xcolor}
\usepackage[colorlinks,allcolors=NavyBlue]{hyperref}
\usepackage{microtype}
\usepackage{listings} 
\usepackage{natbib}
\usepackage{float}

\definecolor{codegreen}{rgb}{0,0.6,0}
\definecolor{codegray}{rgb}{0.5,0.5,0.5}
\definecolor{codepurple}{rgb}{0.58,0,0.82}
\definecolor{backcolour}{rgb}{0.95,0.95,0.92}

\lstdefinestyle{mystyle}{
    backgroundcolor=\color{backcolour},
    commentstyle=\color{codegreen},
    keywordstyle=\color{magenta},
    numberstyle=\tiny\color{codegray},
    stringstyle=\color{codepurple},
    basicstyle=\ttfamily\footnotesize,
    breakatwhitespace=false,
    breaklines=true,
    captionpos=b,
    keepspaces=true,
    numbers=left,
    numbersep=5pt,
    showspaces=false,
    showstringspaces=false,
    showtabs=false,
    tabsize=2
}
\usepackage{aas_macros}

\begin{document}

\preprint{}

\title{A Multimessenger Analysis of the High-Energy Milky Way: Source Populations Contribute Significantly to IceCube's Galactic Neutrino Flux}

\author{Alisha Roberts$^{1}$}
\thanks{\scriptsize anroberts2@wisc.edu, 
https://orcid.org/0000-0003-1498-2888}

\author{Ilias Cholis$^{2}$}
\thanks{\scriptsize cholis@oakland.edu, http://orcid.org/0000-0002-3805-6478}

\author{Dan Hooper$^{1}$}
\thanks{\scriptsize dwhooper@wisc.edu, http://orcid.org/0000-0001-8837-4127}

\author{Samyak Jain$^{1}$}
\thanks{\scriptsize samyak@icecube.wisc.edu, 
https://orcid.org/0009-0000-7455-782X}

\affiliation{$^1$University of Wisconsin-Madison, Department of Physics, Madison, Wisconsin 53706, USA}
\affiliation{$^2$Oakland University, Department of Physics, Rochester, Michigan 48309, USA}

\date{\today}

\begin{abstract}

We perform a joint analysis of the high-energy neutrino emission observed from the Galactic Plane by IceCube and the diffuse ultra-high-energy gamma-ray emission measured by LHAASO. We compare this data to models that include diffuse emission from cosmic-ray interactions in the interstellar medium, unresolved TeV halos, and unresolved Galactic neutrino sources. We find that the gamma-ray emission can be explained by a combination of diffuse processes and unresolved TeV halos. The observed neutrino emission cannot be generated by cosmic-ray interactions in the interstellar medium alone, but requires contributions from one or more unresolved source populations. Across a wide range of assumptions about Galactic cosmic-ray transport, we find that Galactic neutrino sources contribute significantly to the neutrino flux observed from the Galactic Plane and are likely responsible for most of this emission.

\end{abstract}

\maketitle

\section{Introduction}

In 2023, the IceCube Collaboration reported the first detection of neutrino emission from the Galactic Plane~\cite{icecube_collaboration_observation_2023}. These neutrinos, with energies in the range of $E_{\nu} \sim 1-100 \,{\rm TeV}$, are thought to be produced through cosmic-ray (CR) interactions with gas in the interstellar medium (ISM), along with possible contributions from individual CR accelerators, including supernova remnants~\cite{Drury:1993pd,Alvarez-Muniz:2002con,Kappes:2006fg,Ahlers:2013xia,Ahlers:2009ae}, pulsar wind nebulae~\cite{Bednarek:1997cn,Amato:2003kw,Bednarek:2003cv,Guetta:2002hv,Liu:2019iga}, and microquasars~\cite{Romero:2005fr,Distefano:2002qw,Levinson:2001as,Bednarek:2005gf}. The data currently available from IceCube cannot, on its own, distinguish between scenarios in which individual sources or diffuse processes produce the majority of the neutrinos observed from the Galactic Plane (see, for example, Ref.~\cite{Desai:2023cmq}).

Over a similar range of energies, $E_{\gamma} \sim 10 -1000 \, {\rm TeV}$, the Large High Altitude Air Shower Observatory (LHAASO) has measured the diffuse gamma-ray emission from the Galactic Plane~\cite{cao_measurement_2023} (see also, Refs.~\cite{HAWC:2023wdq,HESS:2014ree}). This flux of gamma rays exceeds that predicted from CR interactions in the ISM~\cite{Zhang_2023}, and thus appears to require contributions from individual sources, such as the TeV halos~\cite{dekker_diffuse_2024} that are known to surround young and middle-aged pulsars~\cite{Hooper:2017gtd,Sudoh:2019lav,HAWC:2021dtl,sudoh_highest_2021,Linden_2017}.

In this study, we consider the high-energy neutrino and gamma-ray emission observed from the Galactic Plane and attempt to constrain the origin of these particles. To this end, we employ a series of models describing CR transport in the ISM, unresolved neutrino sources, and the gamma-ray emission from unresolved TeV halos. Comparing the predictions of these models to the data collected by IceCube and LHAASO, we identify viable scenarios in which the Galactic diffuse ultra-high-energy gamma-ray flux is dominated by diffuse processes, and others in which this emission is dominated by unresolved TeV halos, especially in the direction of the inner Galactic Plane. Across all of the models considered, however, we find a consistent statistically significant ($\Delta \chi^2 = 25.5-38.2$) preference for neutrino emission from one or more source populations, in addition to the emission from CR interactions in the ISM. For our best-fit model parameters, these sources generate $\sim$60-90\% of the total neutrino emission observed from the Galactic Plane, and we can constrain this fraction to be $\gtrsim 20\%$ at the 95\% confidence level. We thus conclude that unresolved sources are responsible for a significant fraction of the neutrino emission observed by IceCube from the Galactic Plane (for earlier work in this direction, see Refs.~\cite{Gagliardini:2024een,Vecchiotti:2023ill,Fang:2023ffx,Fang:2023azx,Lipari:2026kpk,Prevotat:2025ktr,DeLaTorreLuque:2025zsv,Lipari:2024pzo}).

The remainder of this paper is structured as follows. In Sec.~\ref{sec:Modeling}, we describe our models for CR transport and interactions in the ISM, for unresolved TeV halos, and for unresolved Galactic sources of high-energy neutrinos. In Sec.~\ref{sec:Results}, we fit our model to the combined data of IceCube and LHAASO and present constraints on the respective model parameters. In Sec.~\ref{Sec:Summary} we summarize our results and their implications.

\section{Neutrino and Gamma-Ray Emission from the Galactic Plane}
\label{sec:Modeling}

\subsection{Cosmic-Ray Transport and Diffuse Emission}
\label{subsec:CR_propagation_models}

\begin{table*}[t]
    \begin{tabular}{ccccccccccc}
         \hline
           Cosmic-Ray & $\delta$ & $z_{L}$ & $D_{0} \times 10^{28}$ & $v_{A}$ & $dv_{c}/d|z|$
              & $\alpha_{1}$ & $R_{br_{1}}$ & $\alpha_{2}$ & $R_{br_{2}}$  & $\alpha_{3}$\\
               Model & &  (kpc) & (cm$^2$/s) & (km/s) & (km/s/kpc) & H/He & H/He (GV) & H/He & H/He (GV) & H/He\\
            \hline \hline
            A  & 0.33 & 5.7 & 6.70 & 30.0 & 0.0 & 1.74/1.70 & 6.0/7.4 & 2.04/2.16 & 14.0/21.5 & 2.41/2.39 \\
            B &  0.37 & 5.5 & 5.50 & 30.0 & 2.0 & 1.72/1.74 & 6.0/8.0 & 2.00/2.14 & 12.4/21.0 & 2.38/2.375 \\
            C &  0.40 & 5.6 & 4.85 & 24.0 & 1.0 & 1.69/1.65 & 6.0/6.7 & 2.00/2.13 & 12.4/20 & 2.38/2.355 \\
            D &  0.45 & 5.7 & 3.90 & 24.0 & 5.5 & 1.69/1.68 & 6.0/7.0 & 1.99/2.12 & 12.4/18.7 & 2.355/2.34 \\
            E &  0.50 & 6.0 & 3.10 & 23.0 & 9.0 & 1.71/1.68 & 6.0/7.2 & 2.02/2.14 & 11.2/17.5 & 2.38/2.33 \\
            F &  0.43 & 3.0 & 1.85 & 20.0 & 2.0 & 1.68/1.74 & 6.0/10.5 & 2.08/2.09 & 13.0/21.0 & 2.41/2.33 \\
            \hline
        \end{tabular}
    \caption{The parameters for each of the six cosmic-ray transport models adopted in this study. These models span a wide range of physical assumptions and yield predictions that are in good agreement with measurements of the cosmic-ray spectrum by AMS-02 and Voyager 1, as well as with GeV-scale gamma-ray measurements along the Galactic Disk and Inner Galaxy by Fermi. For further details, see the main text and Ref.~\cite{cholis_return_2022}.}    
    \label{tab:ISM_models}
\end{table*}

The interactions of CR protons with gas in the ISM produce charged pions, and thus neutrinos through their subsequent decays. Gamma rays are produced by CR protons via neutral pion production, as well as by CR electrons through the processes of inverse Compton scattering and bremsstrahlung.

To model the propagation and interactions of CRs in the Milky Way, we use the publicly available \texttt{GALPROP} code \cite{porter_galprop_2022}, which numerically solves the differential equations governing CR transport in the Galaxy, including the processes of diffusion, diffusive reacceleration, convection, energy losses, spallation, radioactive decay, and secondary particle production.

In this study, we consider a set of six CR propagation models whose parameters are given in Table~\ref{tab:ISM_models}, and which were originally introduced in Ref.~\cite{cholis_return_2022}. These models span a broad range of physical assumptions and were constructed to yield predictions that are in good agreement with measurements of the CR spectrum by AMS-02, including those of protons, helium, carbon, oxygen, beryllium-to-carbon, boron-to-carbon, and carbon-to-oxygen~\cite{AMS:2015azc, AMS:2017seo, AMS:2018tbl}, as well as measurements by Voyager 1~\cite{Cummings:2015qvx, Potgieter:2013mcc}. 

In the ISM (excluding those regions surrounding TeV halos), we take CR diffusion to be isotropic and homogeneous and parameterize it in terms of the following diffusion coefficient:
\begin{equation}
    \label{eqn:CRdiffusion}
    D(R) = \beta D_{0} \left(\frac{R}{4 \textrm{GV}}\right)^{\delta},
\end{equation}
where $R$ is the CR rigidity, $\delta$ is the diffusion index, and $D_{0}$ normalizes the diffusion coefficient at $R=4 \, {\rm GV}$. We take the diffusion zone to be a cylinder of radius of 20 kpc and half-height, $z_L$. Outside of this region, CRs are not magnetically confined and are thus allowed to freely escape from the Galaxy. The processes of diffusive reacceleration and convection are characterized by the parameters $v_A$ and $dv_c/d|z|$, respectively. Finally, the injected CR spectra are described by the spectral indices $\alpha_1$, $\alpha_2$, and $\alpha_3$, corresponding to the energy ranges below, between, and above spectral breaks at rigidities $R_{br_1}$ and $R_{br_2}$, respectively. In each model, these spectral parameters were chosen to best reproduce the measured spectra of CR protons and helium nuclei.

The six CR transport models adopted in this study were selected to encompass the dominant uncertainties associated with the Galactic CR population. In addition to reproducing the CR spectrum, these models have been shown to provide good agreement with GeV-scale gamma-ray measurements of the Galactic Disk and Inner Galaxy, as made by the Fermi Large Area Telescope~\cite{cholis_return_2022}. For further details, see Ref.~\cite{cholis_return_2022}.

To obtain energy-dependent maps of the gamma-ray emission from pion production in the ISM, we take the CR distribution predicted by each of the six models described here, convolve them with the measured gas distribution, and integrate the predicted emissivity over the line-of-sight. In the TeV-PeV energy range, the diffuse gamma-ray emission is dominated by $\pi^0$ production, with inverse Compton scattering and bremsstrahlung contributing at only the $\mathcal{O}(10^{-2})$ level. In our calculations we take into account all CR species up to and including nickel.

While \texttt{GALPROP} can generate energy-dependent maps of gamma-ray emission from pion production, inverse Compton scattering, and bremsstrahlung, it does not produce maps of the corresponding neutrino emission. To produce such maps, we take the component of gamma-ray emission from neutral pion production as predicted by \texttt{GALPROP} (in units of $E^2 dN/dE$), shift it downward in energy by a factor of 0.57 and increase its normalization by a factor of 1.1. This procedure yields results that are in good agreement with the all-flavor neutrino spectrum parameterized in Ref.~\cite{kelner_energy_2006}.

\begin{figure*}[ht]
\centering
    \includegraphics[width = 0.48\linewidth]{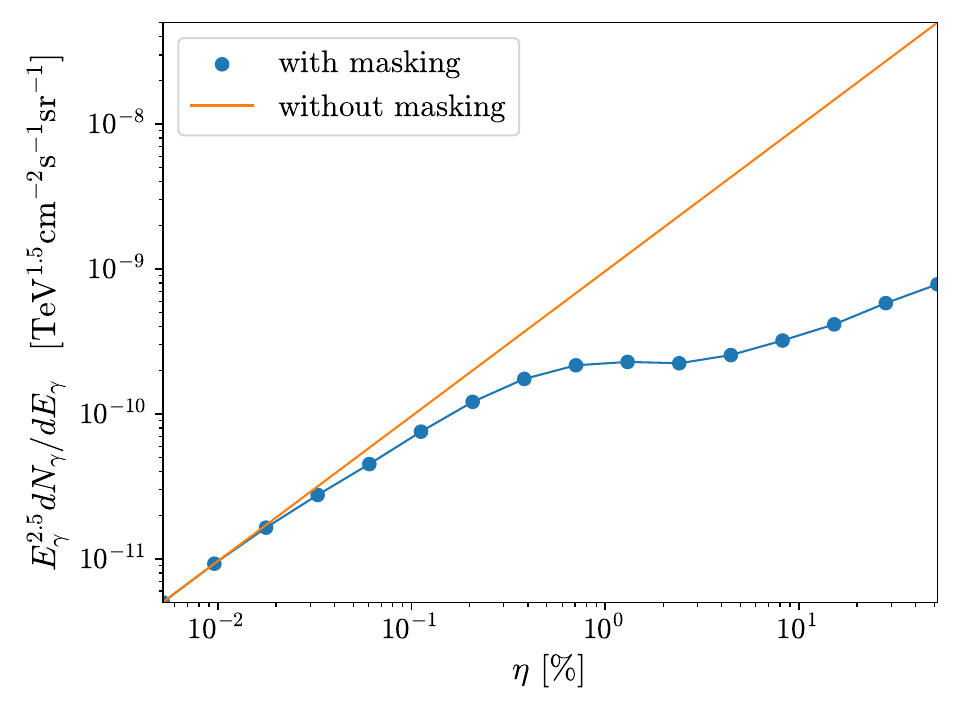}
    \hspace{0.2in}
  \includegraphics[width = 0.48\linewidth]{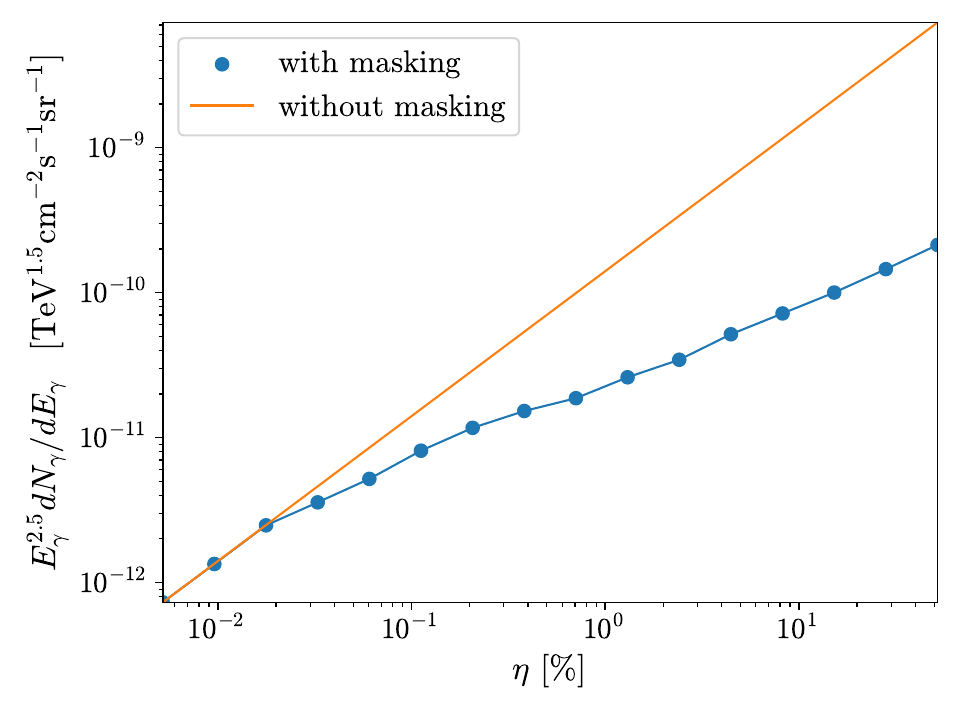}
\caption{The diffuse gamma-ray flux (evaluated at $E_{\gamma}= 10 \, {\rm TeV}$) from TeV halos, before and after masking, as a function of the gamma-ray efficiency, $\eta$, and for the case of $\alpha_{\gamma}=3.1$. Results in the left and right frames correspond to the inner and outer regions of the Galactic Plane, respectively. For details, see Sec.~\ref{subsec:TeV halo model}. }
\label{efficiency_masking_fig}
\end{figure*}

\subsection{Unresolved TeV Halos}
\label{subsec:TeV halo model}

Observations by HAWC~\cite{HAWC:2020hrt,abeysekara_extended_2017,HAWC:2024scl}, LHAASO~\cite{LHAASO:2023rpg,LHAASO:2021crt,LHAASO:2024flo}, and HESS~\cite{HESS:2018pbp,HESS:2017lee} have identified bright, multi-TeV emission from the regions surrounding young and middle-aged pulsars~\cite{Hooper:2017gtd,Sudoh:2019lav,HAWC:2021dtl,sudoh_highest_2021,Linden_2017}. The spectra and intensity of the observed TeV halos indicate that on the order of $\sim 10\%$ of their pulsars' spindown power is converted into very high-energy electron-positron pairs, which go on to produce gamma-ray emission through the process of inverse Compton scattering. These sources are typically $\sim 20 \, {\rm pc}$ in extent, indicating that CR propagation is far less efficient in the regions surrounding these pulsars than elsewhere in the ISM~\cite{Hooper:2017gtd,Hooper:2017tkg,Johannesson:2019jlk,DiMauro:2019hwn}. In Ref.~\cite{dekker_diffuse_2024}, it was argued that TeV halos are likely to provide the dominant contribution to the very-high-energy diffuse gamma-ray emission from the direction of the inner Galactic Plane. As we will show here, however, this conclusion is somewhat sensitive to the CR transport model that is adopted.

To model the Milky Way's TeV halo population and their contribution to the diffuse gamma-ray flux, we utilize the Monte Carlo simulations introduced in Ref.~\cite{dekker_diffuse_2024}. In this, we
take the gamma-ray luminosity of a given TeV halo to be
proportional to the corresponding pulsar's spindown power, $L_\gamma = \eta \dot{E}_{\rm rot}$, where $\eta$ is the gamma-ray efficiency (we define $L_{\gamma}$ as the gamma-ray luminosity integrated above 1 TeV). The spindown power of a pulsar can be expressed as 
\begin{align}
    \label{eqn:spindown}
    \dot{E}_{\rm rot} &= -\frac{4\pi^2I\dot{P}}{P^3}\nonumber\\&= -\frac{4\pi^2I}{P_0^2(n-1)\tau}\left( \frac{t}\tau +1\right)^{-(n+1)/(n-1)},
\end{align}
where $I$ is the neutron star's moment of inertia, $P$ is its period, $P_0$ is its period at birth, and $\tau = 3c^3IP_0^2/4\pi^2B^2r^6$ is its spindown timescale, in which $B$ and $r$ are the neutron star's surface magnetic field strength and radius, respectively. The pulsar's braking index, $n$, is defined as $\dot{P} \propto P^{-n+2}$. Throughout this study, we adopt $n=3$, corresponding to the case in which spindown is dominated by magnetic dipole breaking.

To produce a simulated population of TeV halos, we assign each pulsar a random age (between $t =  2 \times 10^4  \text{ and } 10^8 $ years) and a spindown power given by $\dot{E}_{\rm rot} = -1.1 \times 10^{34} \, {\rm erg/s} \times (t/10^6 \, {\rm yr})^2$~\cite{bitter_constraining_2022}. We normalize the Milky Way's pulsar population to a birth rate of 0.7 per century and draw from the following spatial distribution~\cite{lorimer_galactic_2003}:
\begin{equation}
    n_{\rm pulsar} \propto R^{2.35} \, e^{-R/1.53 \,\text{kpc}} \, e^{- \vert z \vert/z_s},
\end{equation}
where $R$ and $z$ are the location in cylindrical coordinates, and the vertical scale height is taken to be 
$z_s = 70 \, {\rm pc} + 180 \, {\rm pc} \times (t/10^6 \, {\rm yr})$, up to a maximum value of 
1 kpc~\cite{dekker_diffuse_2024}. This expression for $z_s$ is intended to account for the underlying distribution of progenitors~\cite{2004A&A...425.1009M} 
and the impact of natal kicks~\cite{Hansen:1997zw}. 
We take the Solar System to be located at $R=8.12$ and $z=0.021$ kpc~\cite{collaboration_detection_2018, bennett_vertical_2019}. 
We parameterize the gamma-ray spectrum of these sources in terms of a power-law, 
$dN_\gamma/dE_\gamma \propto E_{\gamma}^{-\alpha_{\gamma}}$, and take each TeV halo to have a radius of 10 pc~\cite{abeysekara_extended_2017}. 
In the energy range measured by LHAASO, individual TeV halos have been observed to feature spectral indices in the range of 
$\alpha_{\gamma} \sim 2-3$~\cite{collaboration_multiple_2020,sudoh_highest_2021,HAWC:2024scl}; for this reason, 
we limit the value of this parameter to $\alpha_{\gamma} \le 3.5$ in our fits. 

At the highest energies of interest, gamma rays can be attenuated via pair production. We account for this scattering, including with the cosmic microwave background, starlight, and Galactic infrared emission~\cite{dekker_diffuse_2024}. Following Ref.~\cite{dekker_diffuse_2024}, we mask the brightest of the simulated TeV halos, removing regions of radius $2.5 \times (\sigma_{\rm halo}^2 +\sigma_{\rm PSF}^2)^{1/2}$ around any TeV halos whose flux at 10 TeV exceeds $5\times 10^{-15} \, {\rm TeV}^{-1} {\rm cm}^{-2} {\rm s}^{-1}$, where $\sigma_{\rm halo}$ is the angular radius of the TeV halo and $\sigma_{\rm PSF}=0.6^{\circ}$ is the point spread function of LHAASO. This procedure yields results that are consistent with the data provided in Ref.~\cite{cao_measurement_2023}. 

The masking is found to have a significant impact on our fits. In Fig.~\ref{efficiency_masking_fig}, we show, for a value of $\alpha_{\gamma} = 3.1$, how the contribution from TeV halos to the Galactic diffuse flux varies as a function of $\eta$. Due to this masking, we obtain similar contributions from TeV halos for a relatively wide range of $\eta$.

\subsection{Unresolved Neutrino Point Sources}
\label{subsec:PS model}

Whereas many Galactic sources of TeV-PeV gamma rays have been detected, no such neutrino sources have been identified to date. This is, in part, due to the significantly smaller number of neutrinos that have been detected, as well as to IceCube's limited angular resolution to neutrino-induced cascades, $\Delta \theta \sim 5-20^{\circ}$ (see Fig.~S5 of Ref.~\cite{icecube_collaboration_observation_2023}). These factors leave open the possibility that many of the neutrinos detected from the Galactic Plane could originate from unresolved point sources, such as supernova remnants, pulsar wind nebulae, or microquasars.

As we did with the gamma-ray emission from TeV halos, we parameterize the spectrum of the neutrino emission from unresolved point sources as a power-law, $dN_{\nu}/dE_{\nu} =  A_{\nu} \times (E_{\nu}/100\,{\rm TeV})^{-\alpha_{\nu}}$. Following Ref.~\cite{verberne_2021}, we adopt the following model for the spatial distribution of Galactic neutrino sources:
\begin{equation}
    n \propto e^{-R/3.35 \, {\rm kpc}} \, e^{ -\vert z\vert/0.2 \, {\rm kpc}}, 
\end{equation}
where $R$ and $z$ once again define the location in cylindrical coordinates. The IceCube Collaboration has reported the neutrino flux from three regions of the Milky Way, corresponding to the inner ($25^{\circ} < l < 100^{\circ}$), outer ($50^{\circ} < l < 200^{\circ}$), and entire ($0^{\circ} < l < 360^{\circ}$) Galactic Plane (each with $|b| < 5^{\circ}$). After integrating the source distribution described above over the line-of-sight, these three regions yield neutrino fluxes (per solid angle) with relative proportions of 1.66, 0.62, and 1.0, respectively.

\section{Results}
\label{sec:Results}

In this section, we compare the predictions of our model to the data reported by the IceCube and LHAASO Collaborations. For the neutrino data, we use the Galactic Plane emission as reported in Ref.~\cite{icecube_collaboration_observation_2023}. These results are presented for fits to three spatial/spectral templates, known as the $\pi^0$, KRA$_\gamma^{5}$, and KRA$_\gamma^{50}$ models. While each of these templates was derived from gamma-ray observations, they differ in their spatial distribution and spectrum. Namely, the $\pi^0$ template traces the gamma-ray template from neutral pion production and adopts a power-law spectrum of the form, $dN_{\nu}/dE_{\nu} \propto E_{\nu}^{-2.7}$~\cite{ackermann_fermi-lat_2012}. The KRA$_\gamma$ (Kraichnan) models, in contrast, feature spectral variations across different regions of the sky with an average spectrum of approximately $dN_{\nu}/dE_{\nu} \propto E_{\nu}^{-2.5}$ up to a maximum energy associated with the limit of CR acceleration, which is taken to be either 5 or 50 PeV in these two models, respectively~\cite{gaggero_gamma-ray_2015}. In broad terms, the KRA$_{\gamma}$ models predict more neutrino emission from the inner Galactic Plane, while the $\pi^0$ model yields a somewhat more uniform distribution of neutrinos from throughout the Plane~\cite{icecube_collaboration_observation_2023}. Throughout this study, we take IceCube's measurement based on the $\pi^0$ template to be our default model, but also consider the KRA$_\gamma^{5}$ and KRA$_\gamma^{50}$ models in Appendix~\ref{sec:AppendixA} of this paper.

For the diffuse ultra-high-energy gamma-ray emission, we use measurements provided by the LHAASO Collaboration, which cover the energy range of 10 TeV to 1 PeV~ \cite{cao_measurement_2023}. In that analysis, the regions surrounding bright point-like and extended gamma-ray sources were masked, and thus the reported flux represents the contribution from diffuse emission and faint (i.e.~unresolved) sources.

\begin{table*}[t]
\centering
\begin{tabular}{c | c || ccc | cccc | c}
\hline\hline
 {Cosmic-Ray} & \texttt{GALPROP} Normalization & \multicolumn{3}{c|}{TeV Halos} & \multicolumn{4}{c|}{Neutrino Sources} & Total \\
{Model} & $A_{\rm G}$ & $\eta$ (\%) & $\alpha_{\gamma}$ & $\chi^2_{\gamma}$ & $A_{\nu}$ & $\alpha_{\nu}$  & $\chi^2_{\nu}$ & $\Delta \chi^2_{\nu}$ &  $\sum \chi^2$ \\
\hline
 A & $1.62^{+0.35}_{-0.12}$ & $6.4^{+2.4}_{-6.1}$ & $3.50^{+0.00}_{-0.11}$ & $22.06$ & $9.95^{+4.91}_{-3.90} \times 10^{-9}$ & $2.72^{+0.15}_{-0.15}$  & $3.72$ & $38.23$ &  \bf{$25.78$} \\
 B & $1.98^{+0.19}_{-0.30}$ & $0.5^{+7.0}_{-0.3}$ & $3.50^{+0.00}_{-0.13}$ & $23.50$ & $9.41^{+5.04}_{-3.85} \times 10^{-9}$ &$2.72^{+0.16}_{-0.16}$  & $3.32$ & $33.89$ &  \bf{$26.83$} \\
 C & $2.55^{+0.20}_{-0.25}$ & $0.4^{+0.4}_{-0.2}$ & $3.50^{+0.00}_{-0.16}$ & $23.48$ & $9.41^{+5.24}_{-4.00} \times 10^{-9}$ &$2.71^{+0.17}_{-0.16}$  & $3.10$ & $31.80$ &  \bf{$26.58$} \\
 D & $2.97^{+0.27}_{-0.27}$ & $0.3^{+0.3}_{-0.2}$ & $3.50^{+0.00}_{-0.23}$ & $25.41$ & $9.41^{+5.44}_{-4.00} \times 10^{-9}$ &$2.70^{+0.17}_{-0.17}$  & $2.95$ & $29.87$ &  \bf{$28.35$} \\
 E & $4.00^{+0.00}_{-0.20}$ & $0.1^{+0.1}_{-0.0}$ & $2.68^{+0.36}_{-0.06}$ & $28.37$ & $1.05^{+0.56}_{-0.42} \times 10^{-8}$ &$2.68^{+0.16}_{-0.17}$  & $3.22$ & $32.10$ &  \bf{$31.59$} \\
 F & $3.72^{+0.28}_{-0.31}$ & $0.3^{+0.3}_{-0.2}$ & $3.50^{+0.00}_{-0.18}$ & $22.24$ & $9.68^{+5.60}_{-4.03} \times 10^{-9}$ &$2.69^{+0.17}_{-0.17}$  & $3.17$ & $30.39$ &  \bf{$25.41$} \\
 \hline
 E$'$ & $5.08^{+0.45}_{-0.78}$ & $0.2^{+0.2}_{-0.1}$ & $3.50^{+0.00}_{-0.77}$ & $26.30$ & $9.68^{+6.02}_{-4.26} \times 10^{-9}$ & $2.66^{+0.18}_{-0.19}$  & $2.53$ & $25.52$ &  \bf{$28.82$} 
 \\
\hline\hline
\end{tabular}

\caption{The results of our joint likelihood, in which we fit the parameters of our model to the high-energy neutrino and ultra-high-energy gamma-ray fluxes measured from the Galactic Plane. Results are shown for each of the cosmic-ray transport models described in Table~\ref{tab:ISM_models}. We have restricted the normalization of the Galactic diffuse emission to $0.25 \le A_G \le 4$, except in the case of Model E$^{\prime}$, for which this prior requirement was not imposed. We have also required $\alpha_{\gamma} \le 3.5$, consistent with the spectral indices observed from individual TeV halos. The errors quoted are at the $1\sigma$ confidence level (corresponding to $\Delta \chi^2=1$). The quantity $\chi^2_\gamma$ is the total $\chi^2$ obtained for our model from LHAASO's measurements of the inner and outer Galactic Plane, evaluated at the best-fit values of $A_G$, $\eta$, and $\alpha_{\gamma}$. Likewise, $\chi^2_{\nu}$ is the $\chi^2$ obtained from IceCube's measurements of the inner and outer Galactic Plane, evaluated at the best-fit values of $A_G$, $A_{\nu}$, and $\alpha_{\nu}$. The quantity $\Delta \chi^2_{\nu}$ represents the change in the value of $\chi^2$ that results if neutrino sources are not included in the fit, reflecting the strong statistical preference for neutrino emission from one or more source populations. The rightmost column provides the sum of $\chi^2_{\gamma}$ and $\chi^2_{\nu}$.}
\label{tab:Ag4results}
\end{table*}

For each of the CR transport models described in Table~\ref{tab:ISM_models}, we perform a fit to the LHAASO data, varying the gamma-ray efficiency ($\eta$) and spectral index ($\alpha_{\gamma}$) of the TeV halos, as well as the normalization of the Galactic diffuse emission, as parameterized by the quantity $A_G$ (such that $A_G=1$ corresponds to the default predictions of \texttt{GALPROP}). We then proceed to fit the normalization ($A_{\nu}$) and spectral index ($\alpha_{\nu}$) of the unresolved neutrino sources to the IceCube data. Note that $A_G$ is intended to account for the fact that, while the local CR spectrum has been used to calibrate \texttt{GALPROP}, the CR densities in other regions of the Milky Way could be somewhat higher or lower than inferred from these measurements. In our default analysis, we allow $A_G$ to take on values between 0.25 and 4.0. As described below, however, we will also consider larger values of $A_G$ in the case of Model E$'$. 

In performing our fits, we calculate the total $\chi^2$ of our model to the gamma-ray measurements of the inner ($\vert b \vert <5^\circ$, $15^\circ < l < 125^\circ$) and outer ($\vert b \vert <5^\circ$, $125^\circ < l < 235^\circ$) Galactic Plane, as reported by the LHAASO Collaboration~\cite{cao_measurement_2023}, including their reported systematic and statistical uncertainties. We also calculate the total $\chi^2$ to the neutrino measurements of the inner ($\vert b\vert <5^\circ$, $25^\circ < l < 100^\circ$) and outer ($\vert b\vert <5^\circ$, $50^\circ < l < 200^\circ$) Galactic Plane, as reported by IceCube~\cite{icecube_collaboration_observation_2023}. As the neutrino flux from the Galactic Plane has been presented as fitted uncertainty bands rather than error bars, we treat this information as a measurement in two energy bins of equal width in $\log E_{\nu}$ in each of the inner and outer regions of the Galactic Plane. Although these two regions of the Galactic Plane overlap, we treat them as independent measurements, as the IceCube Collaboration has not yet provided this data in a more useful form. 

\begin{figure*}[ht]
\includegraphics[width=0.49\textwidth, keepaspectratio]{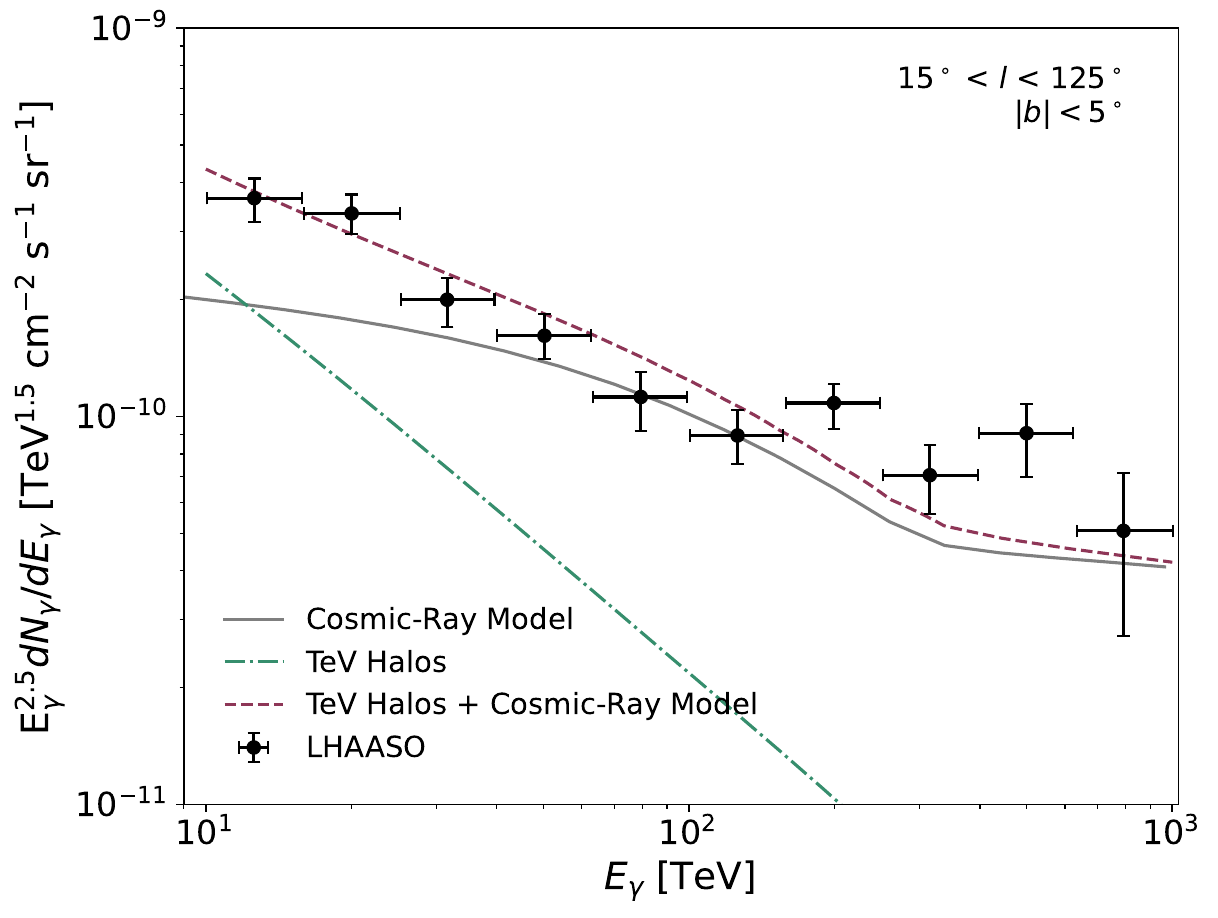}
\includegraphics[width=0.49\textwidth, keepaspectratio]{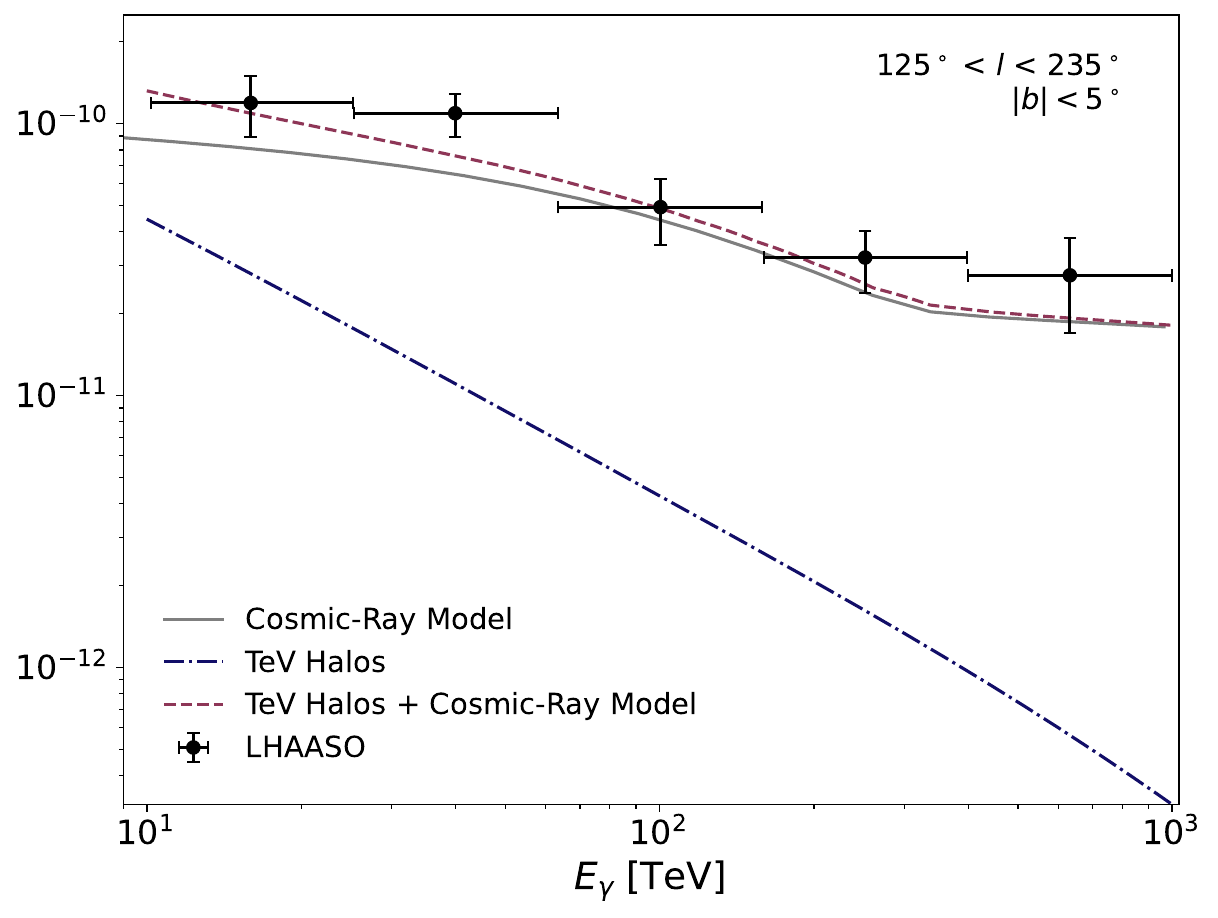}\\
\hspace{0.9mm}
\includegraphics[width=0.48\textwidth, keepaspectratio]{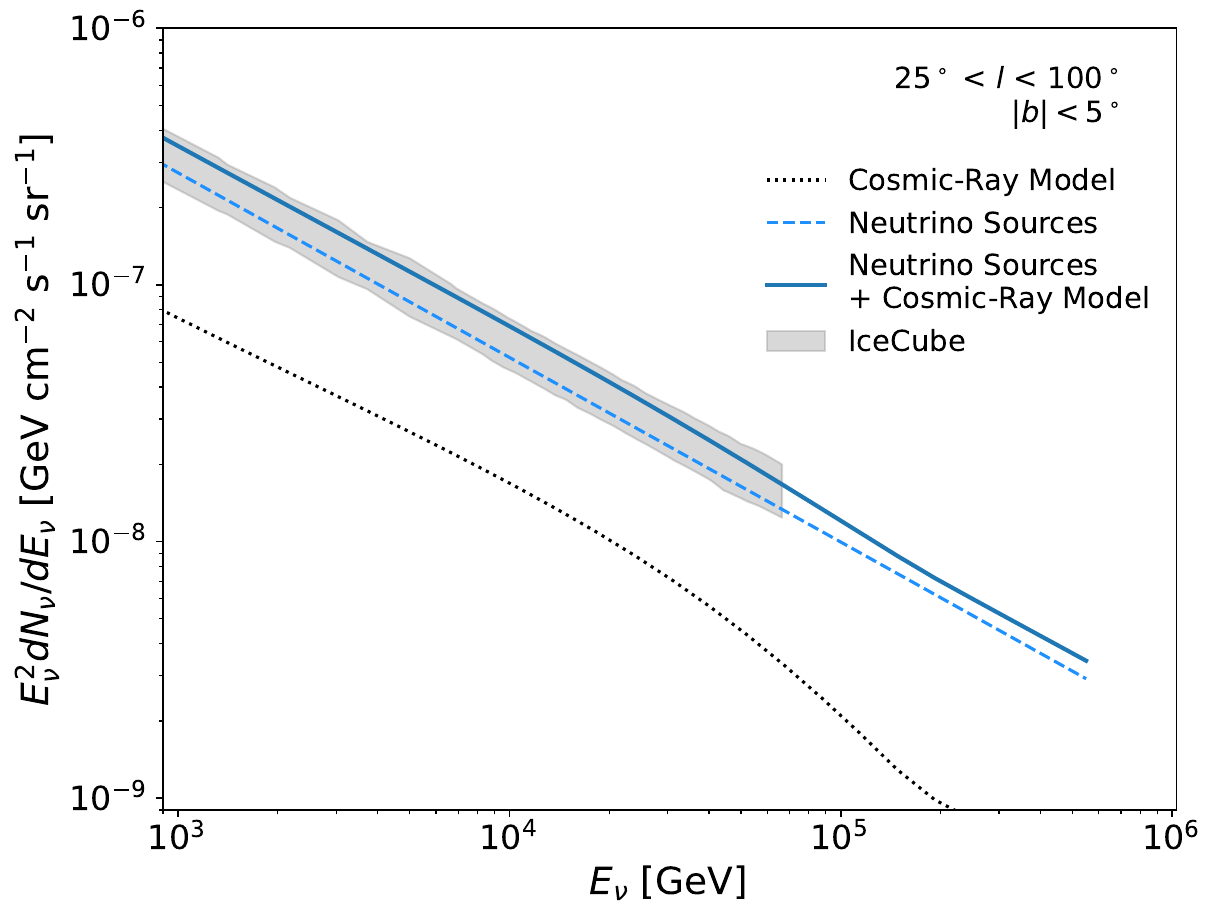}
\hspace{0.6mm}
\includegraphics[width=0.48\textwidth, keepaspectratio]{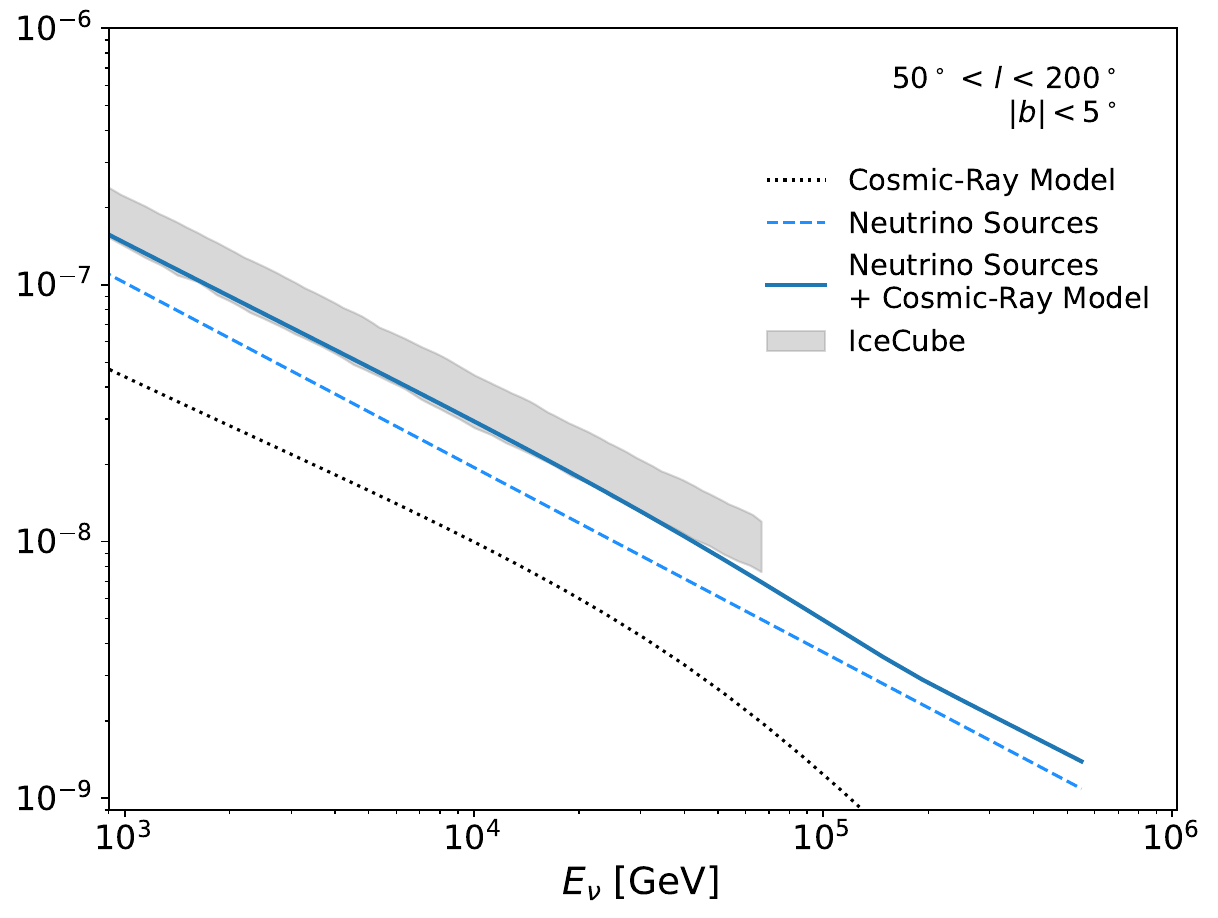}\\
\includegraphics[width=0.48\textwidth, keepaspectratio]{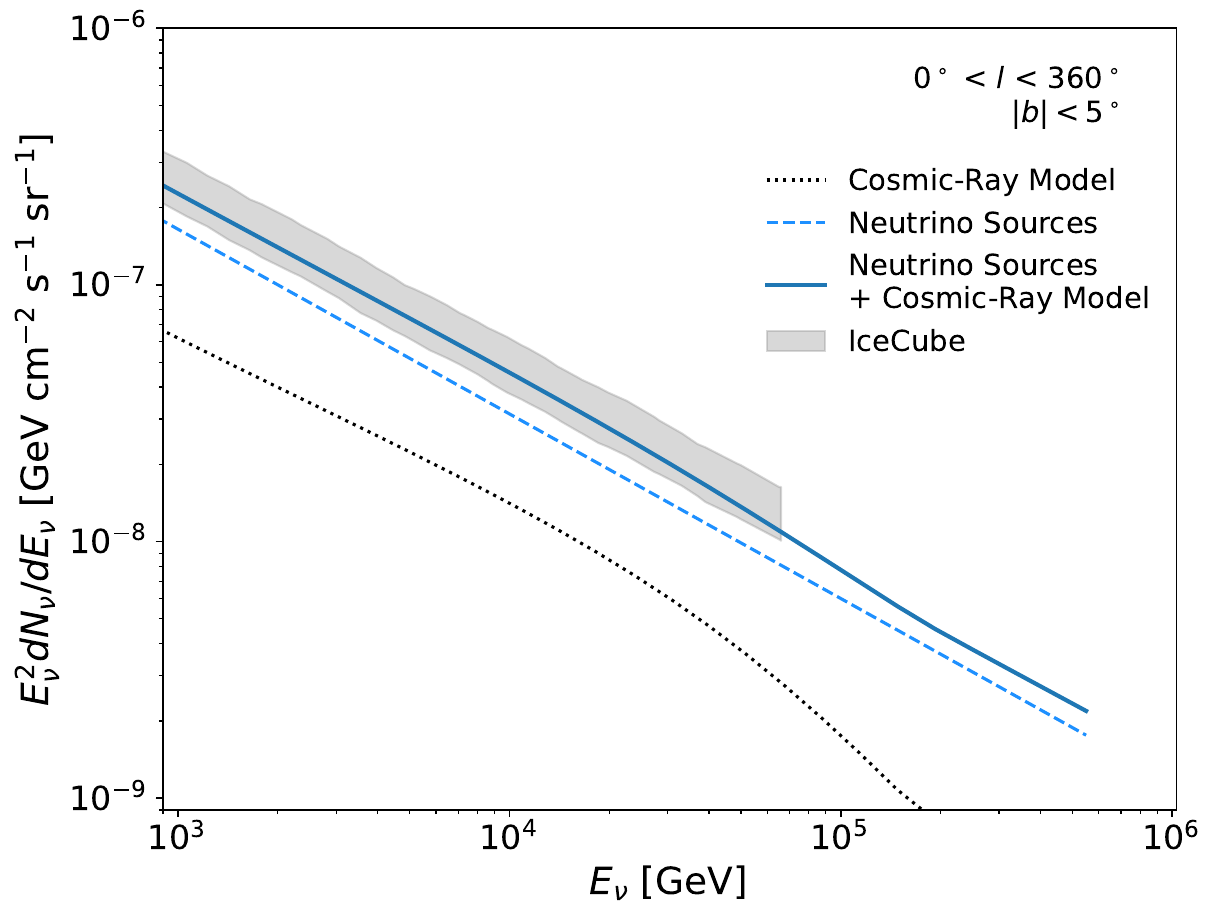}
\caption{The predictions of our model compared to the gamma-ray and neutrino emission observed from the Galactic Plane. In the upper two panels, we show the gamma-ray spectra from the inner and outer Galactic Plane, including contributions from cosmic-ray interactions in the ISM (adopting cosmic-ray model A) and from unresolved TeV halos. These spectra are compared to the LHAASO data from Ref.~\cite{cao_measurement_2023}. In the lower three panels, we show the neutrino spectra from the inner, outer, and entire Galactic Plane, including contributions from cosmic-ray interactions in the ISM (again, adopting cosmic-ray model A) and from neutrino source populations. These spectra are compared to the IceCube data from Ref.~\cite{icecube_collaboration_observation_2023}, as derived using the $\pi^0$ template model. The gamma-ray and neutrino spectra shown were calculated using the best-fit model parameters.}
\label{fig:CR_A_model_panel} 
\end{figure*}

\begin{figure*}[p]

\includegraphics[width=0.49\textwidth, keepaspectratio]{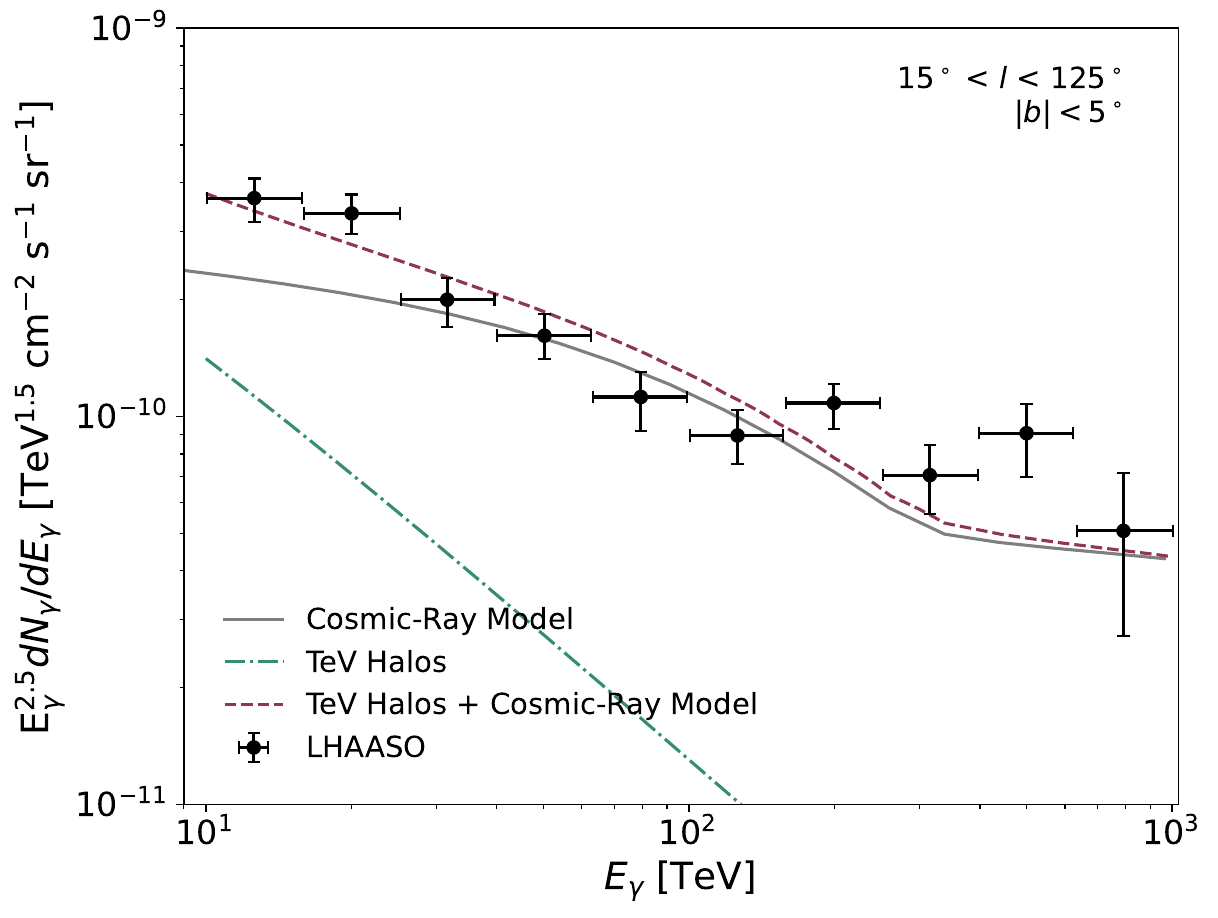}
\includegraphics[width=0.49\textwidth, keepaspectratio]{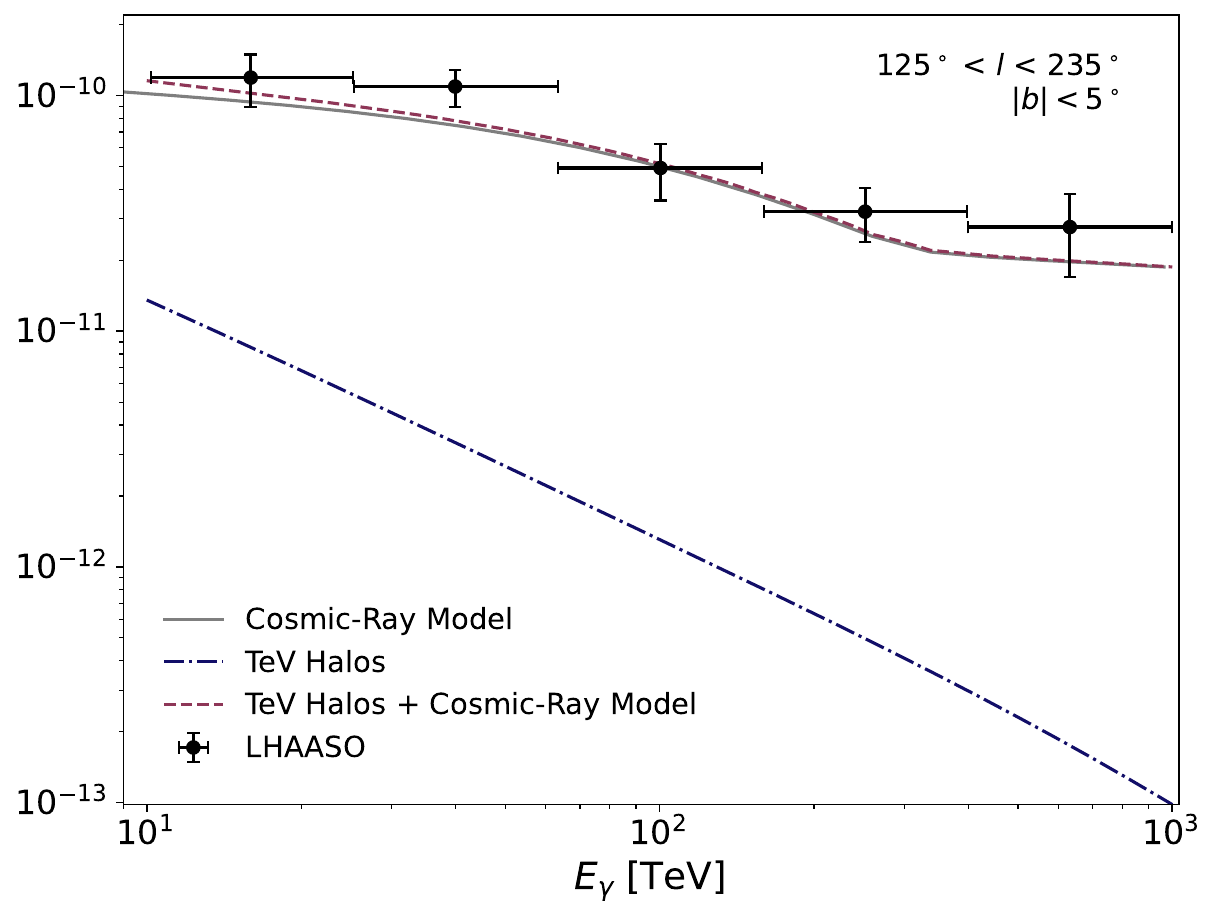}\\
\hspace{0.9mm}
\includegraphics[width=0.48\textwidth, keepaspectratio]{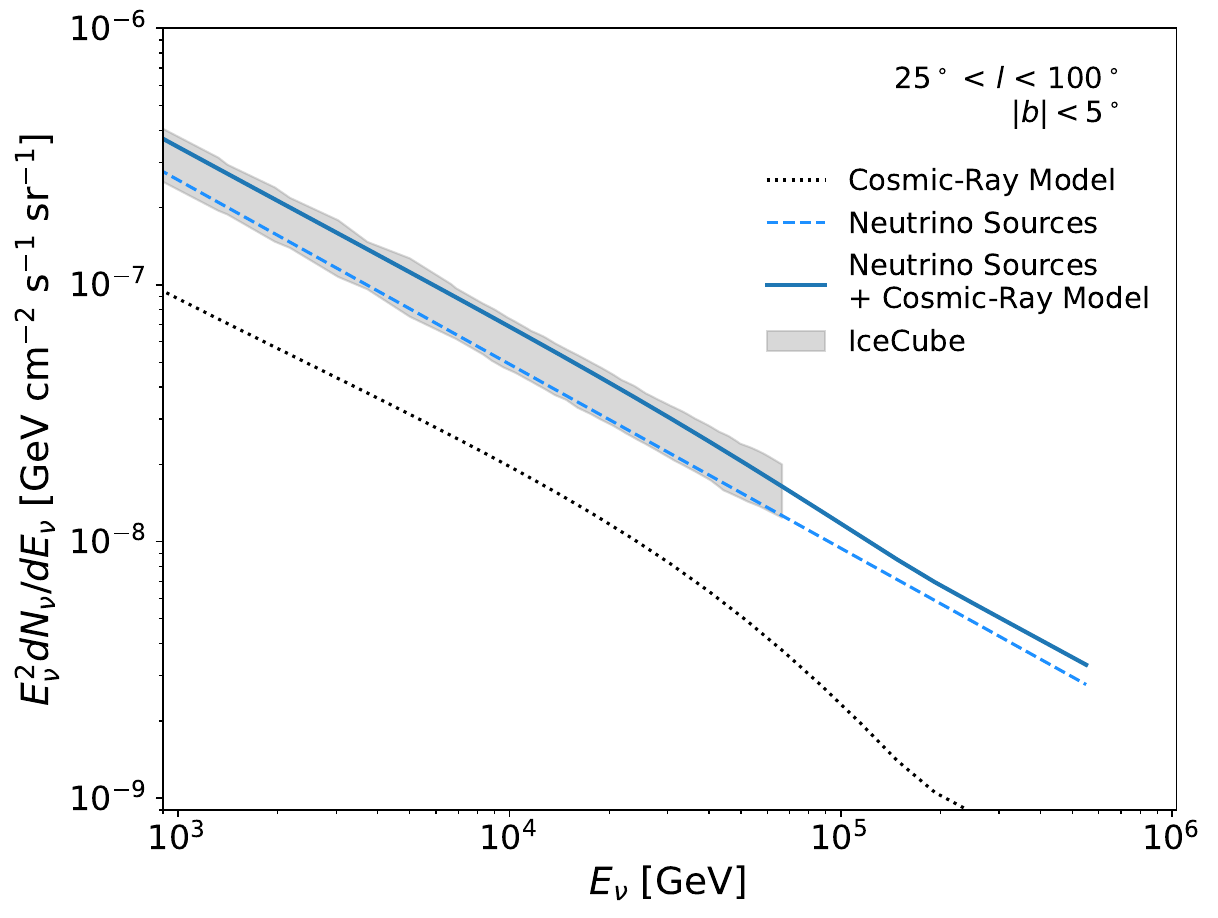}
\hspace{0.6mm}
\includegraphics[width=0.48\textwidth, keepaspectratio]{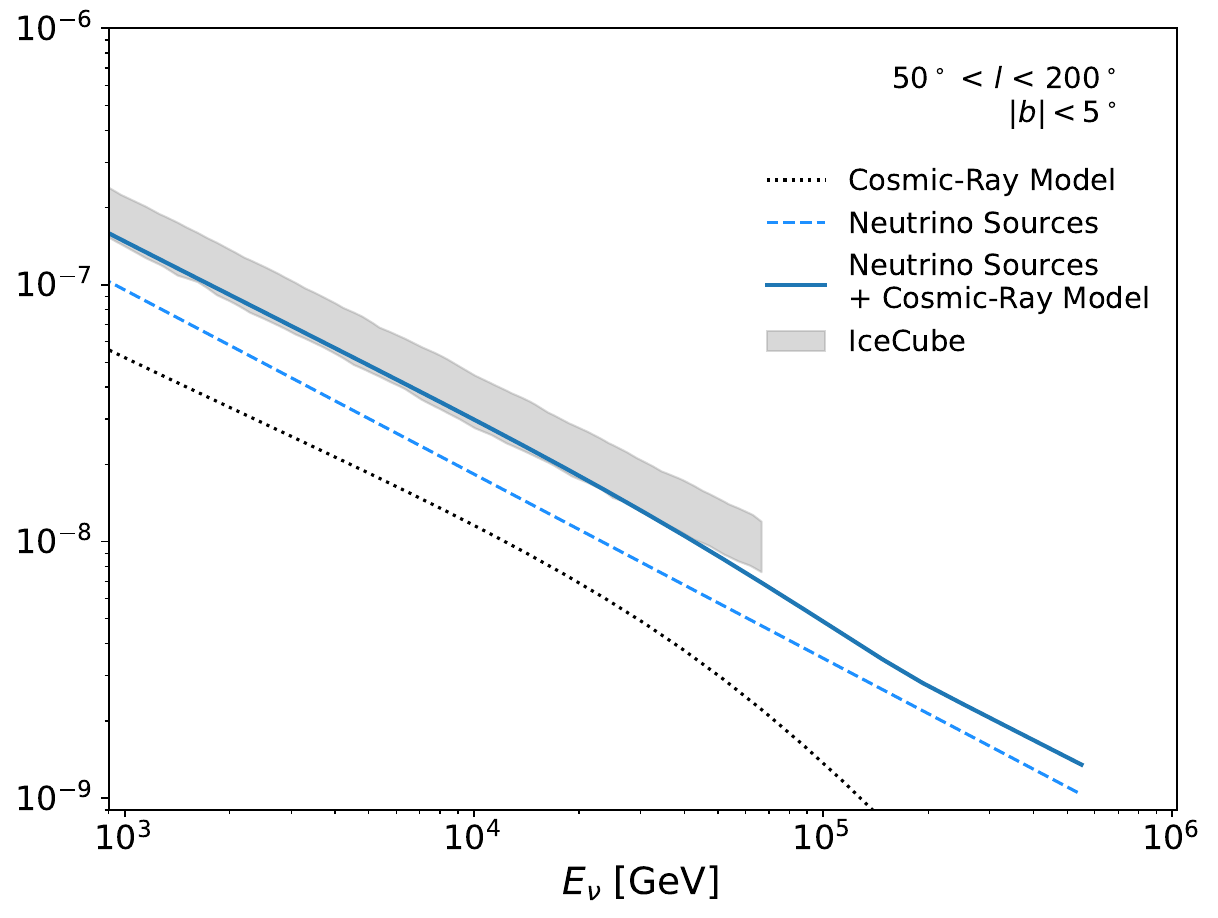}\\
\includegraphics[width=0.48\textwidth, keepaspectratio]{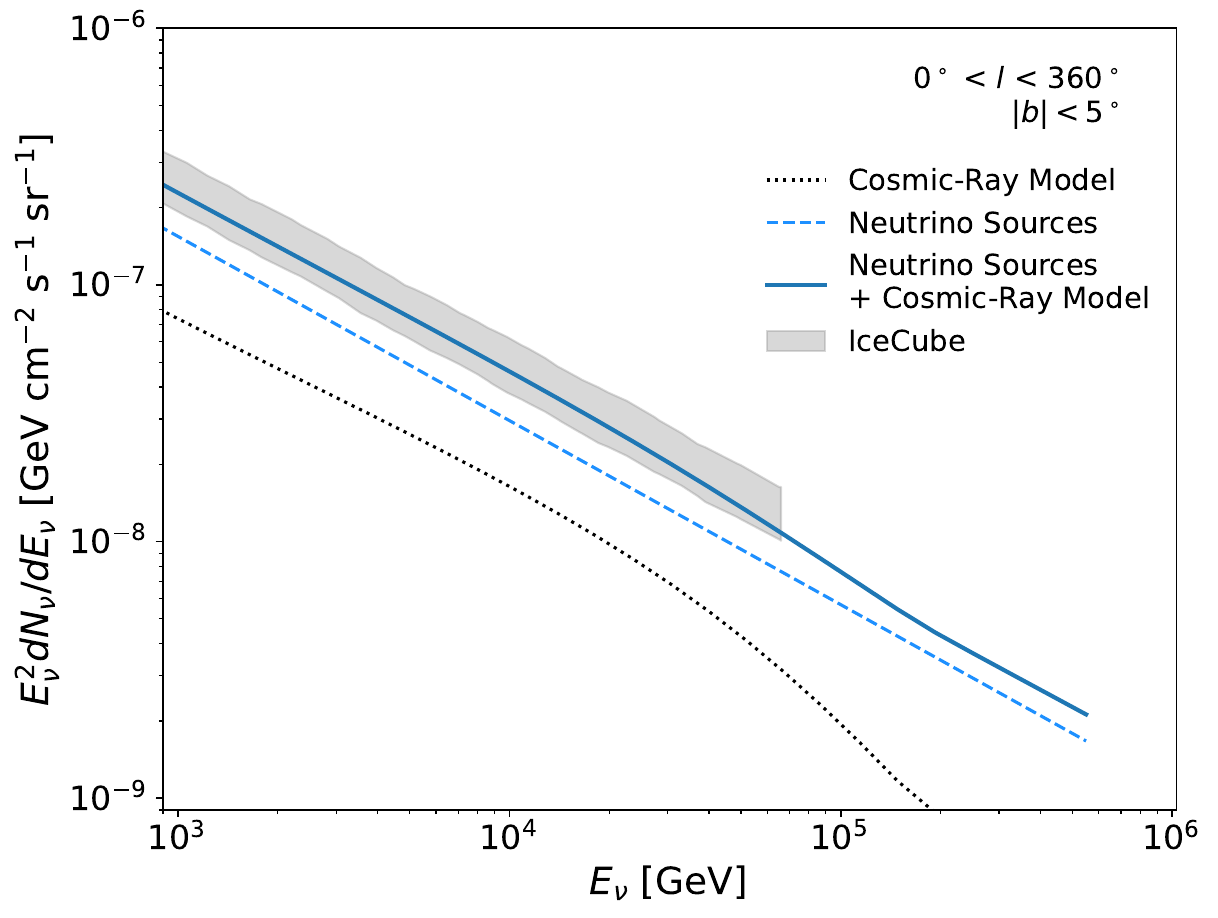}
\caption{
As in Fig.~\ref{fig:CR_A_model_panel}, but adopting cosmic-ray model B.}
\label{fig:CR_B_model_panel} 
\end{figure*}

\begin{figure*}[p]

\includegraphics[width=0.49\textwidth, keepaspectratio]{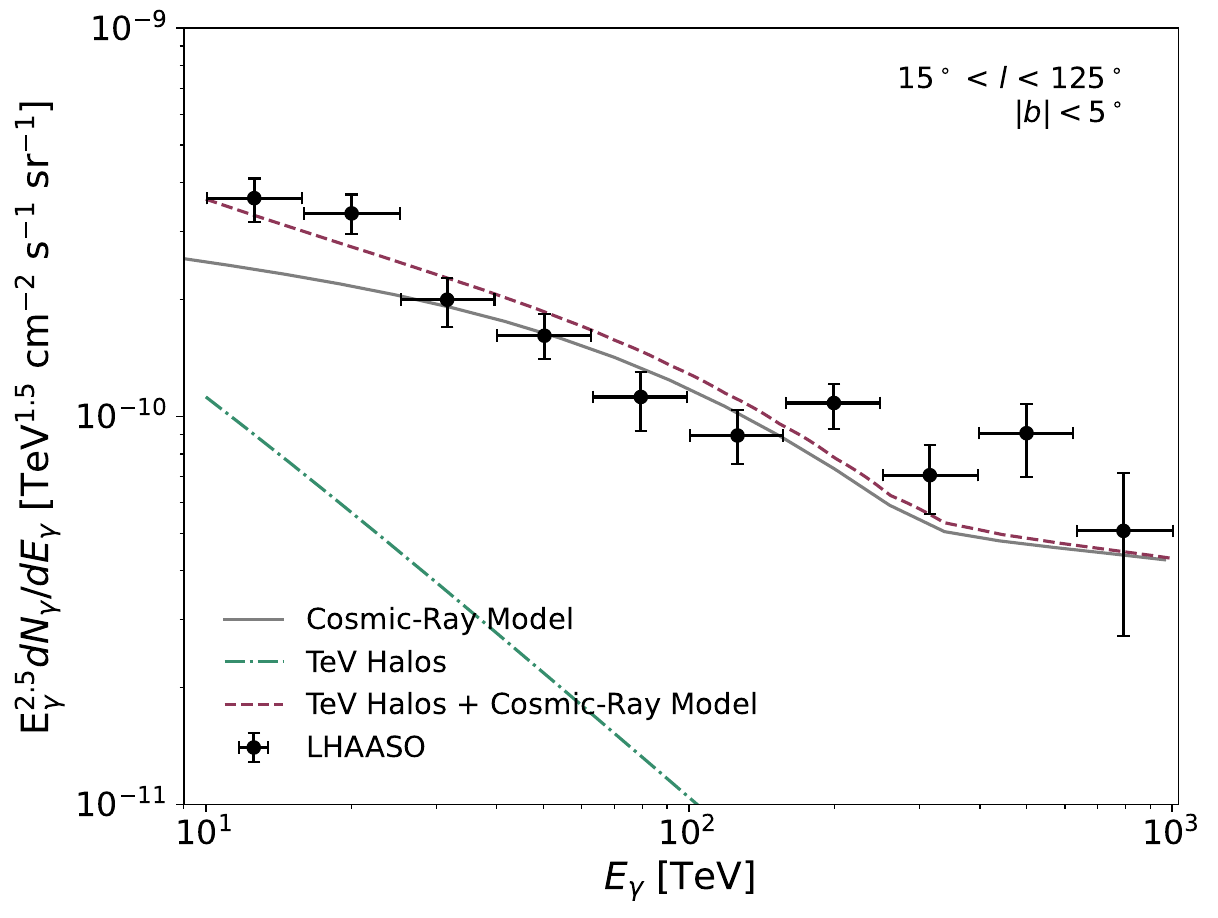}
\includegraphics[width=0.49\textwidth, keepaspectratio]{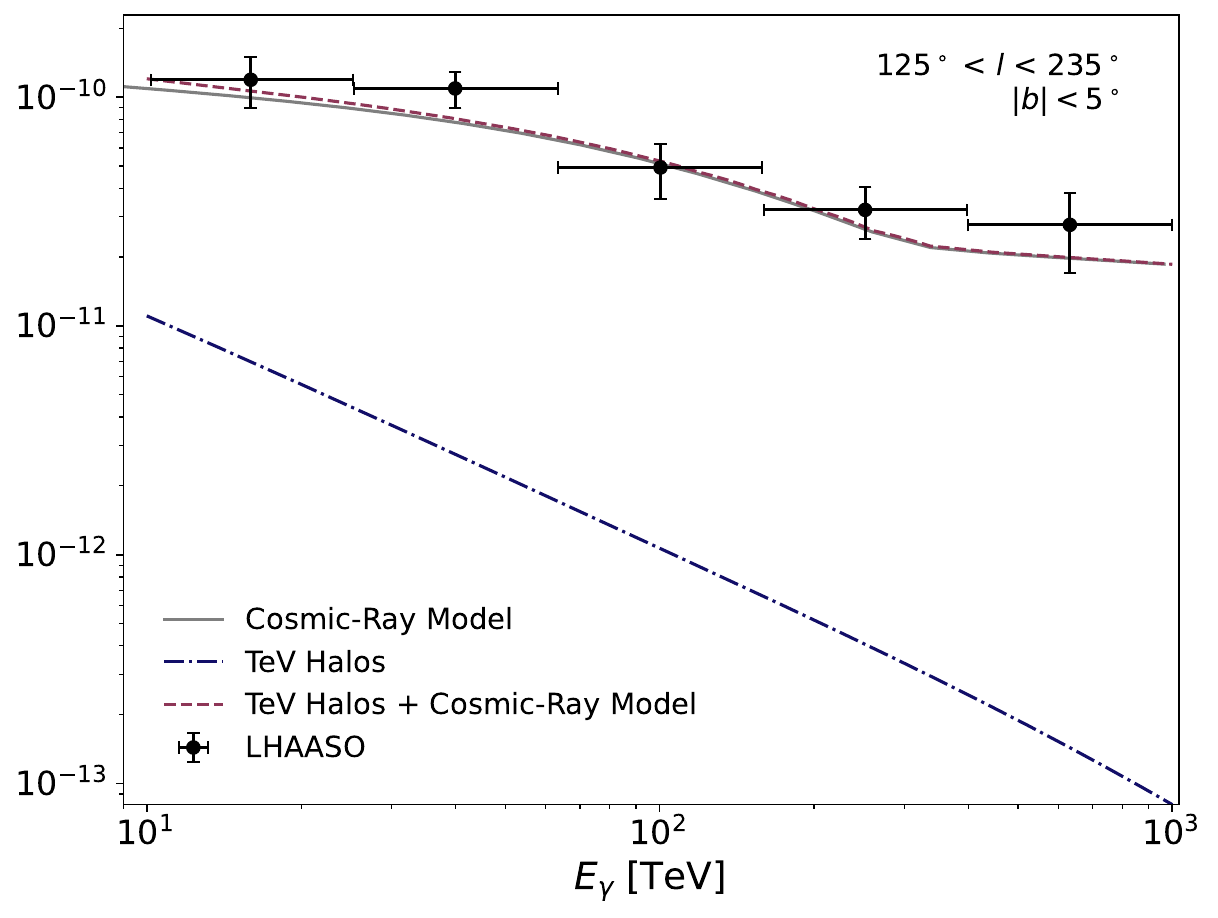}\\
\hspace{0.9mm}
\includegraphics[width=0.48\textwidth, keepaspectratio]{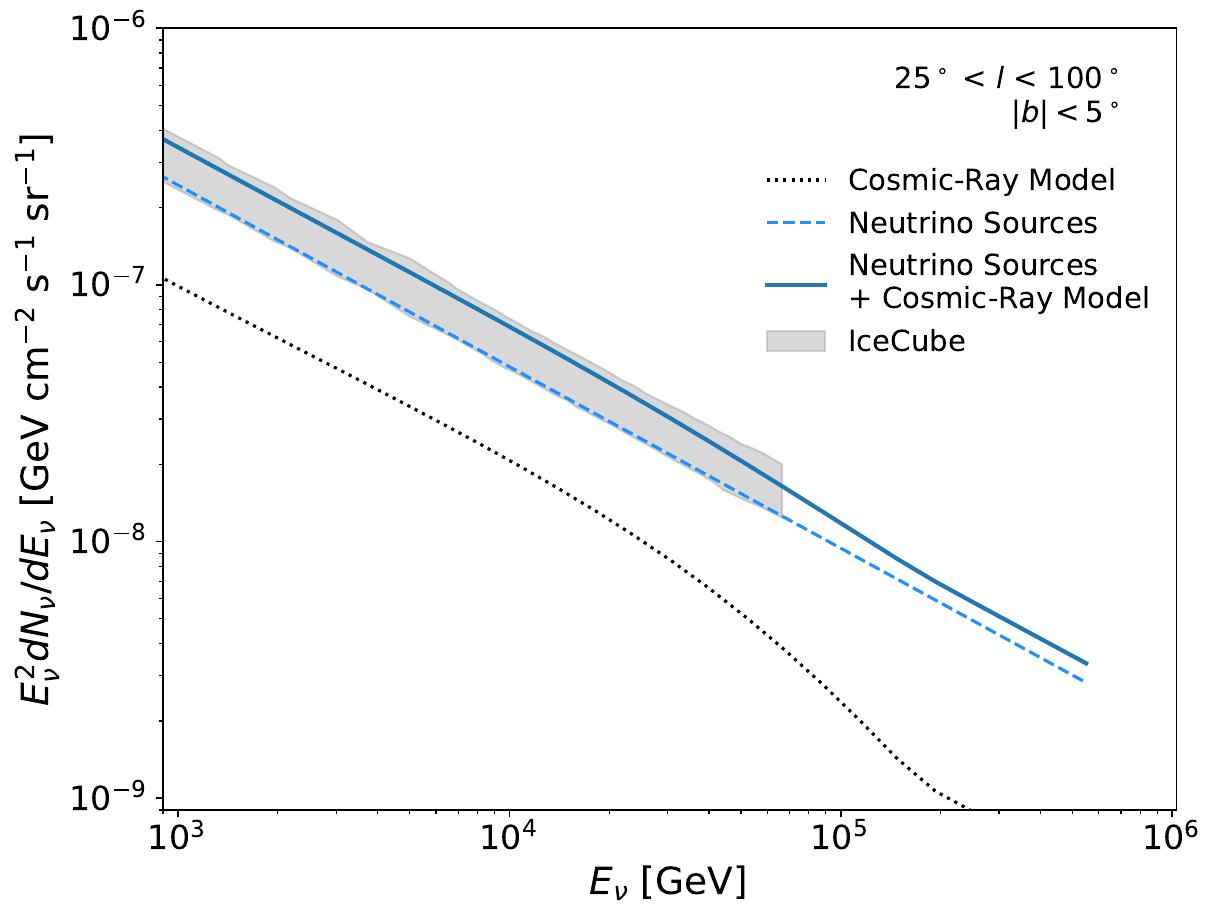}
\hspace{0.6mm}
\includegraphics[width=0.48\textwidth, keepaspectratio]{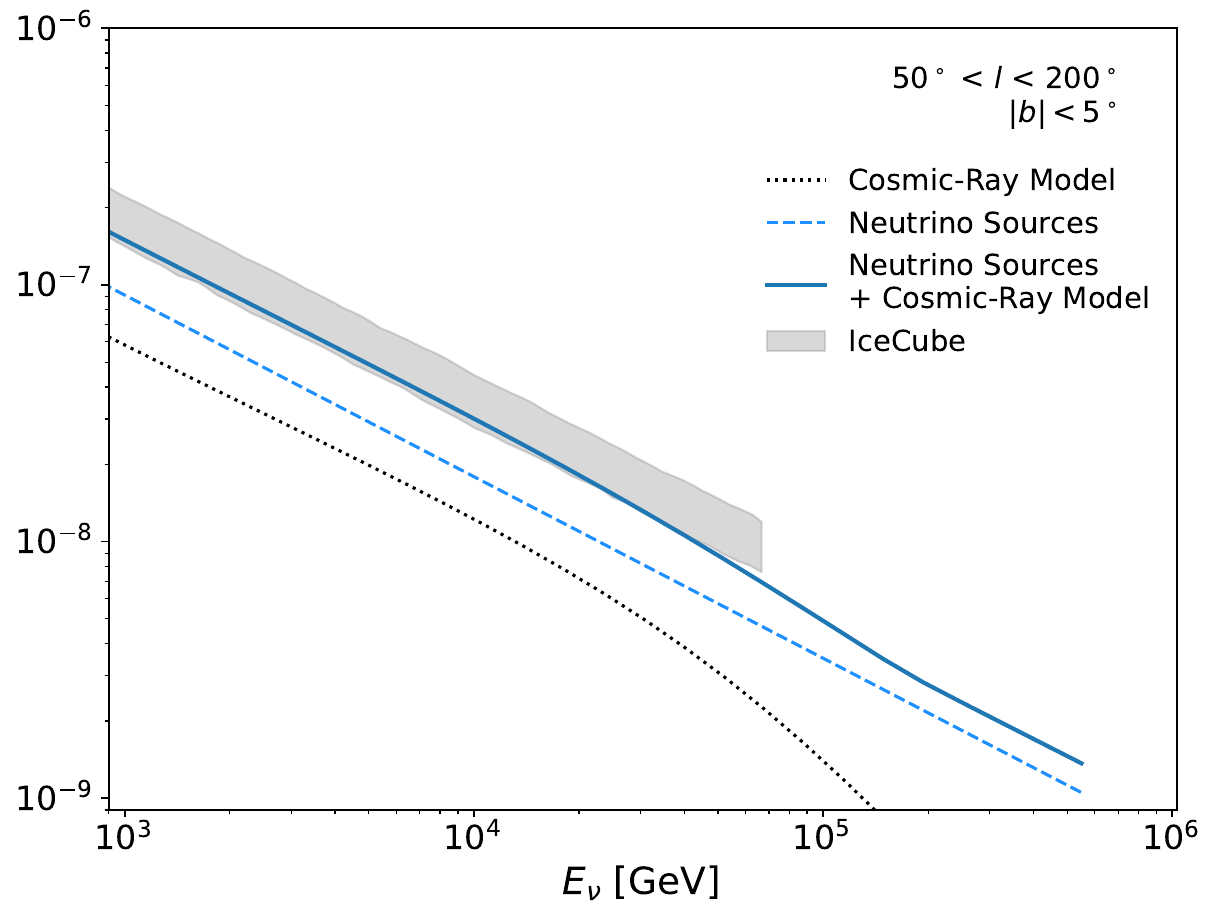}
\includegraphics[width=0.48\textwidth, keepaspectratio]{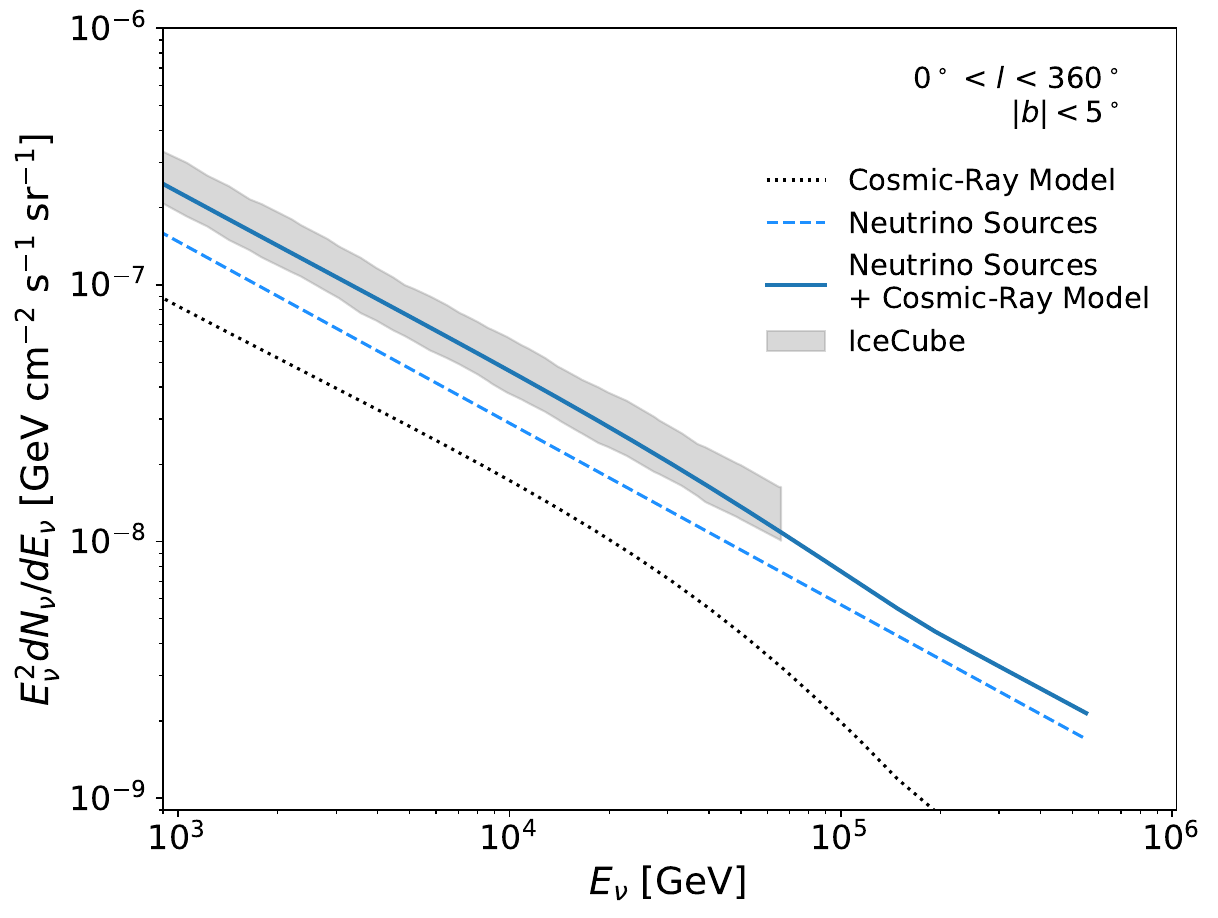}
\caption{
As in Figs.~\ref{fig:CR_A_model_panel}-\ref{fig:CR_B_model_panel}, but adopting cosmic-ray model C.}
\label{fig:CR_C_model_panel} 
\end{figure*}

\begin{figure*}[p]

\includegraphics[width=0.49\textwidth, keepaspectratio]{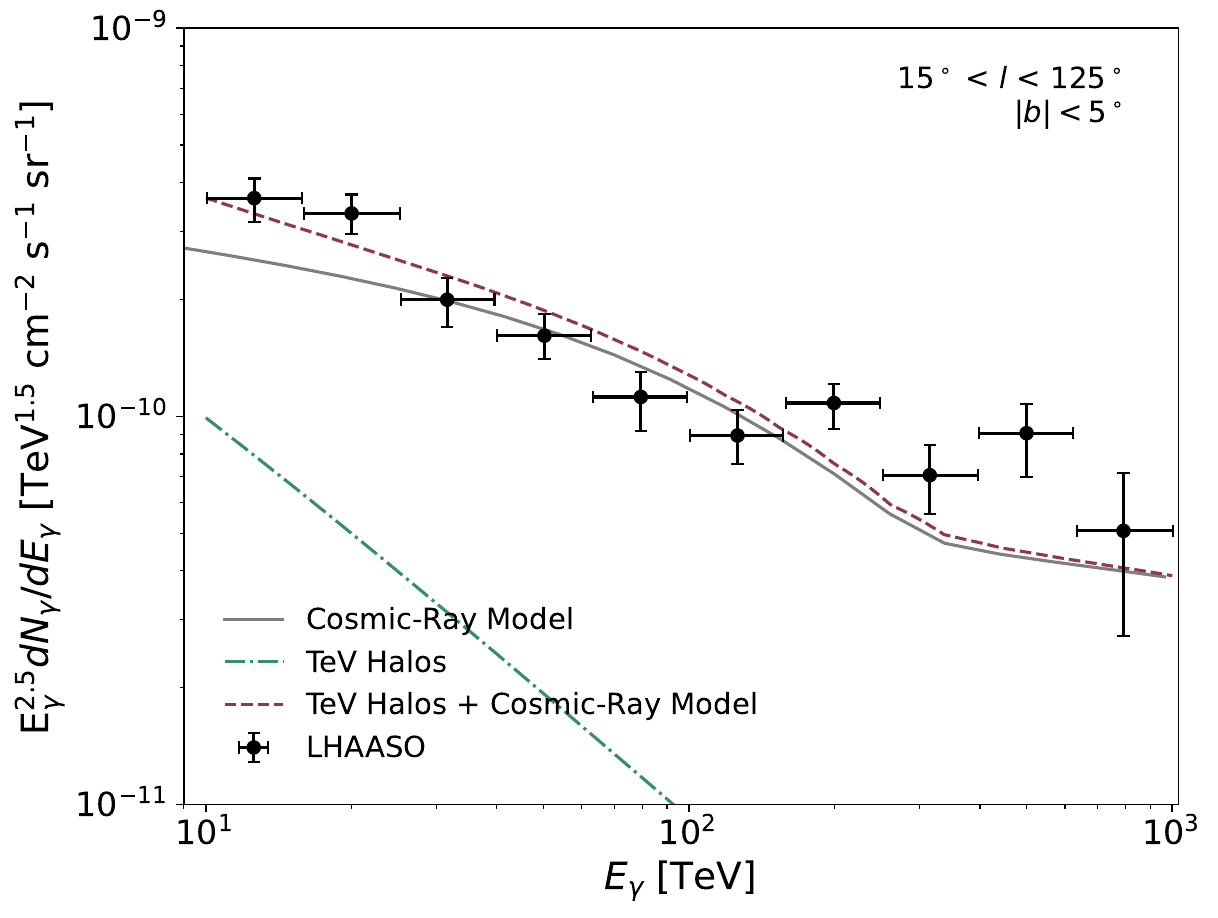}
\includegraphics[width=0.49\textwidth, keepaspectratio]{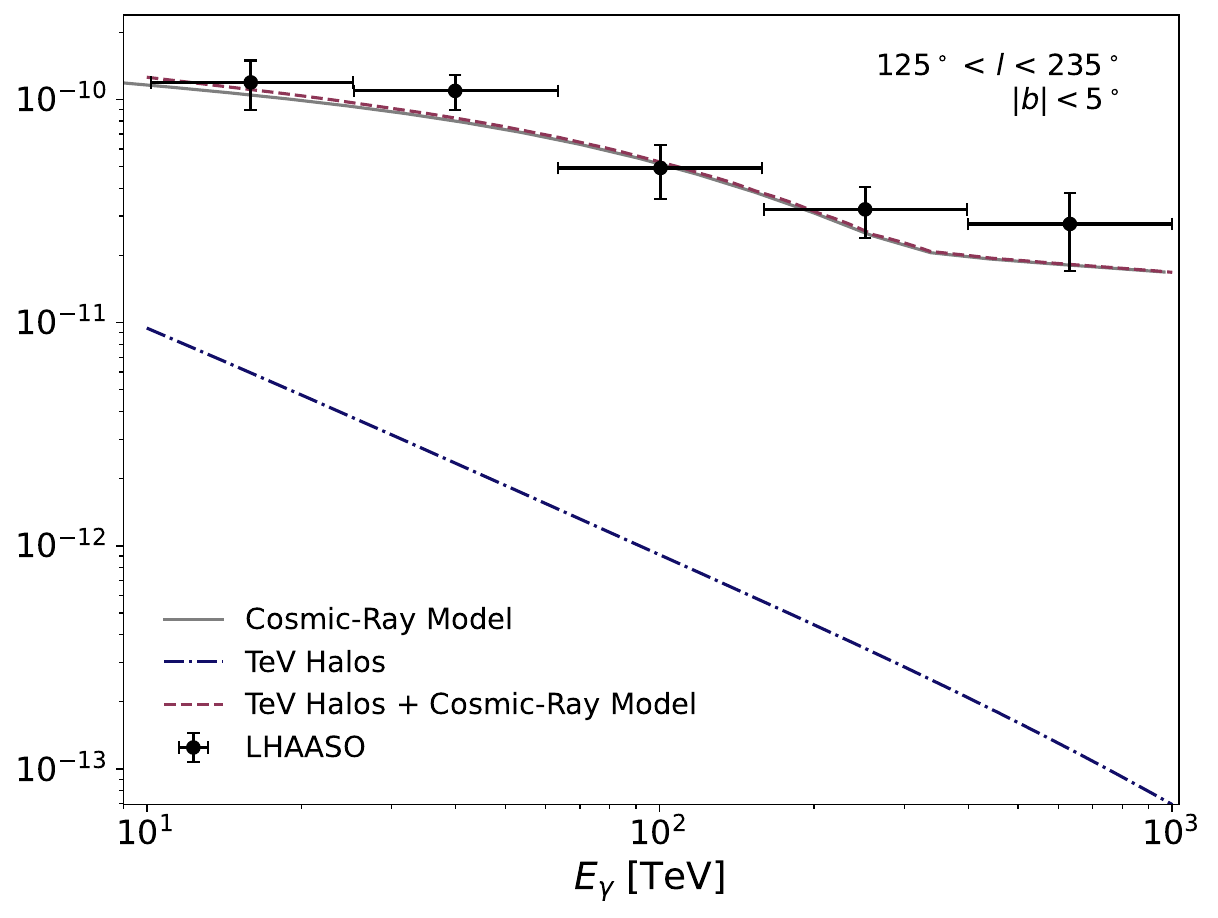}\\
\hspace{0.9mm}
\includegraphics[width=0.48\textwidth, keepaspectratio]{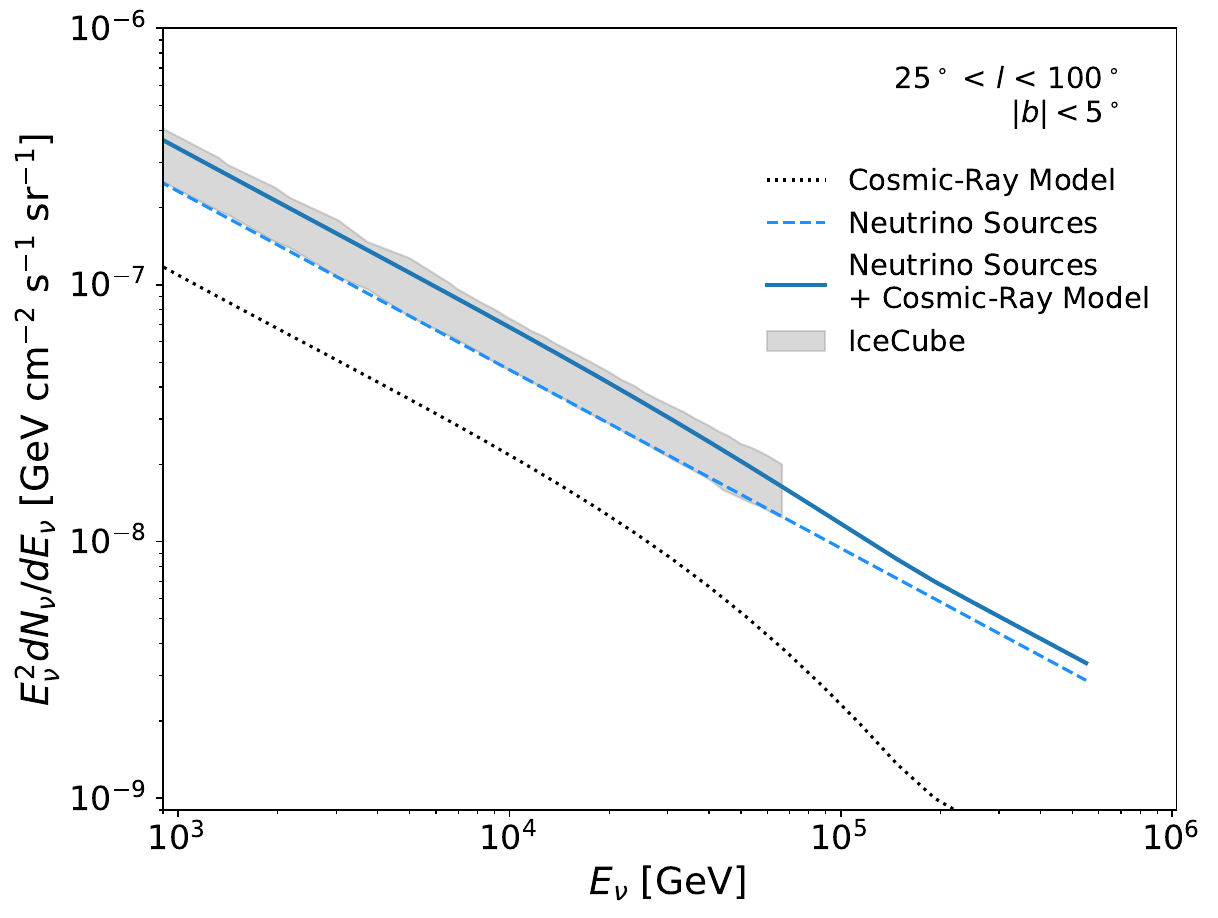}
\hspace{0.6mm}
\includegraphics[width=0.48\textwidth, keepaspectratio]{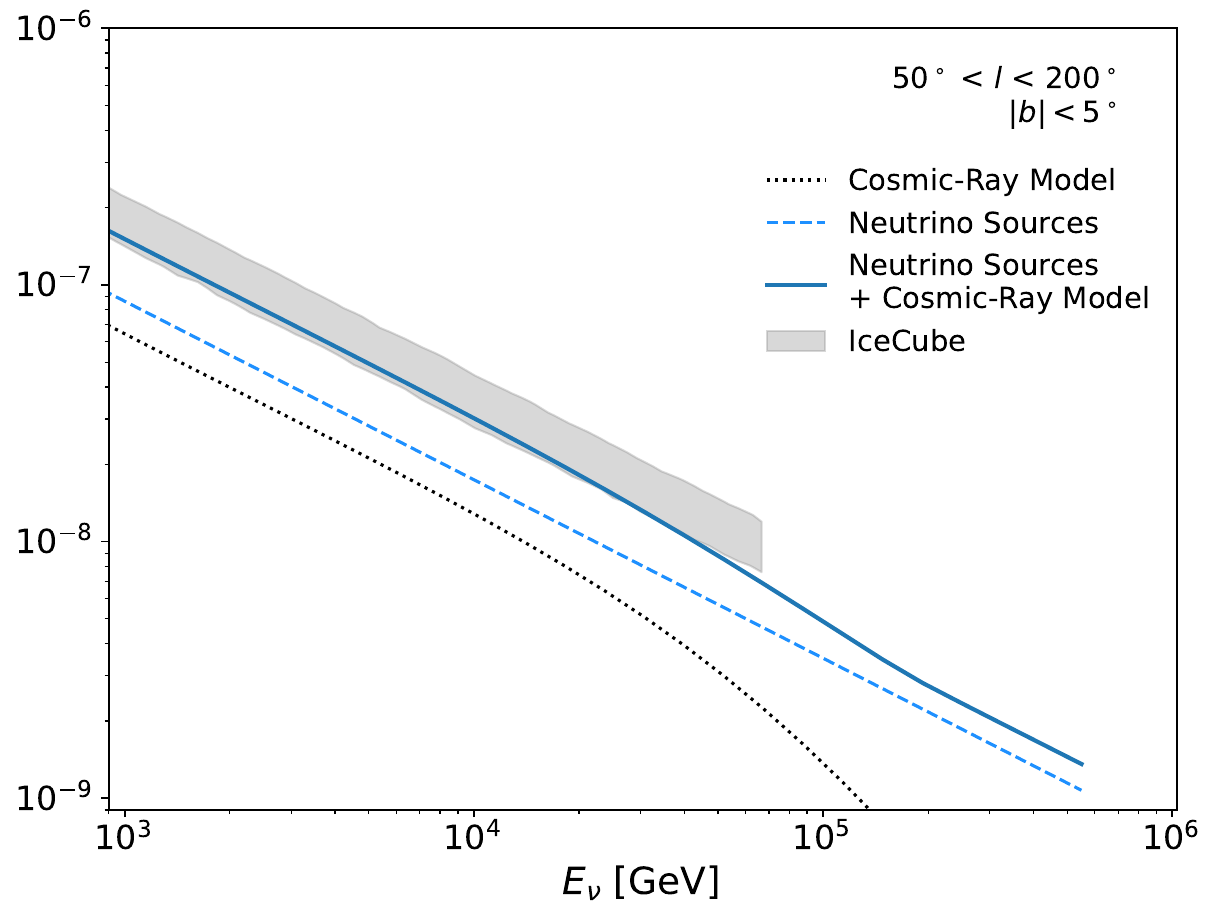}\\
\includegraphics[width=0.48\textwidth, keepaspectratio]{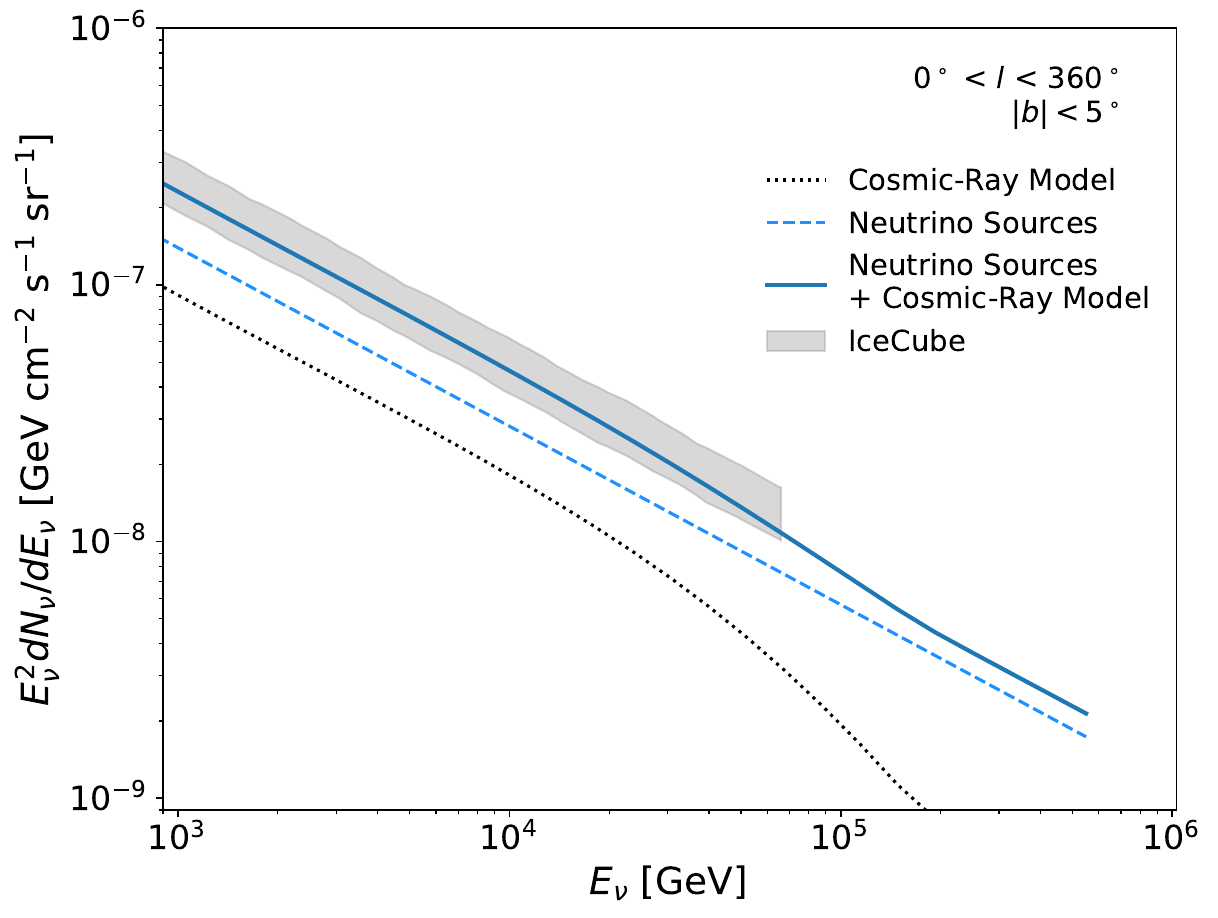}
\caption{
As in Figs.~\ref{fig:CR_A_model_panel}-\ref{fig:CR_C_model_panel}, but adopting cosmic-ray model D.}
\label{fig:CR_D_model_panel} 
\end{figure*}

\begin{figure*}[p]

\includegraphics[width=0.49\textwidth, keepaspectratio]{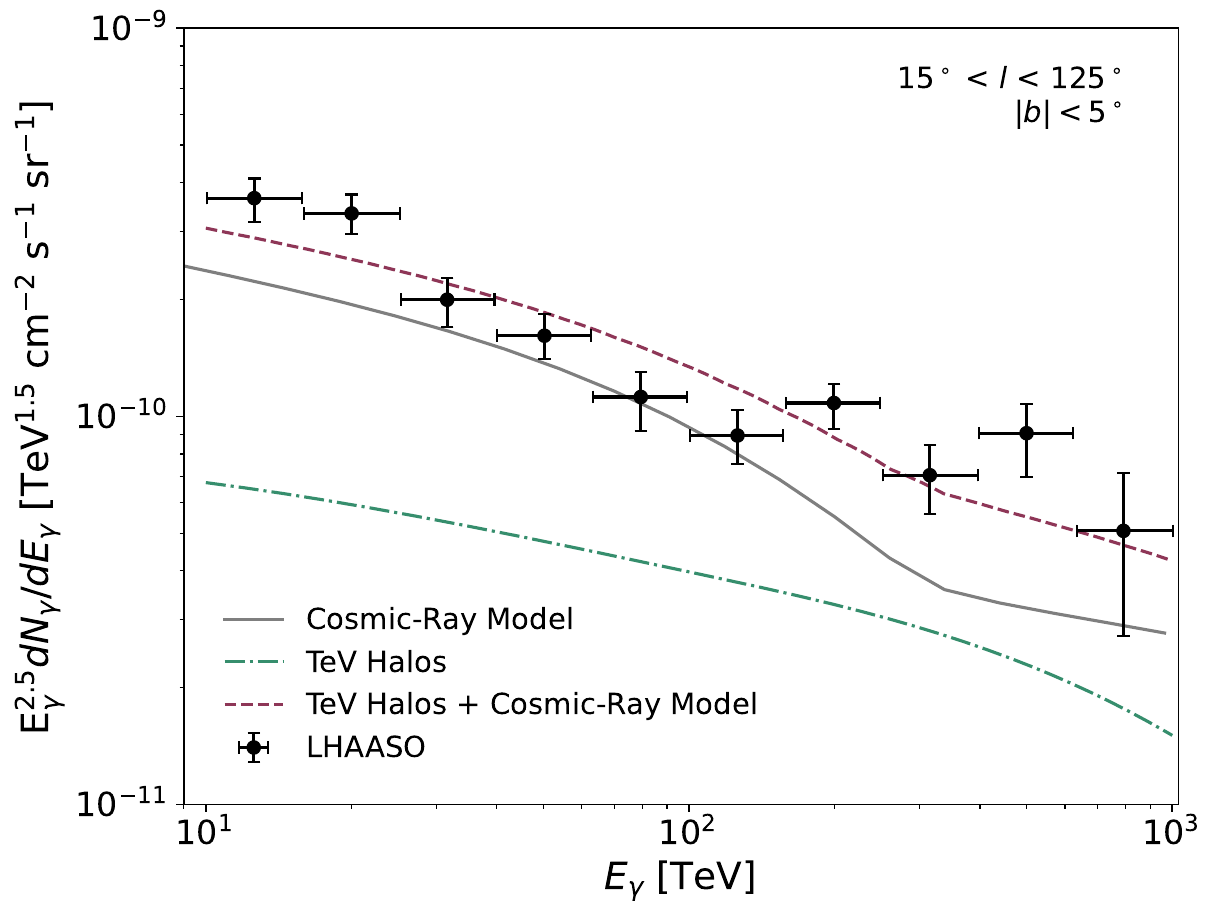}
\includegraphics[width=0.49\textwidth, keepaspectratio]{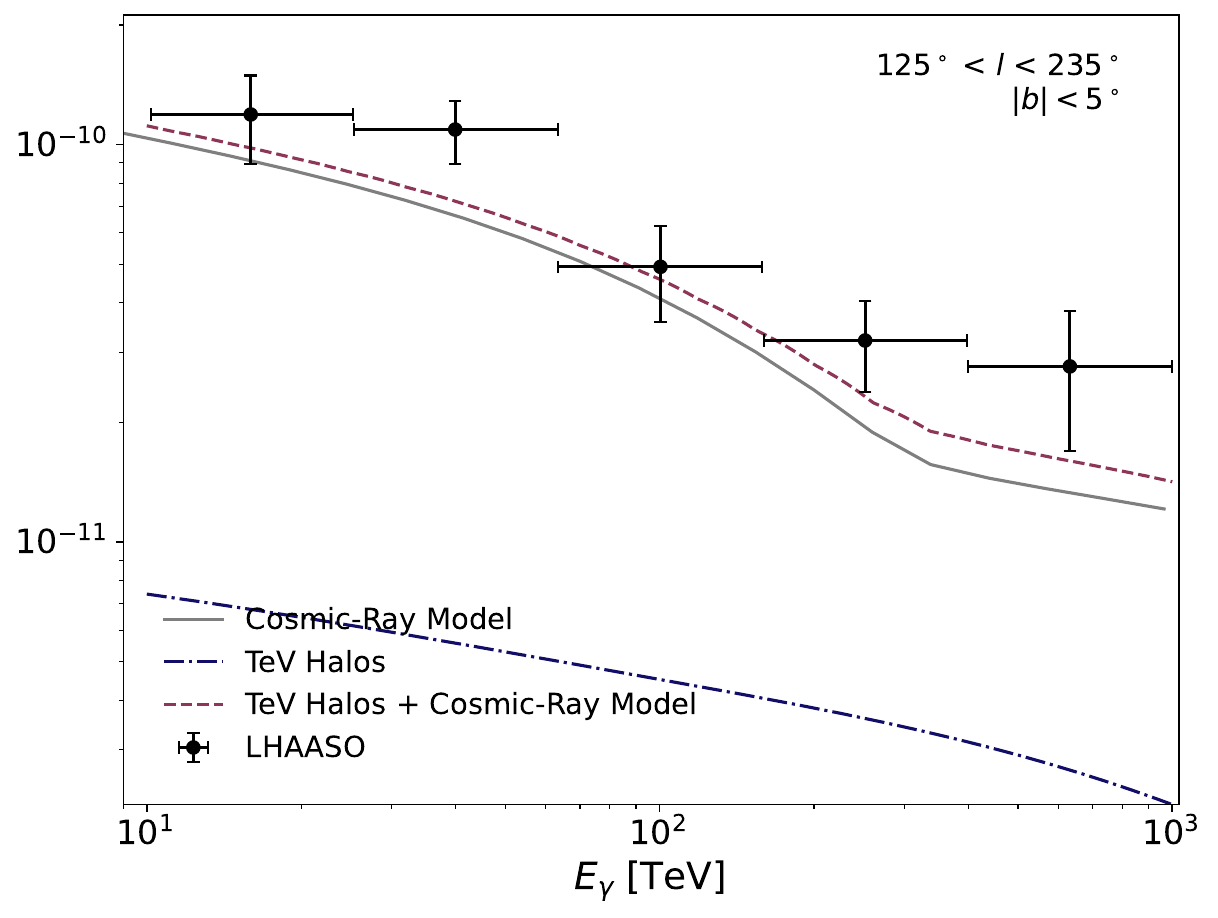}\\
\hspace{0.9mm}
\includegraphics[width=0.48\textwidth, keepaspectratio]{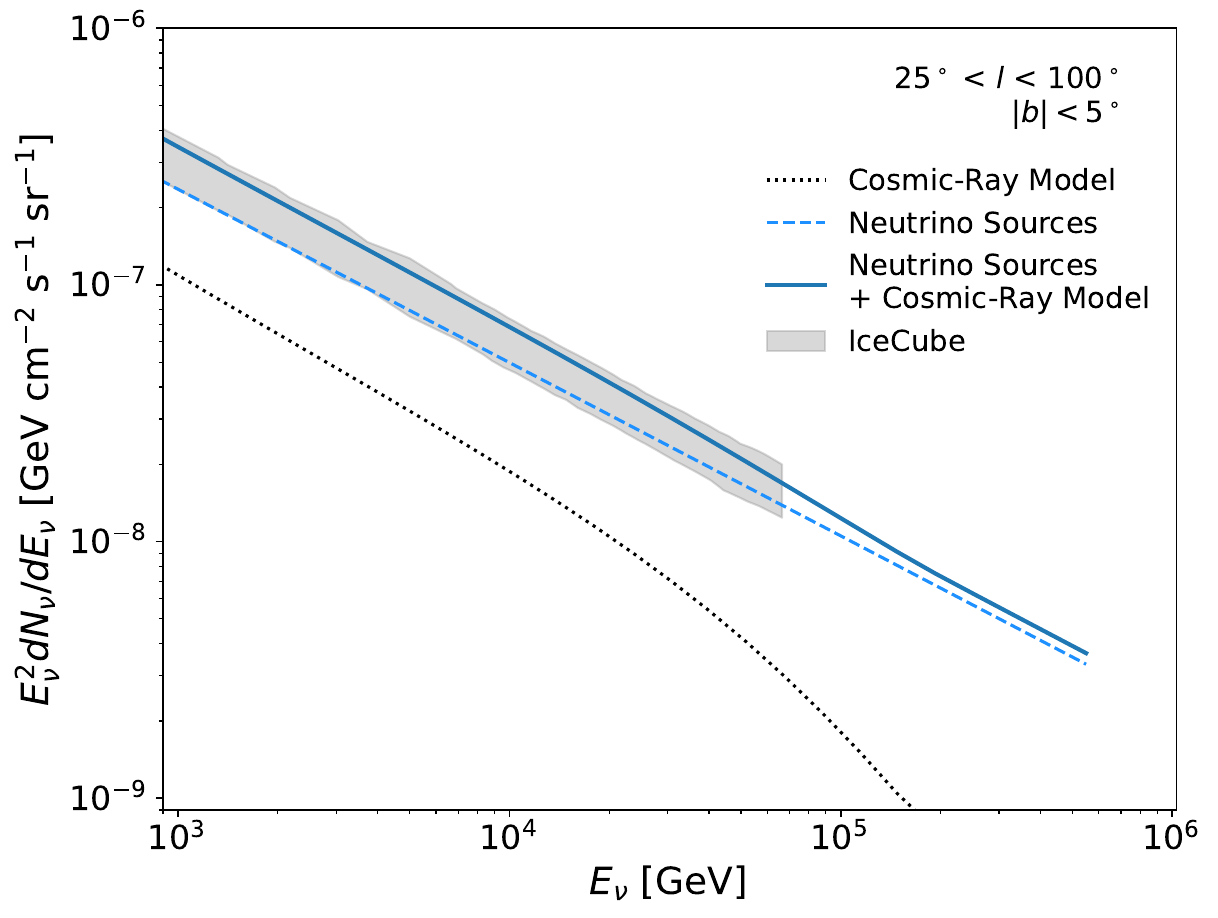}
\hspace{0.6mm}
\includegraphics[width=0.48\textwidth, keepaspectratio]{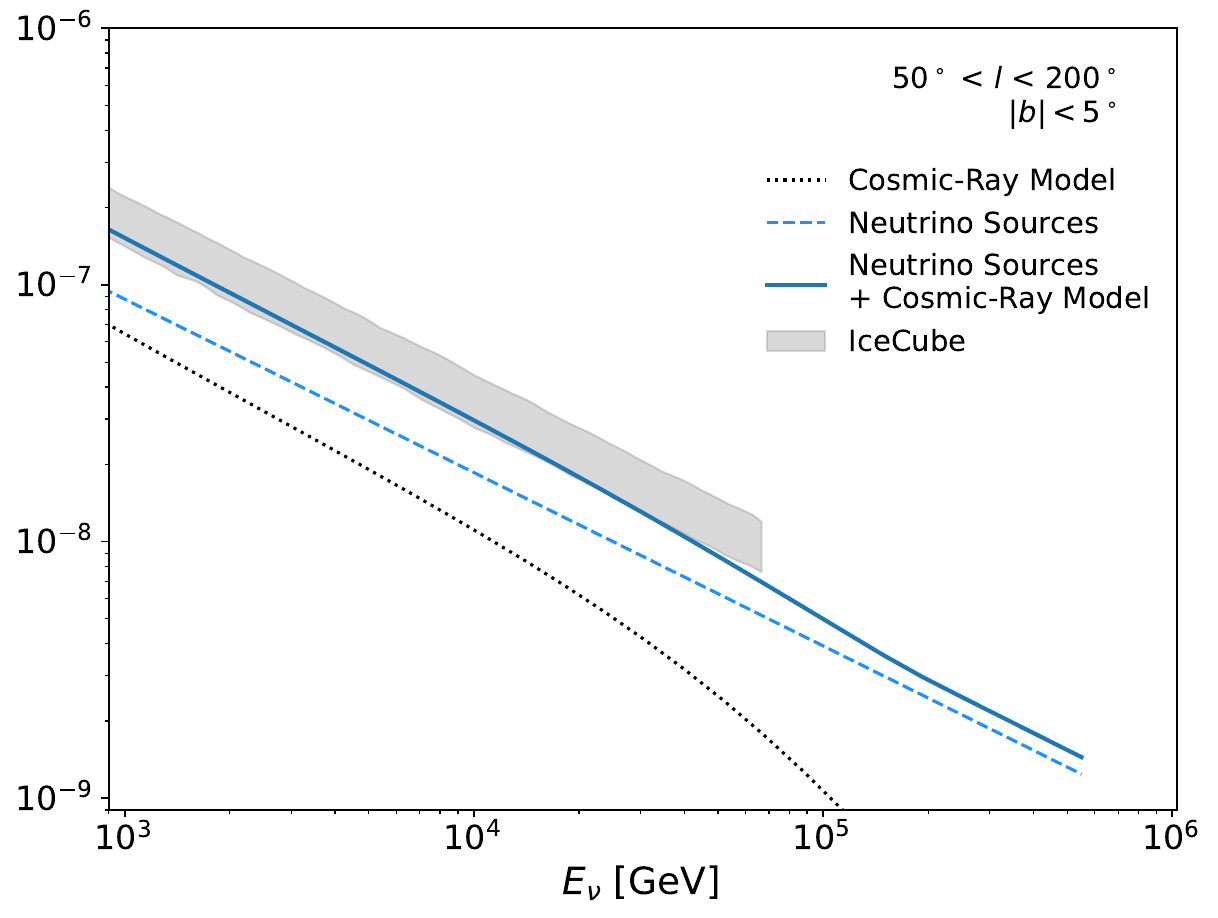}\\
\includegraphics[width=0.48\textwidth, keepaspectratio]{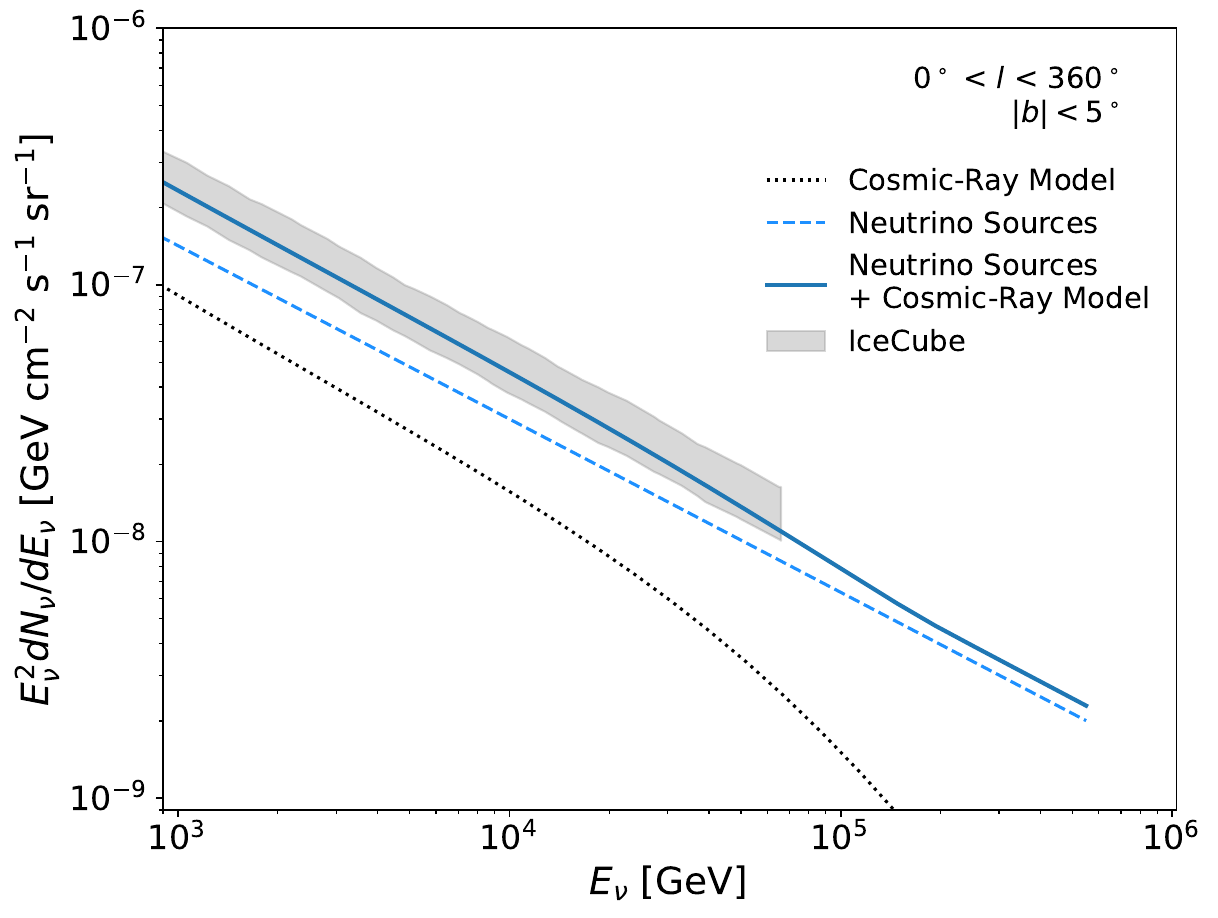}
\caption{ 
As in Figs.~\ref{fig:CR_A_model_panel}-\ref{fig:CR_D_model_panel}, but adopting cosmic-ray model E.}
\label{fig:CR_E_model_panel}
\end{figure*}

\begin{figure*}[p]

\includegraphics[width=0.49\textwidth, keepaspectratio]{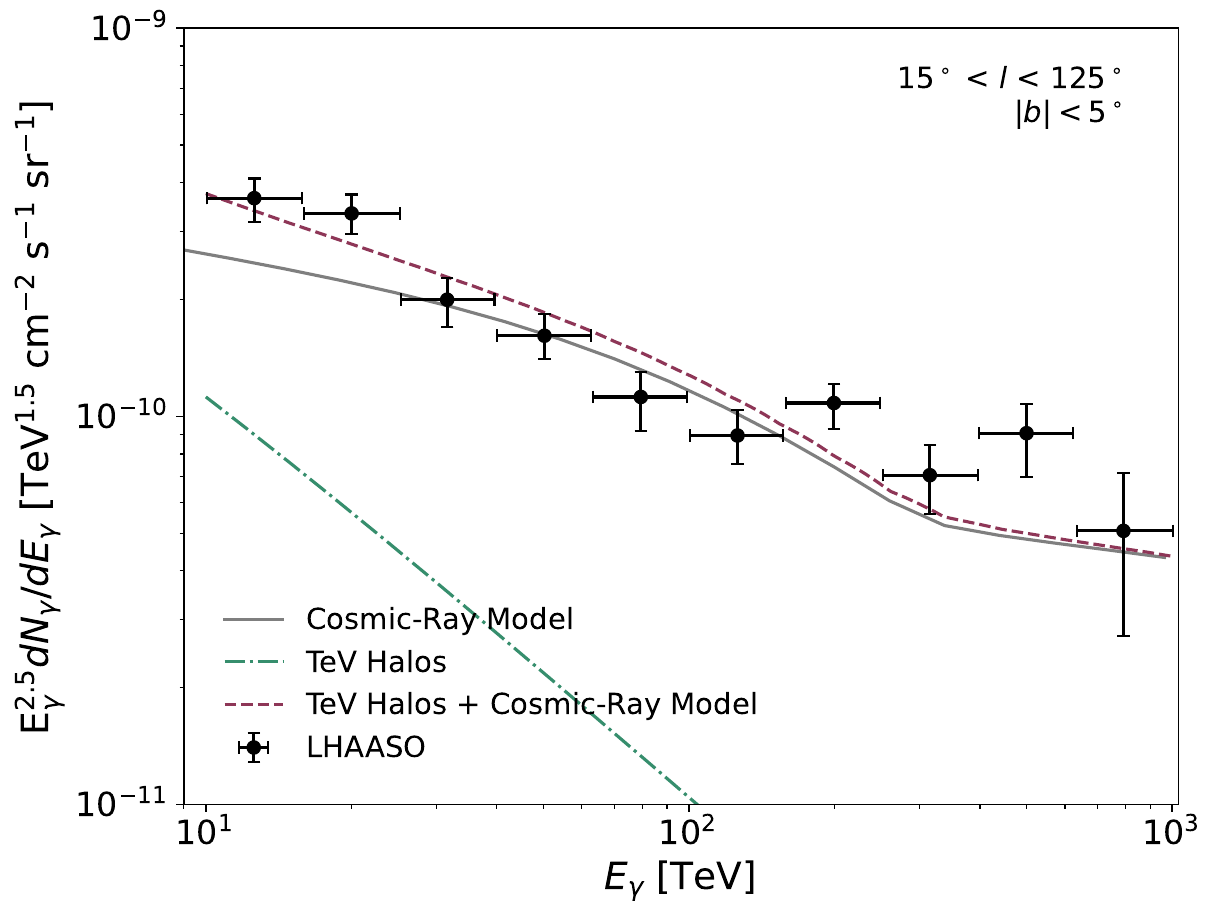}
\includegraphics[width=0.49\textwidth, keepaspectratio]{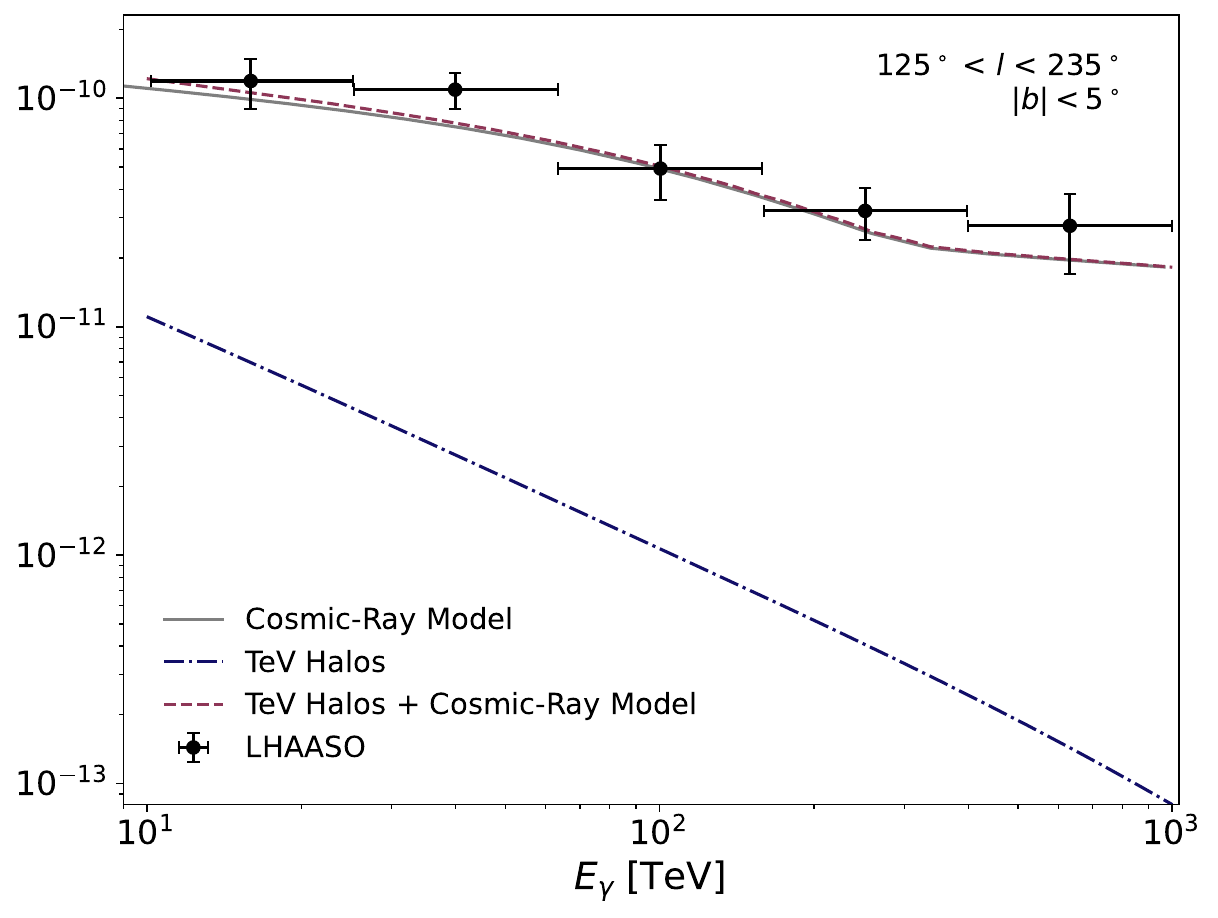}\\
\hspace{0.9mm}
\includegraphics[width=0.48\textwidth, keepaspectratio]{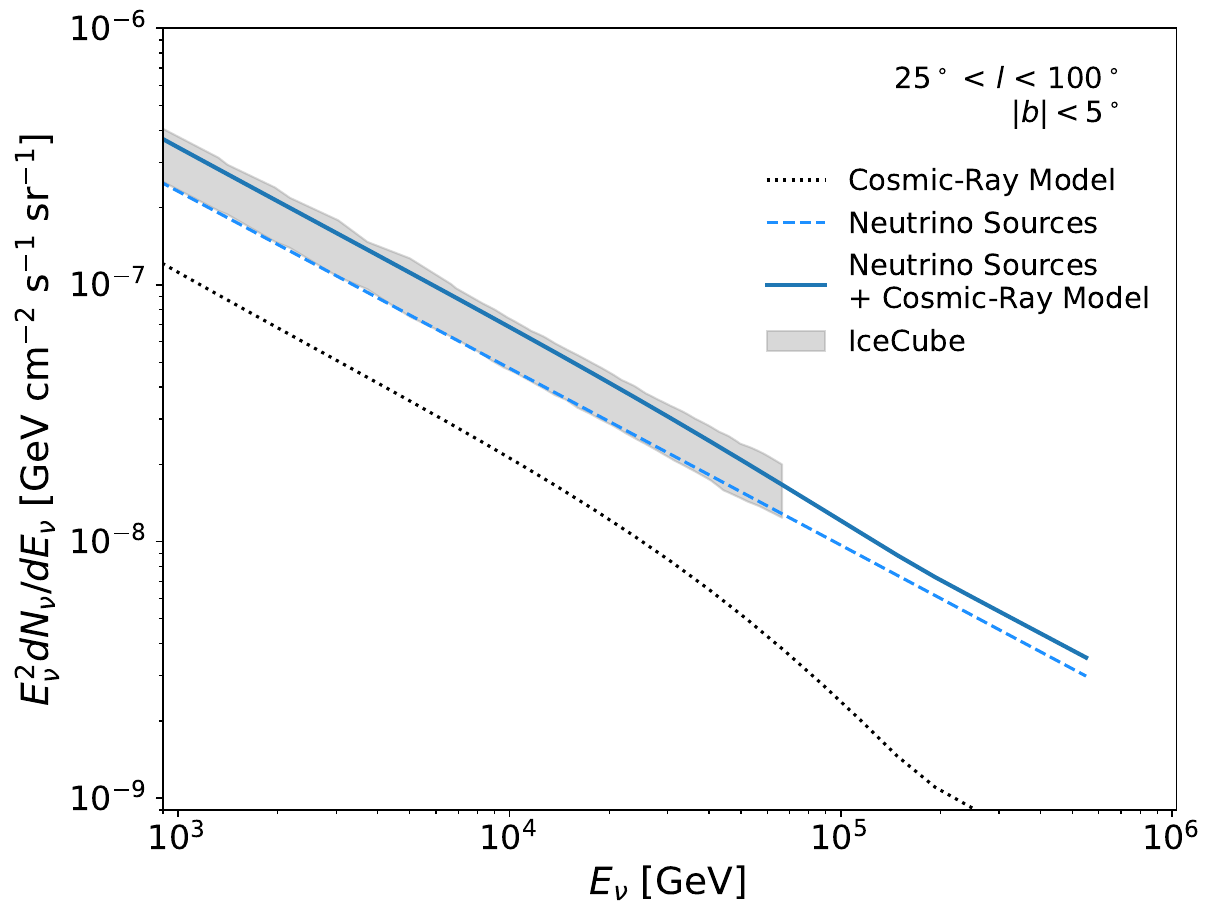}
\hspace{0.6mm}
\includegraphics[width=0.48\textwidth, keepaspectratio]{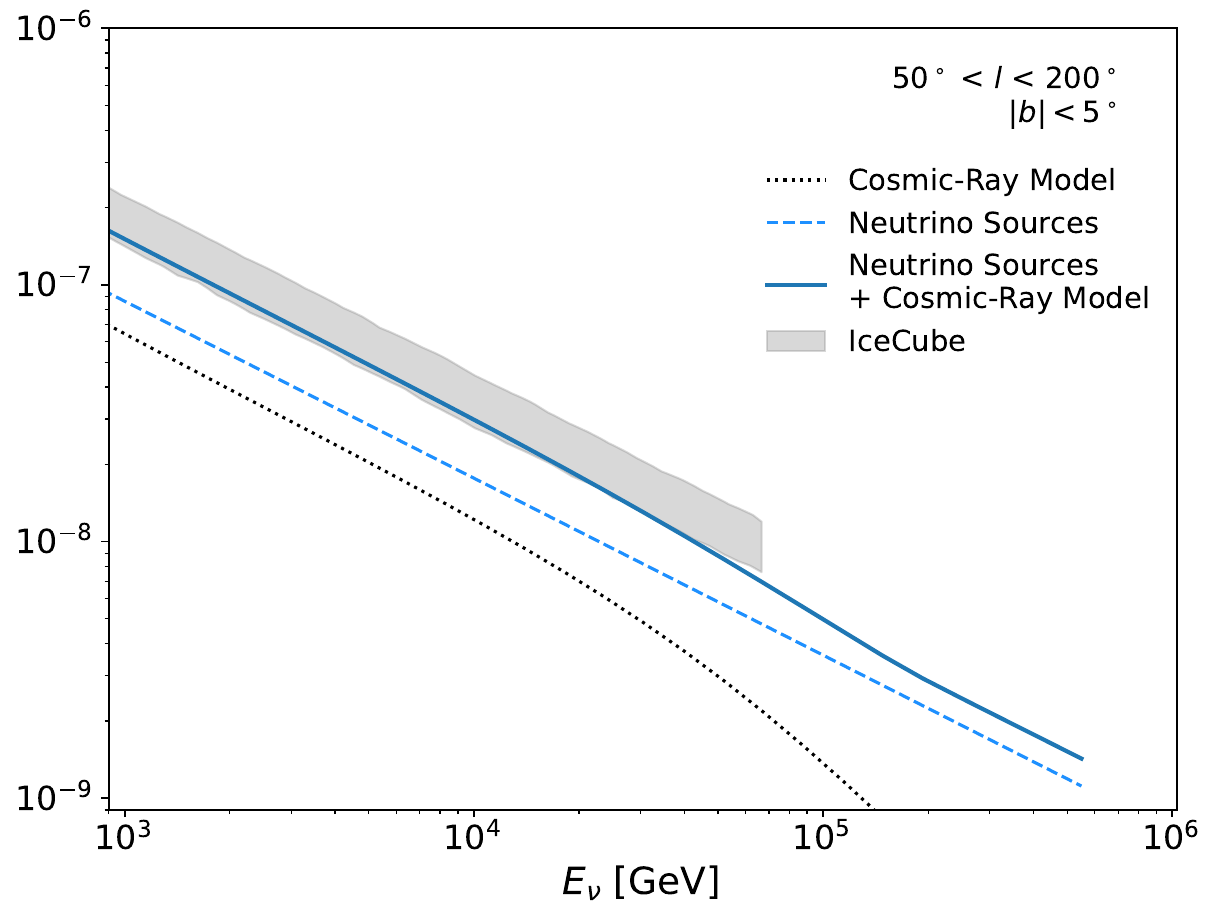}\\
\includegraphics[width=0.48\textwidth, keepaspectratio]{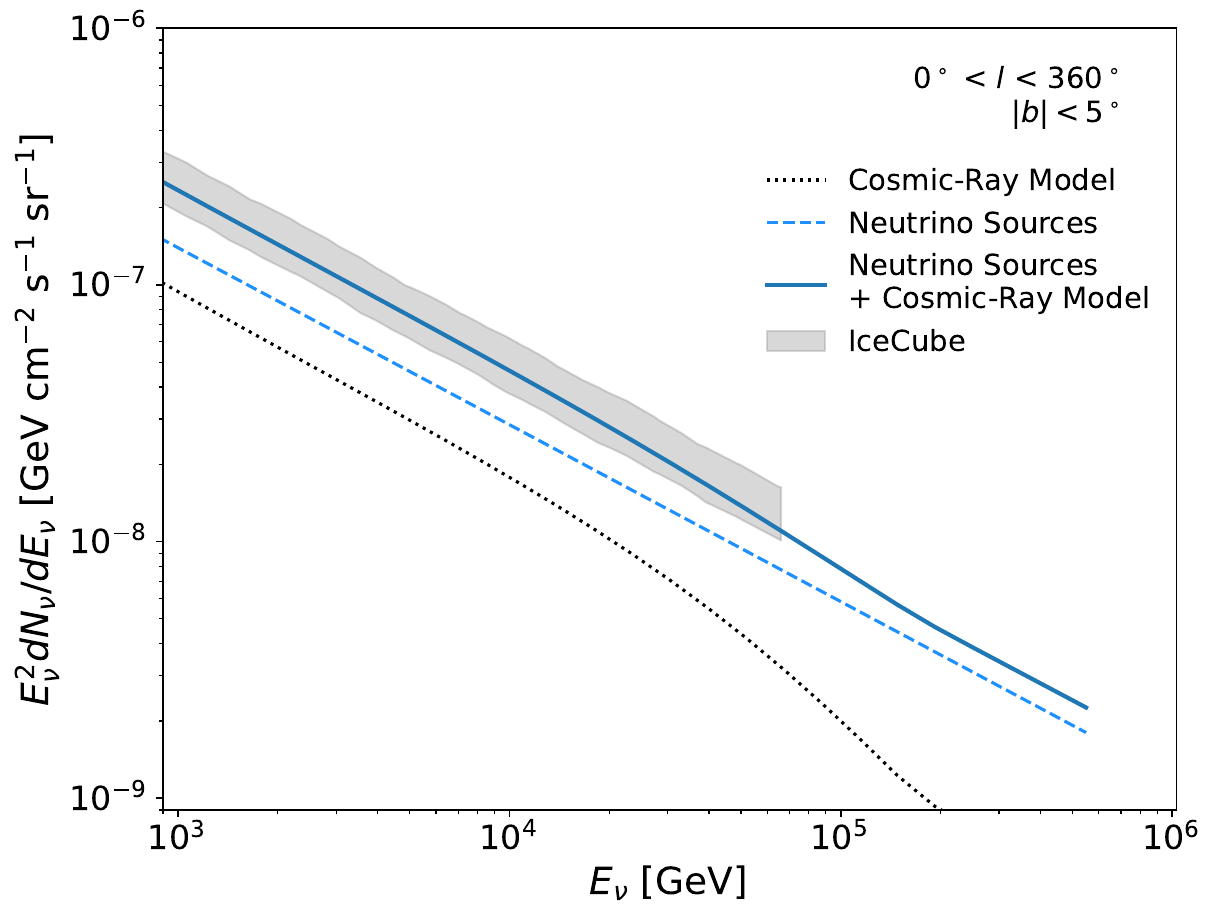}
\caption{
As in Figs.~\ref{fig:CR_A_model_panel}-\ref{fig:CR_E_model_panel}, but adopting cosmic-ray model F.}
\label{fig:CR_F_model_panel} 
\end{figure*}

In Figs.~\ref{fig:CR_A_model_panel}-\ref{fig:CR_F_model_panel}, we show the gamma-ray and neutrino spectra predicted by our model for the best-fit parameter values, and for each of the CR transport models, A-F, as listed in Table~\ref{tab:ISM_models}. In the upper two frames of these figures, we show the contributions to the Galactic gamma-ray flux from cosmic-ray interactions in the ISM and from unresolved TeV halos. In the lower three frames, we compare IceCube's measurement of the Galactic Plane to the neutrino fluxes predicted from cosmic-ray interactions in the ISM and from unresolved source populations. 

In Table~\ref{tab:Ag4results}, we provide the best-fit values of the parameters for each of the CR transport models considered in this study, as well as the $1\sigma$ uncertainties associated with each of these quantities. We also show the contributions to the $\chi^2$ of the fit from LHAASO's measurements of the inner and outer Galactic Plane ($\chi^2_{\gamma}$), from IceCube's measurement of the inner and outer Galactic Plane ($\chi^2_{\nu}$), and the sum of these contributions ($\sum \chi^2$). These models each provide a broadly good description of the diffuse gamma-ray and neutrino emission reported by LHAASO and IceCube, yielding total $\chi^2$ values between 25.4 and 31.6, with $19-5$ corresponding degrees-of-freedom.  

We have repeated these fits without including any contribution from neutrino source populations (i.e.~setting $A_{\nu}=0$). We find, however, that models with $A_{\nu}=0$ provide a far worse fit to the data, at a level of $\Delta \chi^2 \equiv \chi^2_{\nu}(A_{\nu}=0) - \chi^2_{\nu}(A_{\nu}\ne 0) = 29.9-38.2$. This represents highly statistically significant evidence for one or more source populations, with a spectral index of roughly $\alpha_{\nu} \sim 2.7$, which contribute to the neutrino flux observed from the Galactic Plane, in addition to the neutrino emission from CR interactions in the ISM. Across this range of models, we can place a lower limit of $A_{\nu} > 2.57 \times 10^{-9}$ at the 95\% confidence level, corresponding to the requirement that no less than $\sim$20\% of the neutrino flux from the Galactic Plane comes from unresolved source populations. For the best-fit parameters, our model predicts that such sources are responsible for $\sim $60-90\% of the neutrino emission observed from the Galactic Plane.

Recall that in each of these models, we have restricted the normalization of the diffuse emission to fall within a factor of 4 of that predicted by~\texttt{GALPROP}, $ 0.25 < A_{G} < 4$. For five of the six CR models considered, the fit preferred values of $A_G$ in this range, so this restriction had no impact on our results. In the case of model E, however, the fit preferred somewhat larger values of this quantity. To explore this case further, we allowed $A_G$ to take on larger values in the scenario we have called Model E$^{\prime}$, whose results are given in Table~\ref{tab:Ag4results} (see also, Appendix~\ref{sec:AppendixB}).

\begin{figure*}[ht]
\includegraphics[width=0.48\linewidth]{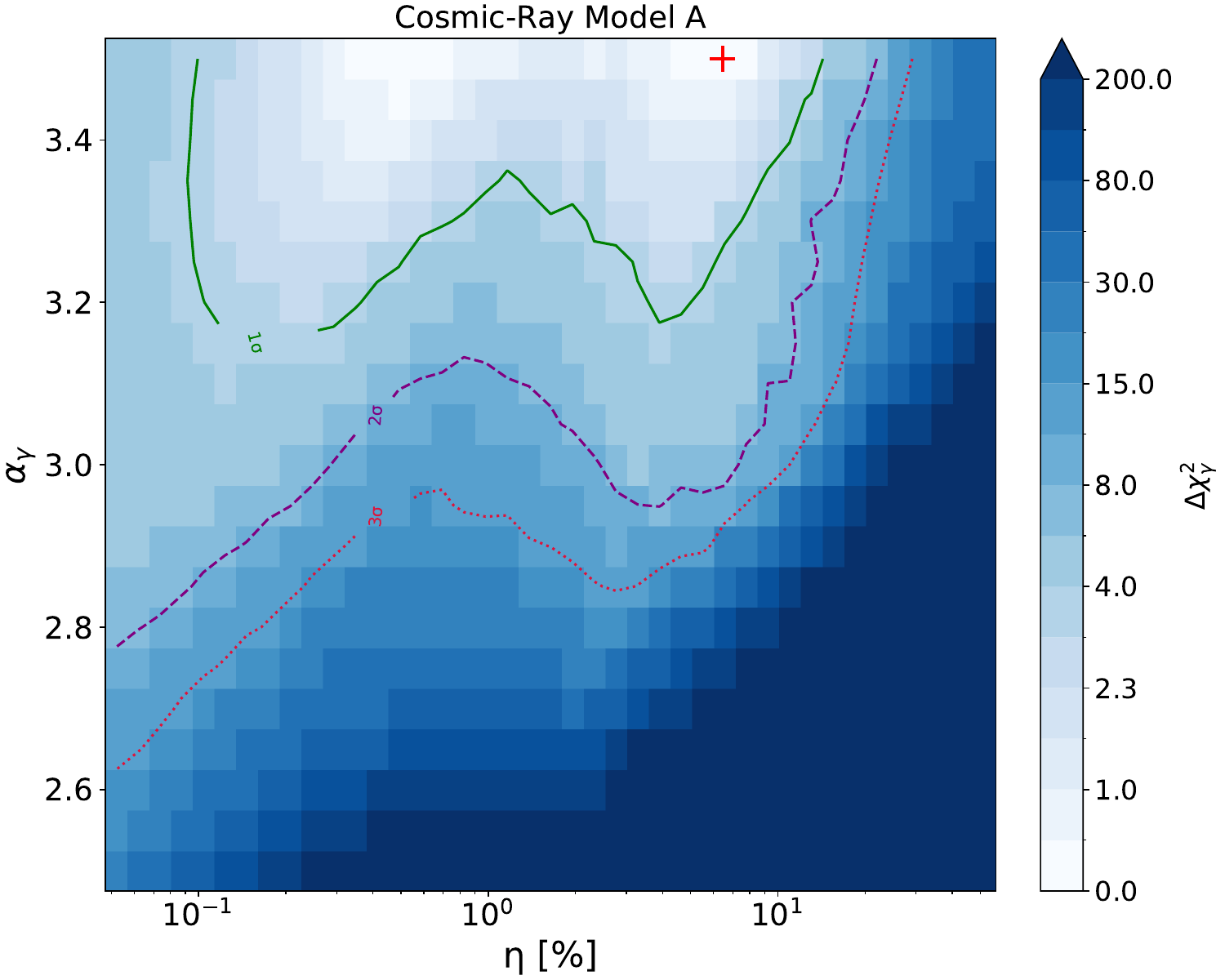}
\hfill
\includegraphics[width=0.48\linewidth]{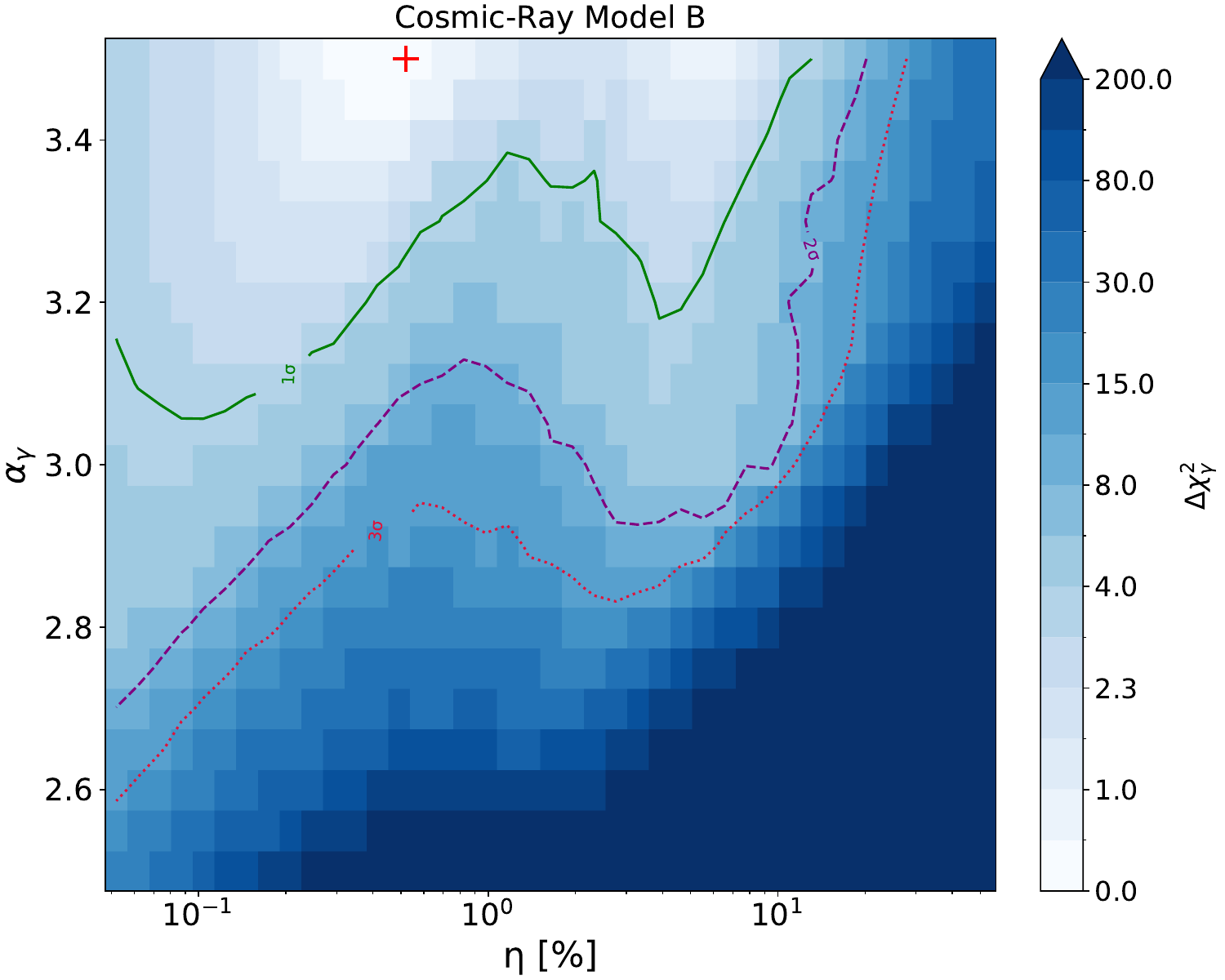}
\hfill \\
\vspace{0.5cm}
\includegraphics[width=0.48\linewidth]{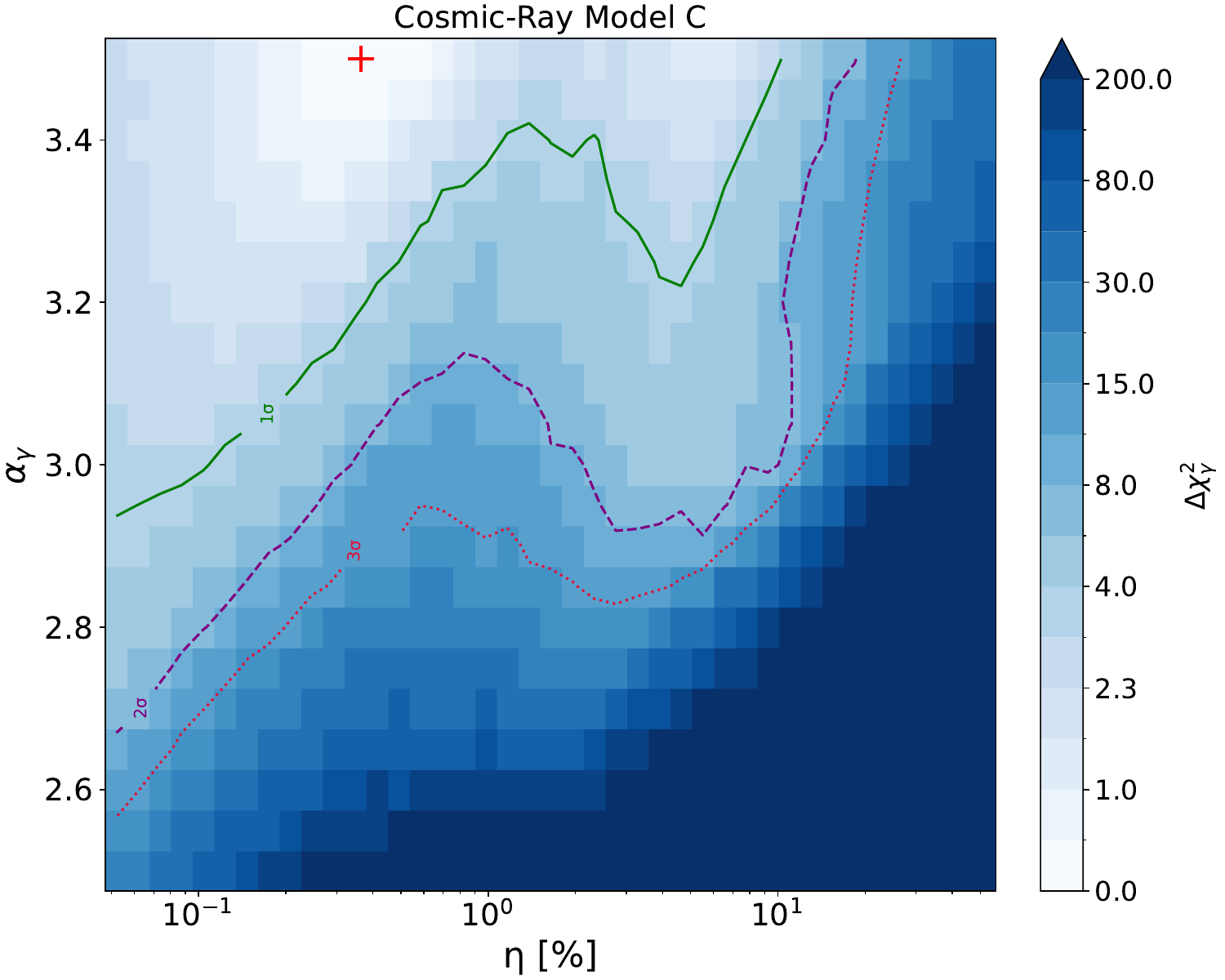}
\hfill
\includegraphics[width=0.48\linewidth]{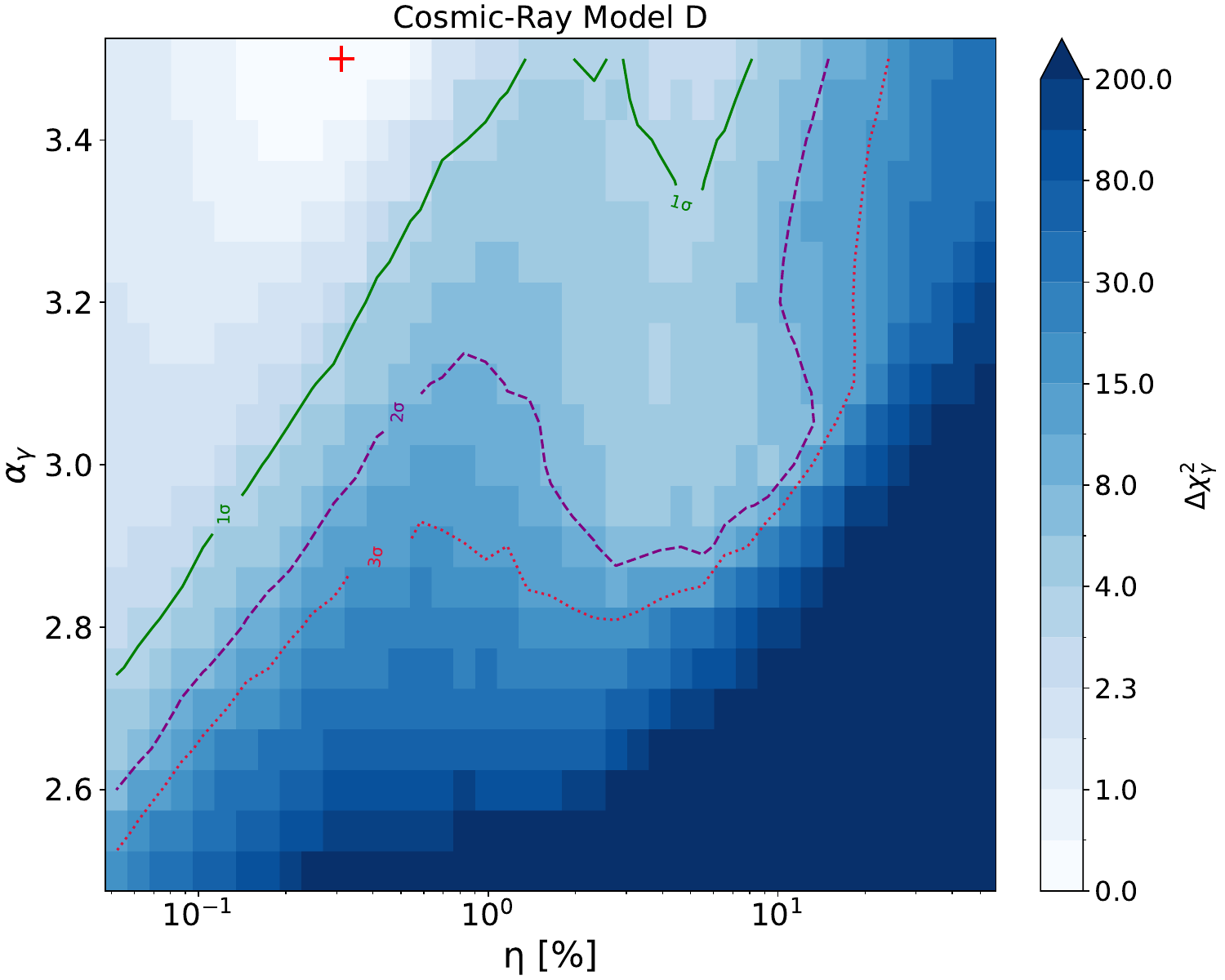}
\hfill \\
\vspace{0.5cm}
\includegraphics[width=0.48\linewidth]{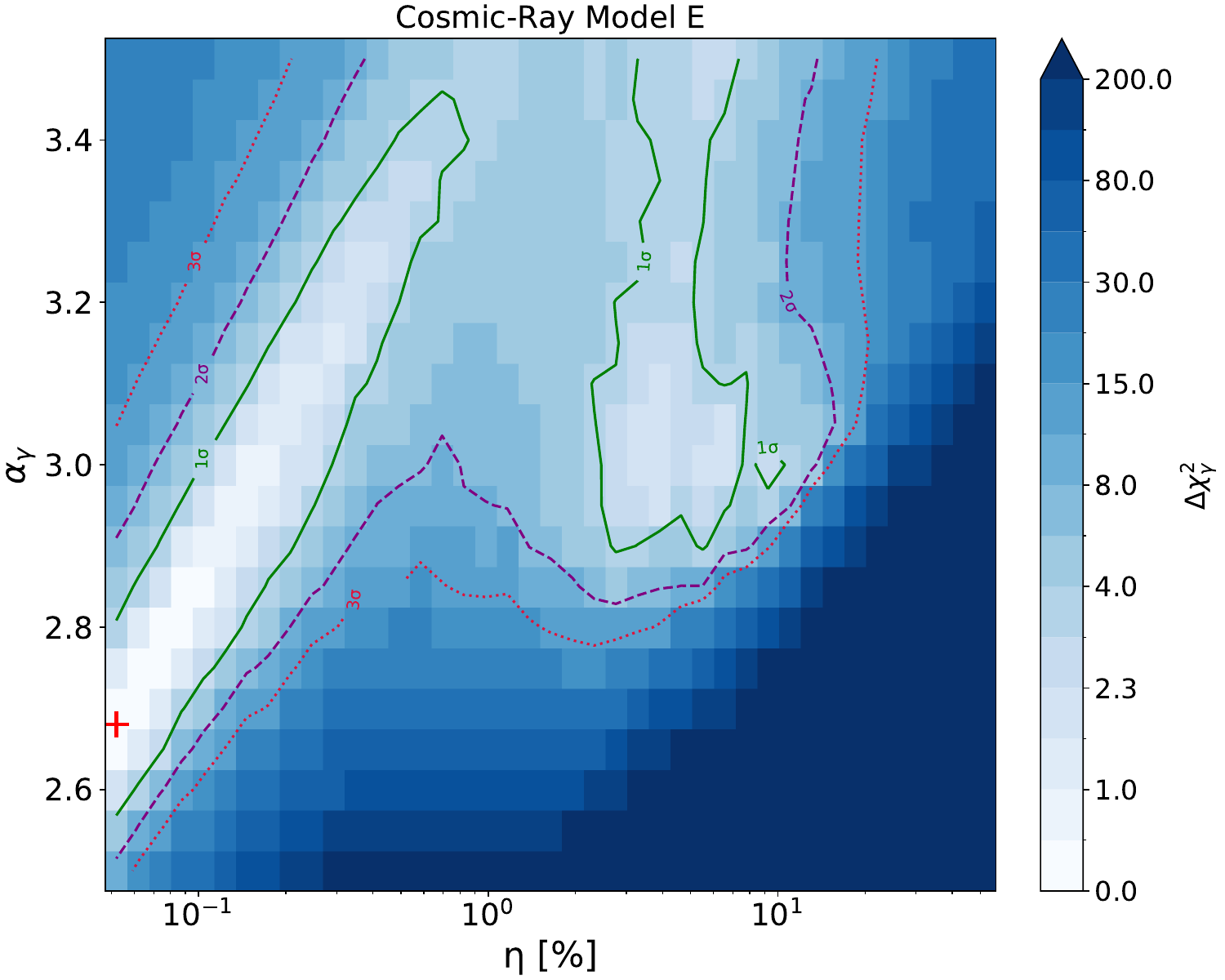}
\hfill
\hspace{0.5cm}
\includegraphics[width=0.48\linewidth]{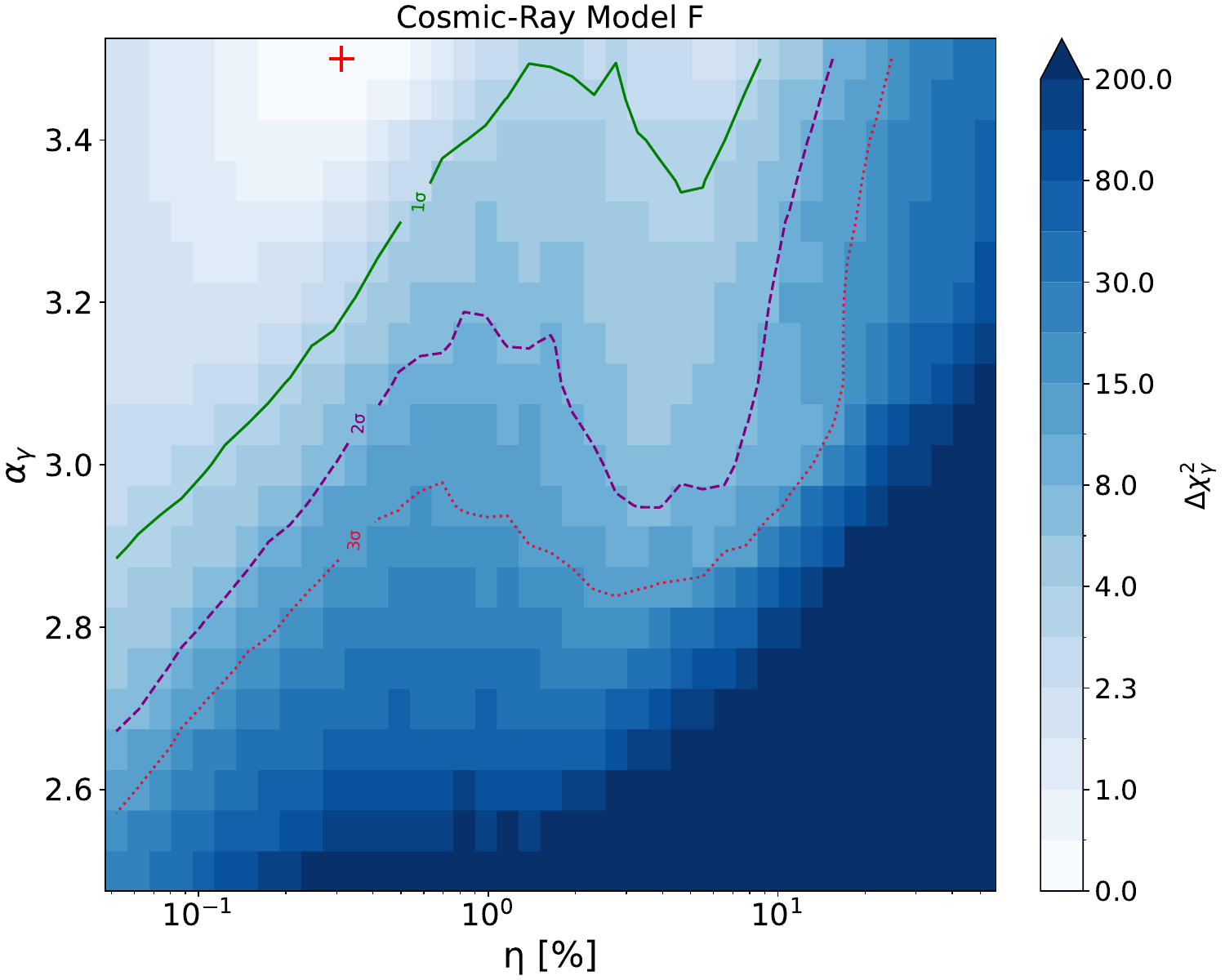}
\hfill

\caption{The values of the spectral index ($\alpha_{\gamma}$) and efficiency ($\eta$) of TeV halos, as favored by our fit to the LHAASO measurements of diffuse gamma-ray emission from the Galactic Plane. Results are shown for each of the six cosmic-ray models considered in this study. The color scale reflects the change in $\chi^2_{\gamma}$ relative to the best-fit parameters, as indicated by a red cross. We have fixed the cosmic-ray model normalization, $A_G$, in each panel to its best-fit value. The contours represent the 1$\sigma$, 2$\sigma$, and 3$\sigma$ confidence regions, as evaluated for the three inferred parameters, $\alpha_{\gamma}$, $\eta$ and $A_G$.}
\label{fig:landscape_constrained_alphaeta}
\end{figure*}

\begin{figure*}[p]
\includegraphics[width=0.48\linewidth]{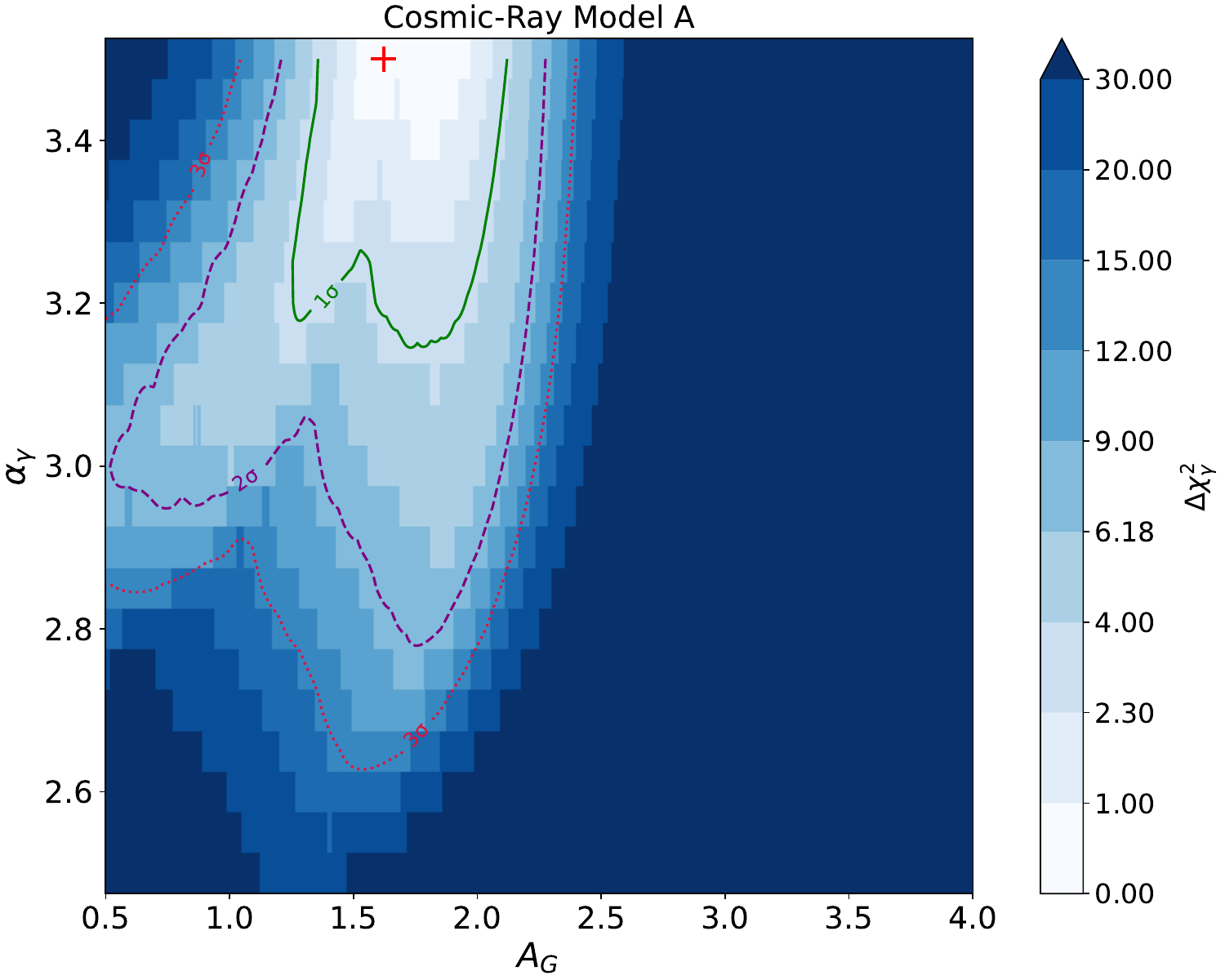}
\hfill
\includegraphics[width=0.48\linewidth]{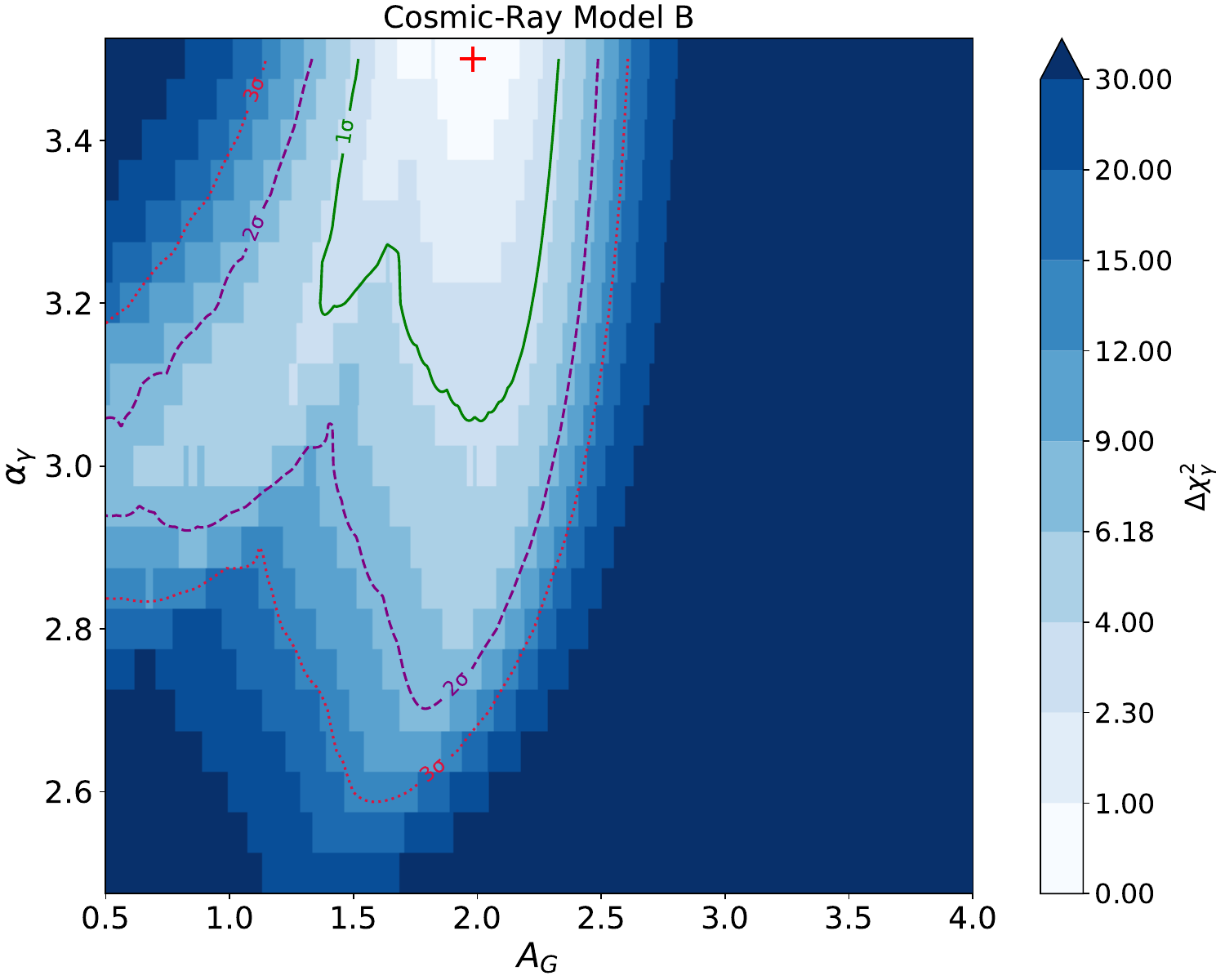}
\hfill \\
\vspace{0.5cm}
\includegraphics[width=0.48\linewidth]{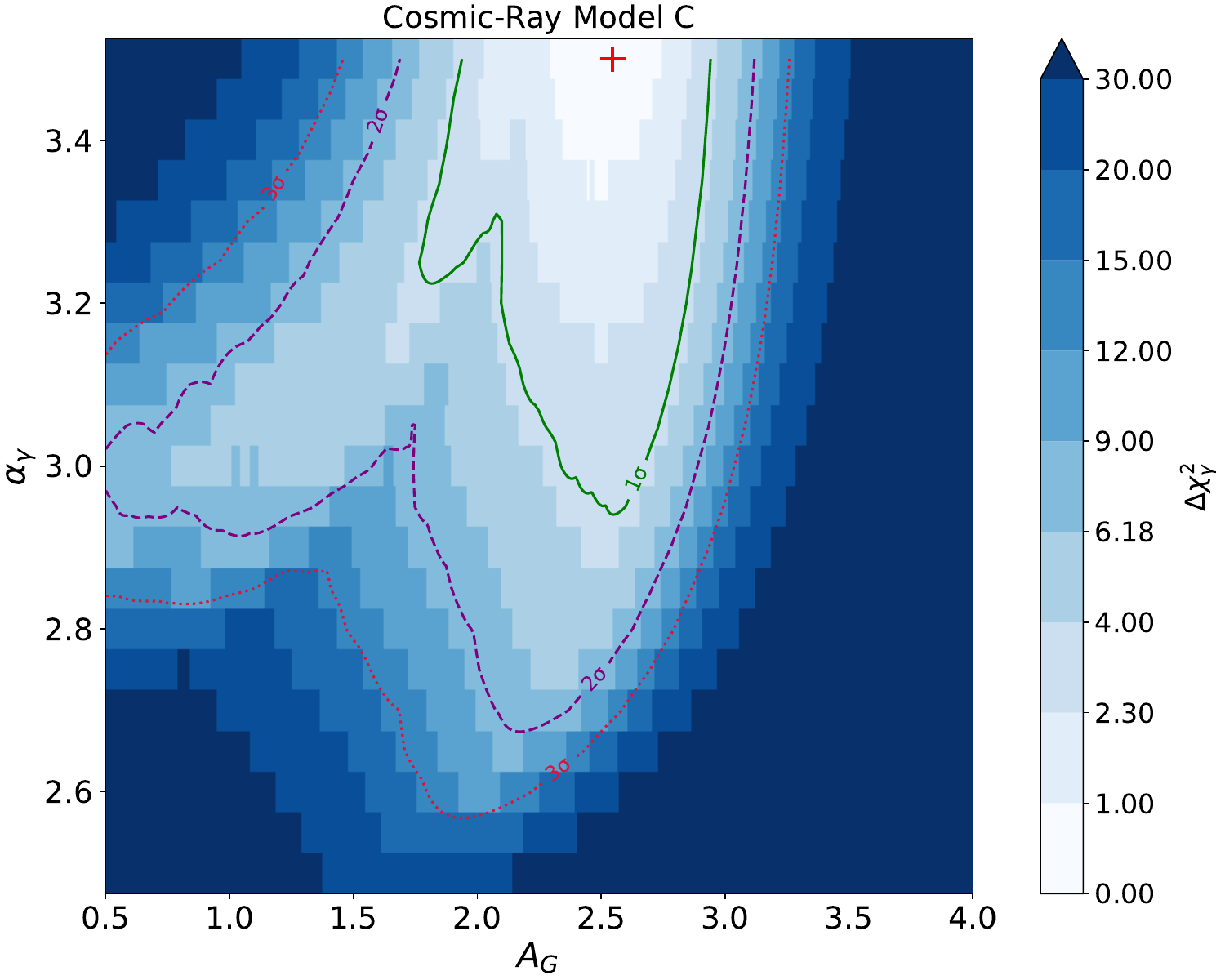}
\hfill
\includegraphics[width=0.48\linewidth]{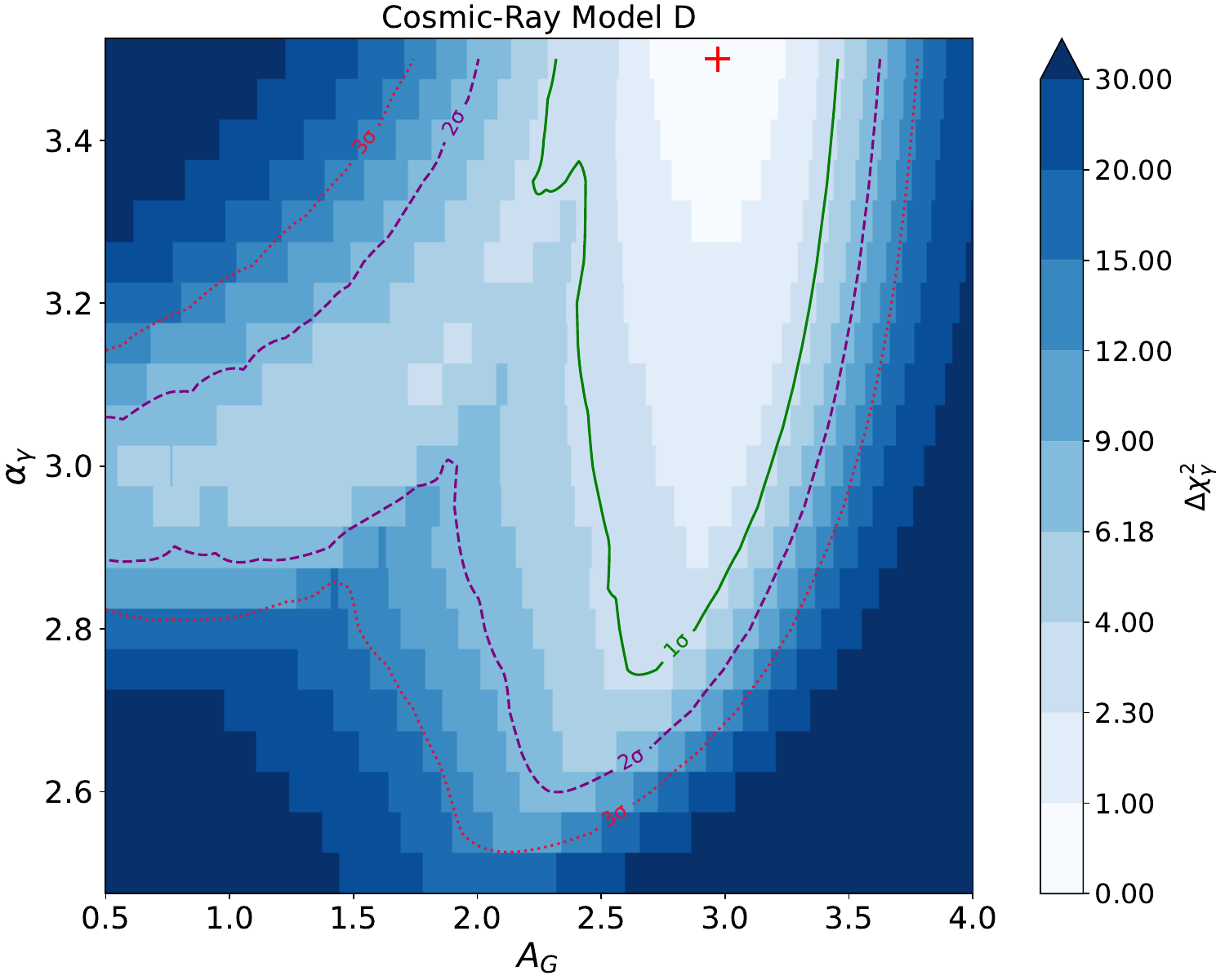}
\hfill \\
\vspace{0.5cm}

\includegraphics[width=0.48\linewidth]{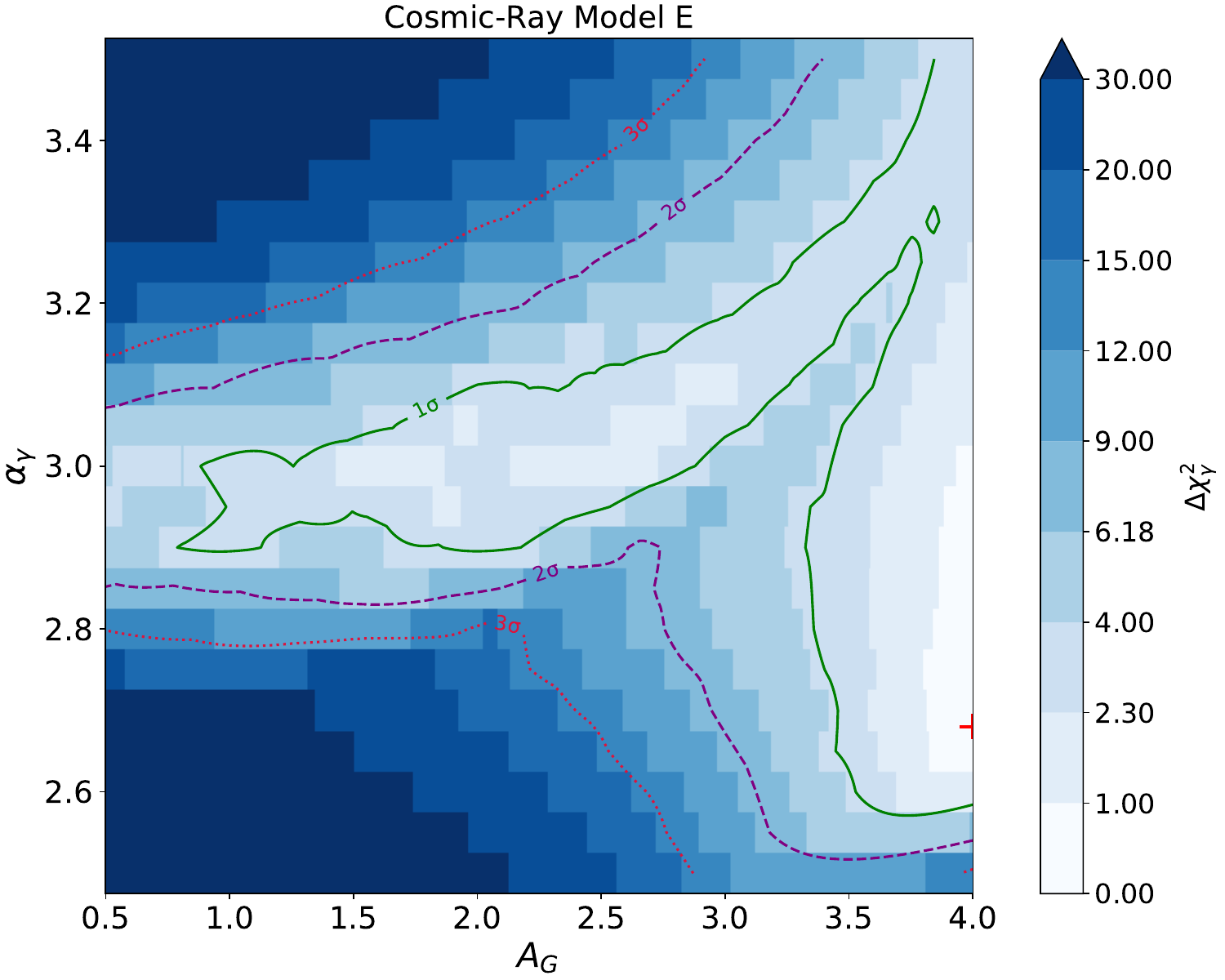}\hfill
\hspace{0.5cm}
\includegraphics[width=0.48\linewidth]{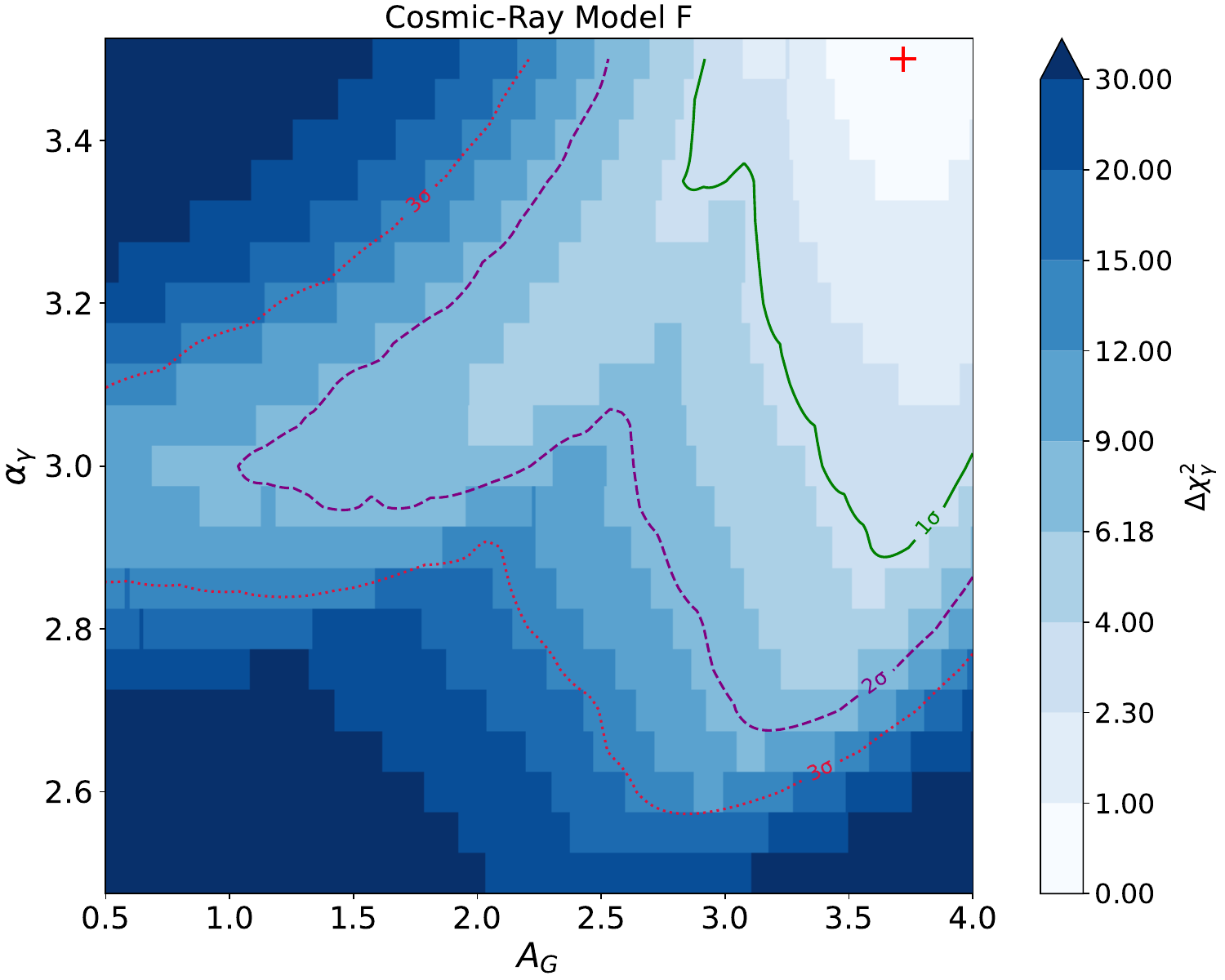}
\hfill
\caption{
As in Fig.~\ref{fig:landscape_constrained_alphaeta}, but showing the values of the TeV halo spectral index ($\alpha_{\gamma}$) and cosmic-ray model normalization ($A_G$) favored by our fit to the LHAASO data. We have fixed the TeV halo efficiency, $\eta$, in each panel to its best-fit value.}
 \label{fig:landscape_constrained_alphaAg}
\end{figure*}

In Figs.~\ref{fig:landscape_constrained_alphaeta} and~\ref{fig:landscape_constrained_alphaAg}, we show the $\eta-\alpha_{\gamma}$ and $A_G-\alpha_{\gamma}$ parameter space that is favored by our fits, for each of the six CR transport models considered in this study. In large part due to the impact of LHAASO's masking in their analysis, these fits allow for a relatively wide range of $\eta$ values. Also note that in comparing these results to those provided in Table~\ref{tab:Ag4results}, the quoted parameter uncertainties in the table have been evaluated based on the one-parameter 1$\sigma$ errors (i.e.~for $\Delta\chi^2=1$), whereas the uncertainty contours shown in Figs.~\ref{fig:landscape_constrained_alphaeta} and~\ref{fig:landscape_constrained_alphaAg} were evaluated for three free parameters, such that contours shown correspond to $\Delta \chi^2_{\gamma}=3.53$, 8.03, and 14.16.

For each of the CR transport models considered in this study, the best-fits parameter values predict that the largest contribution to the diffuse gamma-ray emission from the Galactic Plane comes from CR interactions with gas in the ISM, rather than from unresolved TeV halos (an exception is in Model A, where the gamma-ray emission from TeV halos in the inner Galactic Plane exceeds that from CR interactions at energies below $E_{\gamma} \sim 12 \, {\rm TeV}$). This is in contrast to the conclusion reached in Ref.~\cite{dekker_diffuse_2024}, which favored scenarios in which TeV halos provide the dominant contribution to the diffuse gamma-ray emission from the inner Galactic Plane at energies between $E_{\gamma} \sim 10-100 \, {\rm TeV}$. That being said, the parameters favored in that study ($\eta \sim 5.2\%$, $\alpha_{\gamma} \sim 3.1$) are within the $(1-2)\sigma$ uncertainties of the analysis presented here, as shown in Fig.~\ref{fig:landscape_constrained_alphaeta}.

Our fits tend to favor a soft spectral index for the gamma-ray emission from unresolved TeV halos, near the largest value allowed in our parameter scans, $\alpha_{\gamma} \sim 3.5$. As can be seen in Figs.~\ref{fig:landscape_constrained_alphaeta}-\ref{fig:landscape_constrained_alphaAg}, however, this preference is rather mild and spectral indices as hard as $\alpha_{\gamma} \sim 2.8$ remain entirely consistent with the LHAASO data. The results presented here are compatible with parameter values implied by observations of individual TeV halos, $\alpha_{\gamma} \sim 3$ and $\eta \sim 1-10\%$~\cite{collaboration_multiple_2020,sudoh_highest_2021,HAWC:2024scl}.

\clearpage

\section{Summary and Conclusions}
\label{Sec:Summary}

In this study, we have performed a joint analysis of the high-energy neutrino emission observed from the Galactic Plane by IceCube and the diffuse ultra-high-energy gamma-ray emission measured by LHAASO. To this end, we combined a set of physically motivated models describing cosmic-ray transport and interactions in the interstellar medium, the contribution of unresolved TeV halos, and the emission from unresolved Galactic neutrino sources. By fitting these models to the available gamma-ray and neutrino data, we have investigated the relative roles of diffuse processes and source populations in generating the high-energy emission observed from the Milky Way.

To account for uncertainties associated with Galactic cosmic-ray propagation, we have considered six benchmark cosmic-ray transport models that are capable of reproducing local cosmic-ray measurements and GeV-scale gamma-ray observations. We found that each of these models provides an acceptable description of the combined IceCube and LHAASO data when supplemented by contributions from unresolved TeV halos and unresolved neutrino sources. Although the preferred normalization of the diffuse Galactic component varies significantly among the different cosmic-ray models, our principal conclusions are found to be robust.

For the gamma-ray emission, our fits generally favor scenarios in which cosmic-ray interactions in the interstellar medium provide the dominant contribution to the diffuse flux observed by LHAASO. While unresolved TeV halos can contribute significantly, especially in the inner Galactic Plane, they are not required to dominate the diffuse gamma-ray emission. This conclusion differs somewhat from that reached in Ref.~\cite{dekker_diffuse_2024}, reflecting the impact of the broader range of cosmic-ray propagation models considered here. At the same time, the values of the TeV halo efficiency and spectral index preferred by our fits remain consistent with those inferred from observations of individual TeV halos.

Our most significant result concerns the origin of the Galactic neutrino flux. Across all six cosmic-ray transport models, we find a strong statistical preference for the presence of one or more unresolved neutrino source populations. Models containing only diffuse neutrino emission from cosmic-ray interactions in the interstellar medium provide substantially poorer fits to the IceCube data, with $\Delta\chi^2 = 25.5$-38.2 relative to models that include unresolved sources. This source component exhibits a spectral index of approximately $dN_{\nu}/dE_{\nu} \propto E_{\nu}^{-2.7}$ and is responsible for the majority of the observed Galactic neutrino flux. For the best-fit parameters, unresolved sources account for roughly 60-90\% of the neutrino emission observed from the Galactic Plane. More conservatively, after marginalizing over the uncertainties associated with cosmic-ray transport and source properties, we find that at least $\sim20\%$ of the Galactic neutrino flux must originate from source populations at the 95\% confidence level.

These results provide compelling evidence that the neutrino emission observed by IceCube from the Galactic Plane is not generated solely by diffuse cosmic-ray interactions in the interstellar medium. Instead, a substantial fraction of this flux appears to arise from unresolved Galactic cosmic-ray accelerators, such as supernova remnants, pulsar wind nebulae, microquasars, or other classes of high-energy sources. Further observations by existing and next-generation neutrino telescopes will enable more precise measurements of the neutrino emission from the Galactic Plane, helping to identify the source populations responsible for this emission and further clarifying the origin of the high-energy particles observed throughout the Milky Way.

\vspace{12pt}


{\it Acknowledgements}: AR and DH are supported by the Office of the Vice Chancellor for Research at the University of Wisconsin–Madison with funding from the Wisconsin Alumni Research Foundation. IC acknowledges support from the U.S. Department of Energy, Office of Science, Office of High Energy Physics, under Award No. DE-SC0022352.

\bibliography{references1}

\appendix

\section{Results Using KRA$^5_{\gamma}$ and KRA$^{50}_{\gamma}$ Templates}
\label{sec:AppendixA}

\begin{table*}[t]
\centering

\begin{tabular}{c | cccc c | cccc c}
\hline\hline
Model & \multicolumn{5}{c|}{KRA$^5_{\gamma}$} & \multicolumn{5}{c}{KRA$^{50}_{\gamma}$} \\
 & $A_{\nu}$ & $\alpha_{\nu}$ & $\chi^2_{\nu}$ & $\Delta\chi^2_{\nu}$ & $\sum\chi^2$ & $A_{\nu}$ & $\alpha_{\nu}$ & $\chi^2_{\nu}$ & $\Delta\chi^2_{\nu}$ & $\sum\chi^2$ \\
\hline
 A & $1.26^{+1.22}_{-1.04} \times 10^{-8}$ & $1.31^{+1.07}_{-0.31}$ & $2.50$ & $3.28$ & \bf{$24.56$} & $6.39^{+4.72}_{-4.99} \times 10^{-9}$ & $1.00^{+0.93}_{-0.00}$ & $6.32$ & $1.77$ & \bf{$28.38$} \\

 B & $1.19^{+0.88}_{-1.04} \times 10^{-8}$ & $1.00^{+1.22}_{-0.00}$ & $3.73$ & $1.78$ & \bf{$27.23$} & $4.05^{+4.74}_{-4.05} \times 10^{-9}$ & $1.00^{+2.00}_{-0.00}$ & $11.95$ & $0.72$ & \bf{$35.45$} \\
 C & $1.05^{+0.88}_{-0.96} \times 10^{-8}$ & $1.00^{+1.15}_{-0.00}$ & $5.15$ & $1.38$ & \bf{$28.63$} & $3.34^{+4.75}_{-3.33} \times 10^{-9}$ & $1.00^{+2.00}_{-0.00}$ & $16.49$ & $0.48$ & \bf{$39.97$} \\
 D & $9.41^{+8.87}_{-9.09} \times 10^{-9}$ & $1.00^{+1.15}_{-0.00}$ & $7.04$ & $1.11$ & \bf{$32.45$} & $2.83^{+4.72}_{-2.82} \times 10^{-9}$ & $1.00^{+2.00}_{-0.00}$ & $21.75$ & $0.35$ & \bf{$47.16$} \\
 E & $1.51^{+0.87}_{-1.09} \times 10^{-8}$ & $1.00^{+0.89}_{-0.00}$ & $5.50$ & $2.85$ & \bf{$33.86$} & $6.13^{+4.68}_{-4.81} \times 10^{-9}$ & $1.00^{+0.70}_{-0.00}$ & $17.26$ & $1.64$ & \bf{$45.62$} \\
 F & $1.08^{+0.88}_{-0.95} \times 10^{-8}$ & $1.00^{+1.03}_{-0.00}$ & $6.34$ & $1.48$ & \bf{$28.58$} & $3.58^{+4.74}_{-3.57} \times 10^{-9}$ & $1.00^{+2.00}_{-0.00}$ & $20.13$ & $0.55$ & \bf{$42.37$} \\

 \hline
 E$'$ & $7.14^{+8.78}_{-7.14} \times 10^{-9}$ & $1.00^{+2.00}_{-0.00}$ & $14.39$ & $0.63$ & \bf{$40.69$} & $1.74^{+4.74}_{-1.74} \times 10^{-9}$ & $1.00^{+2.00}_{-0.00}$ & $39.17$ & $0.13$ & \bf{$65.47$} \\

\hline\hline
\end{tabular}
\caption{The results of our joint likelihood similar to Table \ref{tab:Ag4results}, but with the neutrino emission from the Galactic Plane extracted using KRA$^5_{\gamma}$ and KRA$^{50}_{\gamma}$ templates. We have restricted the normalization of the Galactic diffuse emission to $0.25 \le A_G \le 4$, except in the case of Model E$^{\prime}$, for which this prior requirement was not imposed. We have also required $\alpha_{\gamma} \le 3.5$ and $\alpha_{\nu} \ge 1.0$. The errors quoted are at the $1\sigma$ confidence level (corresponding to $\Delta \chi^2=1$).  $\chi^2_{\nu}$ is the $\chi^2$ obtained from IceCube's measurements of the inner and outer Galactic Plane given the respective model, evaluated at the best-fit values of $A_G$, $A_{\nu}$, and $\alpha_{\nu}$. The quantity $\Delta \chi^2_{\nu}$ represents the change in the value of $\chi^2$ that results if neutrino sources are not included in the fit, reflecting the statistical preference for neutrino emission from one or more source populations. The rightmost column provides the sum of $\chi^2_{\gamma}$ and $\chi^2_{\nu}$, where $\chi^2_{\gamma}$ is taken from Table \ref{tab:Ag4results}.}
\label{tab:Ag4results_KRA}
\end{table*}

In the main body of this paper, we adopted the spectrum of neutrino emission from the Galactic Plane as obtained by IceCube using the $\pi^0$ template~\cite{icecube_collaboration_observation_2023}. In Table~\ref{tab:Ag4results_KRA}, we present analogous results obtained using the KRA$^5_{\gamma}$ and KRA$^{50}_{\gamma}$ templates.

These fits favor very hard (and seemingly unphysical) spectral indices for the neutrino source population, and we have imposed the requirement $\alpha_{\nu} \ge 1.0$ in our fits. These fits also provide considerably weaker evidence for any contribution from neutrino sources, yielding only $\Delta \chi^2_{\nu} \sim 1.1-3.3$ for models A-F. 

For several reasons, we consider the results presented in the main text (based on the $\pi^0$ template) to be more likely to reflect the true properties of the Galactic neutrino flux than those obtained using either of the KRA$_{\gamma}$ templates. First, the fits obtained using the KRA$_{\gamma}$ templates exhibit a strong preference for extremely hard and seemingly unrealistic neutrino source spectra, with $\alpha_{\nu} \sim 1$. Second, the softer spectrum implied by the $\pi^0$ template appears to be more consistent with recent observations of $\sim 100 \, {\rm TeV}$ gamma rays by the Tibet Air Shower Array~\cite{icecube_collaboration_observation_2023,TibetASg:2021lni}. Third, the IceCube Galactic Plane analysis itself favors the $\pi^0$ template over either of the KRA$_{\gamma}$ templates, yielding significances of 4.71$\sigma$, 4.37$\sigma$, and 3.96$\sigma$ for the $\pi^0$, KRA$^5_{\gamma}$, and KRA$^{50}_{\gamma}$ templates, respectively.

\section{Results For Cosmic-Ray Model E$'$}
\label{sec:AppendixB}

\begin{figure*}[ht]

\includegraphics[width=0.49\textwidth, keepaspectratio]{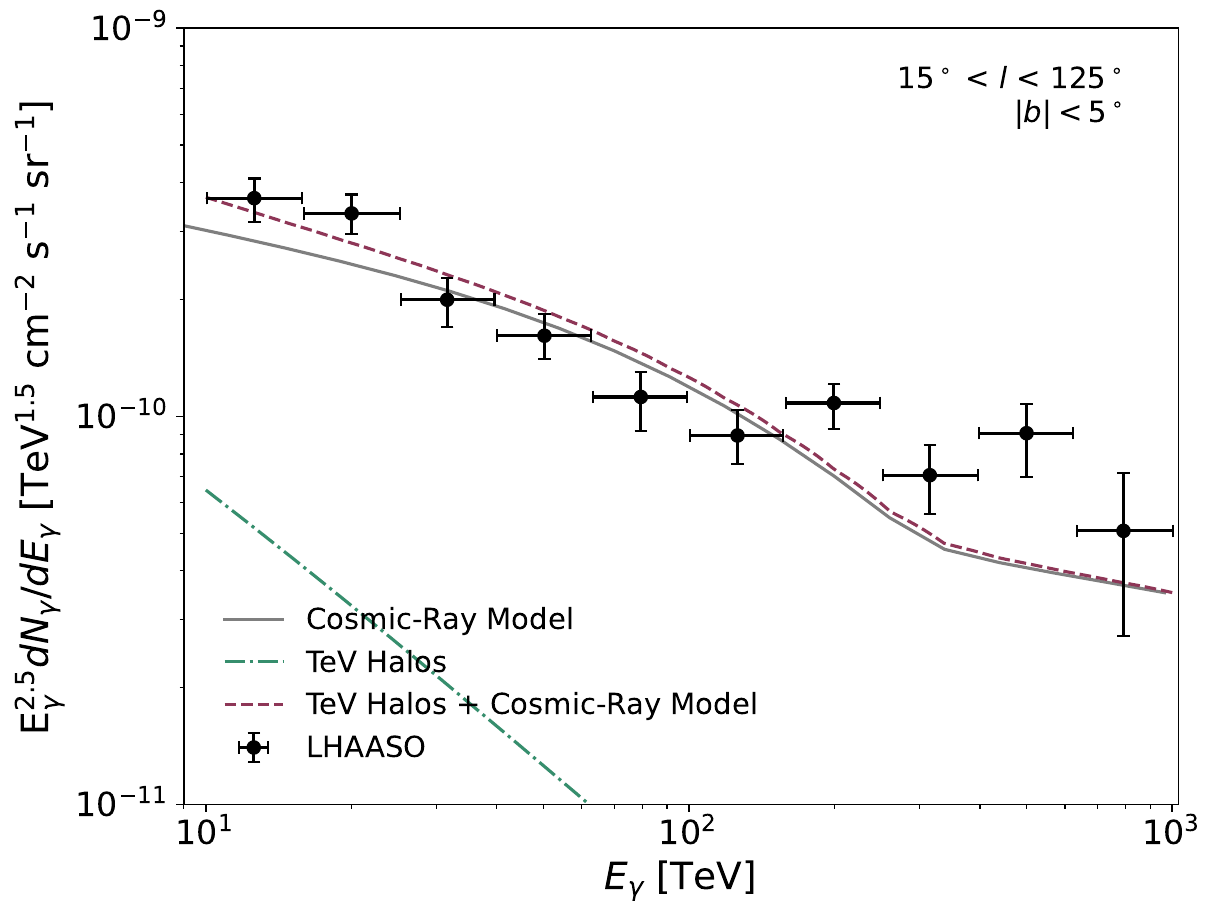}
\includegraphics[width=0.49\textwidth, keepaspectratio]{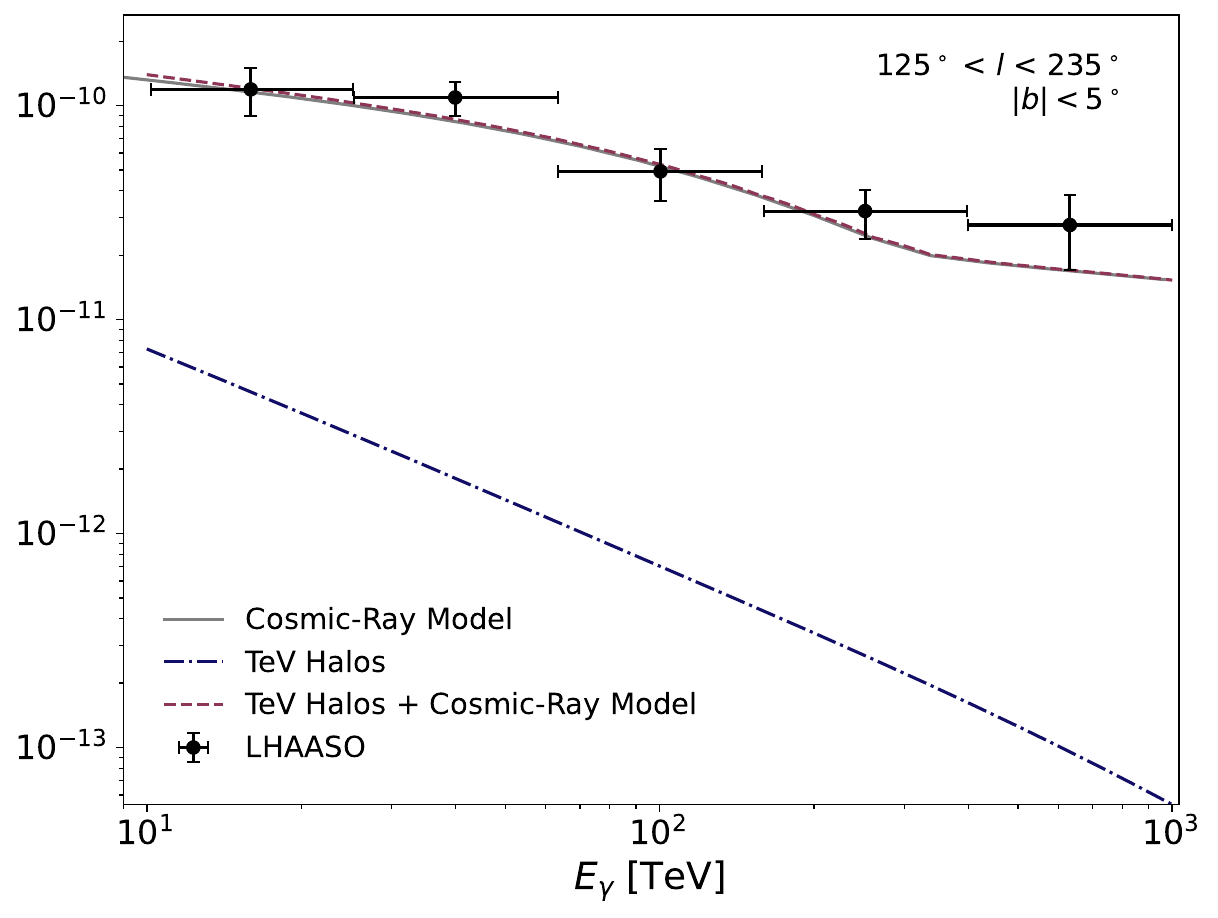}\\
\hspace{0.9mm}
\includegraphics[width=0.48\textwidth, keepaspectratio]{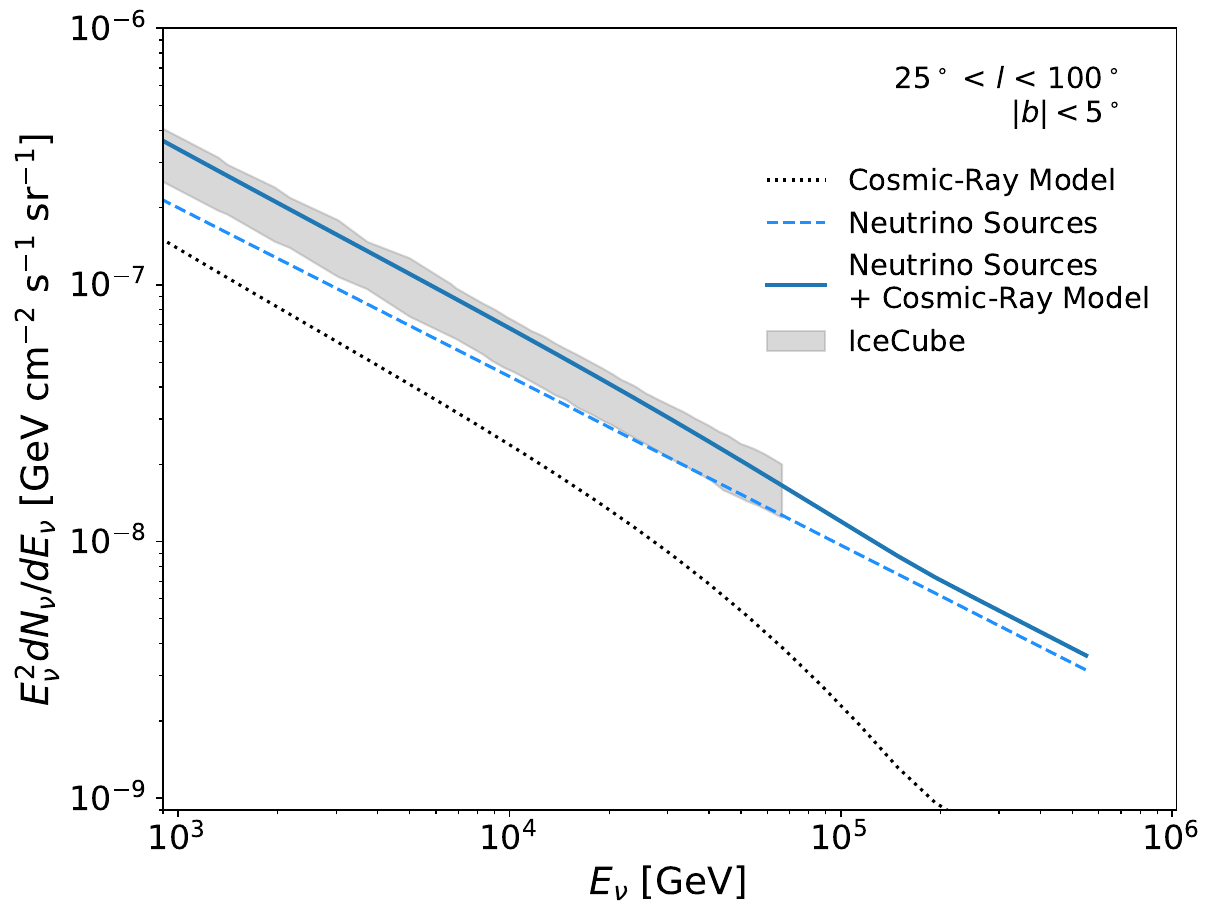}
\hspace{0.6mm}
\includegraphics[width=0.48\textwidth, keepaspectratio]{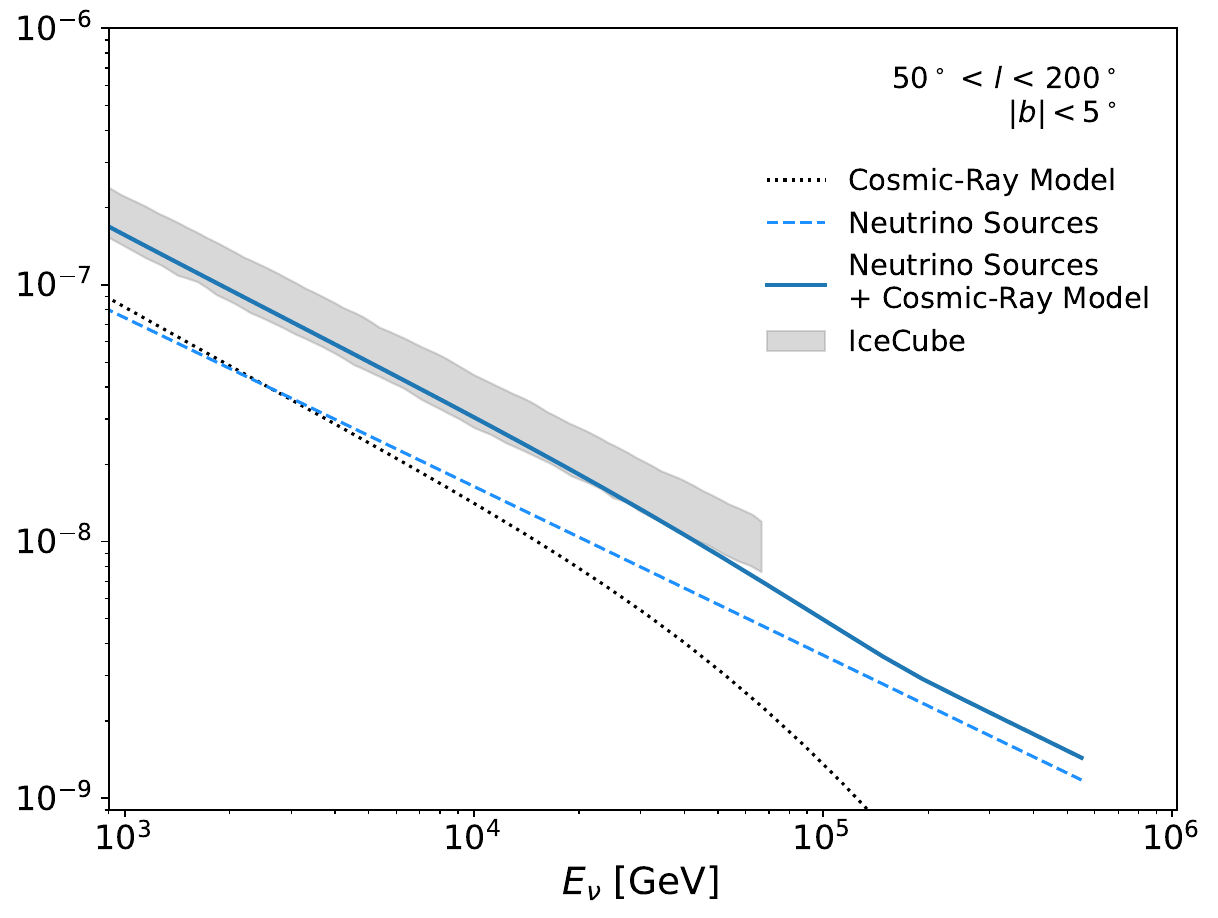}\\
\includegraphics[width=0.48\textwidth, keepaspectratio]{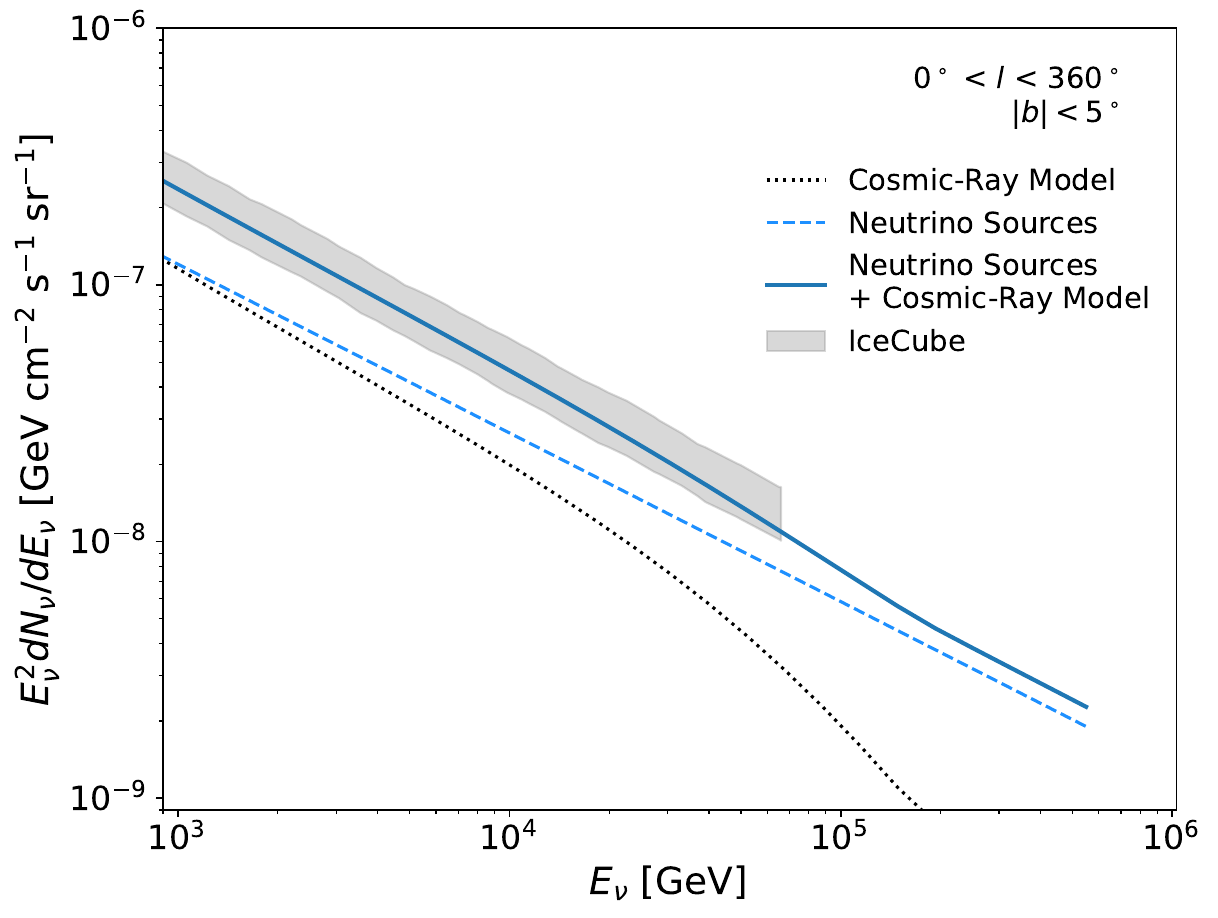}
\caption{As in Figs.~\ref{fig:CR_A_model_panel}-\ref{fig:CR_F_model_panel}, but adopting cosmic-ray model E$'$.}
\label{fig:CR_Eprime_model_panel}
\end{figure*}

\begin{figure*}[ht]
\centering
\includegraphics[width=0.475\linewidth, keepaspectratio]{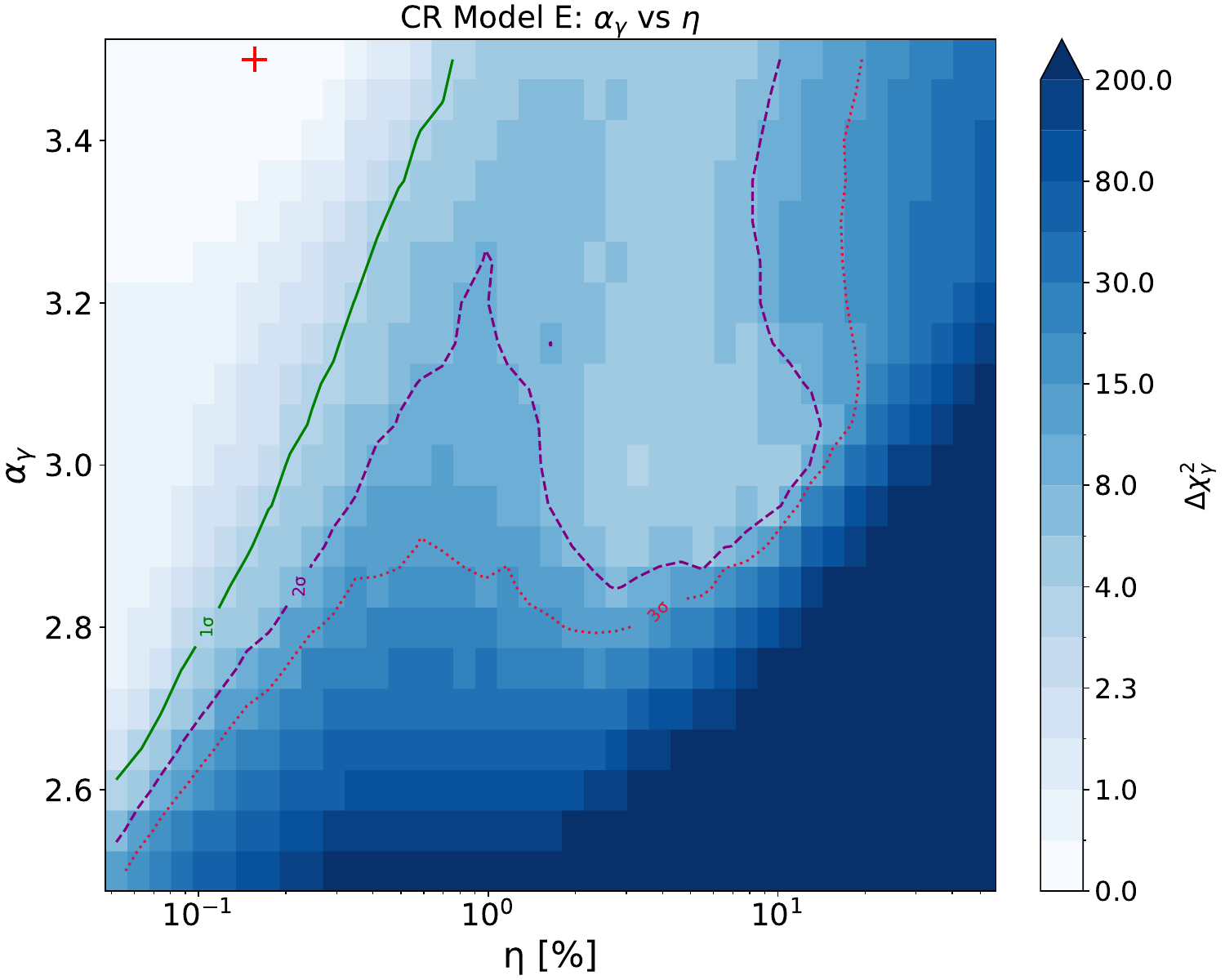}
\hfill
\includegraphics[width=0.475\linewidth, keepaspectratio]{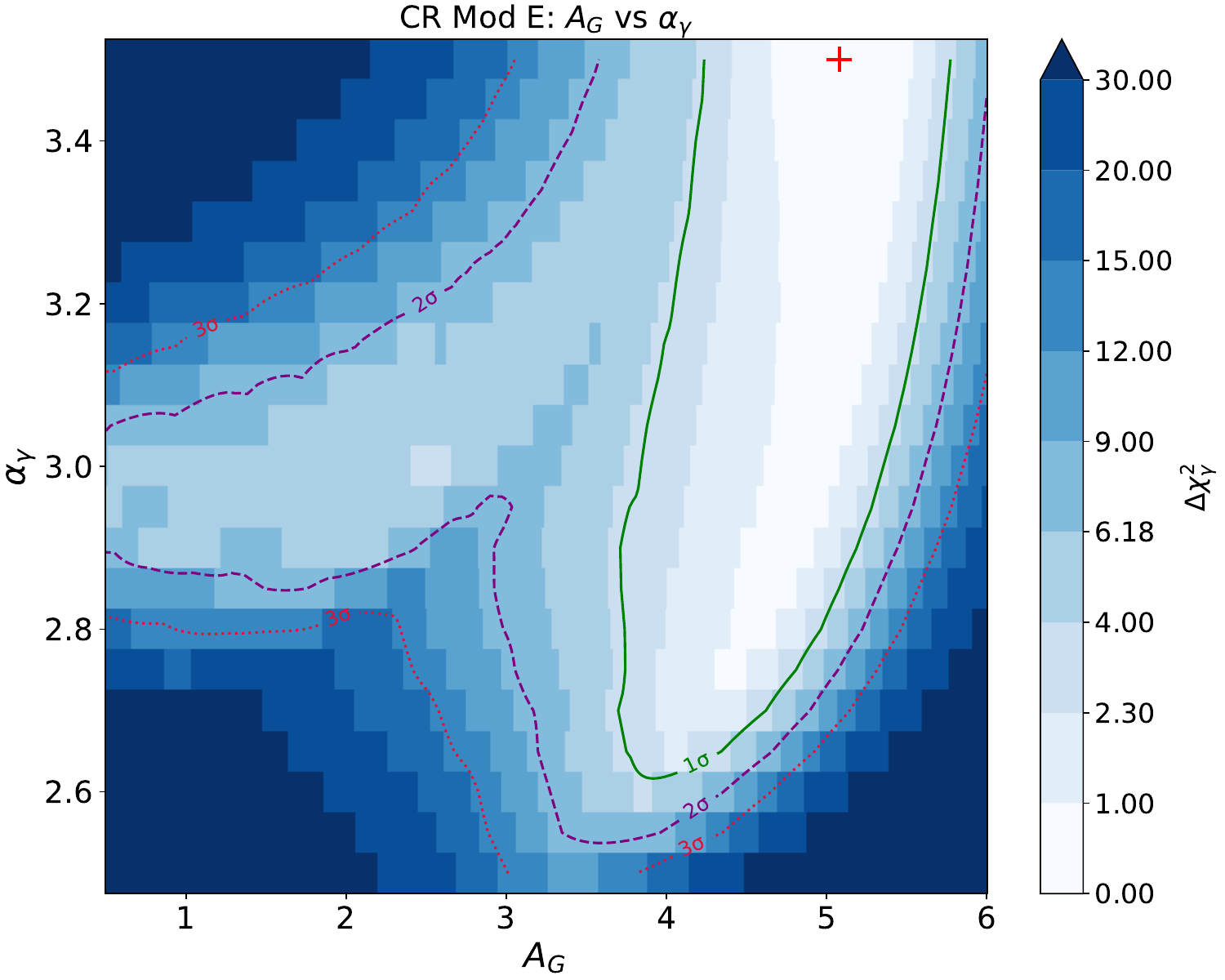}
\caption{As in Figs.~\ref{fig:landscape_constrained_alphaeta} and~\ref{fig:landscape_constrained_alphaAg}, but adopting cosmic-ray model E$'$.}
\label{fig:CR_Eprime_model_landscape}
\end{figure*}

As described in the main text, we allowed the diffuse emission normalization parameter, $A_G$, to take on values between 0.25 and 4.0 in our main analysis. For models A, B, C, D, and F, the fit preferred values of $A_G$ within this range, but in the case of model E the fit favored somewhat larger values of this quantity. To explore this further, we allowed $A_G$ to take on larger values in Model E$^{\prime}$, whose results we present in this appendix. The results for model E$^{\prime}$ are shown in Figs.~\ref{fig:CR_Eprime_model_panel} and~\ref{fig:CR_Eprime_model_landscape}.

\end{document}